\documentstyle[rotating,a4wide,epsfig,twoside,12pt]{article}
\renewcommand{\theequation}{\arabic{section}.\arabic{equation}}

\def\vecgr#1{\mathchoice{\mbox{\boldmath$\mathrm\displaystyle#1$}}
{\mbox{\boldmath$\mathrm\textstyle#1$}}
{\mbox{\boldmath$\mathrm\scriptstyle#1$}}
{\mbox{\boldmath$\mathrm\scriptscriptstyle#1$}}}

\def\vec#1{\mathchoice{\mathrm{\mathbf{\displaystyle#1}}}
{\mathrm{\mathbf{\textstyle#1}}}
{\mathrm{\mathbf{\scriptstyle#1}}}
{\mathrm{\mathbf{\scriptscriptstyle#1}}}}
\newcommand{\be}{\begin{equation}}
\newcommand{\ee}{\end{equation}}
\newcommand{\ba}{\begin{array}}
\newcommand{\ea}{\end{array}}
\newcommand{\bea}{\begin{eqnarray}}
\newcommand{\eea}{\end{eqnarray}}

\newsavebox{\TRS}
\sbox{\TRS}{\hspace{.5em} = \hspace{-1.8em}
                 \raisebox{1ex}{\mbox{\scriptsize TRS}} }

\newsavebox{\defgleich}
\sbox{\defgleich}{\ :=\ }

\newsavebox{\LSIM}
\sbox{\LSIM}{\raisebox{-1ex}{$\ \stackrel{\textstyle<}{\sim}\ $}}

\newsavebox{\GSIM}
\sbox{\GSIM}{\raisebox{-1ex}{$\ \stackrel{\textstyle>}{\sim}\ $}}
\newcommand{\gsim}{\usebox{\GSIM}}

\newcommand{\lk}{\left}
\newcommand{\rk}{\right}
\newcounter{saveeqn}

\newcommand{\ssty}{\scriptstyle}
\newcommand{\sssty}{\scriptscriptstyle}
\newcommand{\ra}{\,\rangle}
\newcommand{\la}{\langle\,}

\newcommand{\lkDk}{\lk(\!\!\lk(}
\newcommand{\rkDk}{\rk)\!\!\rk)}
\newcommand{\ptd}{\partial}
\newcommand{\ptdnd}[1]{\frac{\partial}{\partial #1}}

\newcommand{\ptdxndy}[2]{\frac{\partial #1}{\partial #2}}
\newcommand{\nab}{\vecgr{\nabla}}

\newcommand{\after}{\mbox{\raisebox{0.35ex}{$\ssty\circ$}}}
\newcommand{\cross}{\times}
\newcommand{\prop}{\propto}

\newcommand{\Tr}{\mbox{Tr$\,$}}

\newcommand{\eins}{\mbox{$1 \hspace{-1.0mm}  {\rm l}$}}
\newcommand{\eh}{\mbox{$\frac{1}{2}$}}
\newcommand{\dh}{\mbox{$\frac{3}{2}$}}

\newcommand{\ed}{\mbox{$\frac{1}{3}$}}

\newcommand{\da}{\mbox{$\frac{3}{8}$}}

\newcommand{\edwz}{\mbox{$\frac{1}{\sqrt{2}}$}}

\newcommand{\deld}{\delta^{\sssty (3)}}
\newcommand{\alp}{\alpha}
\newcommand{\bet}{\beta}
\newcommand{\gam}{\gamma}
\newcommand{\Gam}{\Gamma}
\newcommand{\del}{\delta}
\newcommand{\Del}{\Delta}
\newcommand{\eps}{\epsilon}

\newcommand{\lam}{\lambda}

\newcommand{\sig}{\sigma}
\newcommand{\Tet}{\Theta}
\newcommand{\tet}{\theta}

\newcommand{\balp}{{\bar\alpha}}
\newcommand{\bbet}{{\bar\beta}}
\newcommand{\etab}{{\bar\eta}}
\newcommand{\bgam}{{\bar\gamma}}

\newcommand{\hpv}{\mbox{$H_{\sssty\mathrm PV}$}}
\newcommand{\hpve}{\mbox{$H_{\sssty\mathrm PV}^{\sssty (1)}$}}
\newcommand{\hpvz}{\mbox{$H_{\sssty\mathrm PV}^{\sssty (2)}$}}

\newcommand{\heff}{\mbox{$H_{e\!f\!\!f}$}}
\newcommand{\heffn}{\mbox{$H_{e\!f\!\!f, 0}$}}
\newcommand{\heffe}{\mbox{$H_{e\!f\!\!f, 1}$}}

\newcommand{\HINT}{H^{\sssty INT}}

\newcommand{\HEXT}{H^{\sssty EXT}}

\newcommand{\mA}{m_{\sssty \!A}}
\newcommand{\ml}{m_{\ell}}

\newcommand{\qwe}{\mbox{$Q_{\sssty W}^{\sssty (1)}$}}
\newcommand{\qwz}{\mbox{$Q_{\sssty W}^{\sssty (2)}$}}
\newcommand{\qwez}{\mbox{$Q_{\sssty W}^{\sssty (1,2)}$}}
\newcommand{\qwi}{\mbox{$Q_{\sssty W}^{\sssty (i)}$}}

\newcommand{\sint}{\mbox{$\sin^2\theta_{\sssty W}$}}
\newcommand{\theW}{\theta_{\sssty W}}

\newcommand{\cB}{{\cal B}}

\newcommand{\cD}{{\cal D}}
\newcommand{\cE}{{\cal E}}

\newcommand{\cH}{{\cal H}}
\newcommand{\cN}{{\cal N}}
\newcommand{\cO}{{\cal O}}

\newcommand{\cR}{{\cal R}}
\newcommand{\cS}{{\cal S}}
\newcommand{\cT}{{\cal T}}

\newcommand{\deli}{\delta_i}

\newcommand{\dele}{\delta_1}
\newcommand{\delz}{\delta_2}
\newcommand{\delez}{\delta_{1,2}}

\newcommand{\wdele}{\mbox{$\sqrt{\delta_1}$}}

\newcommand{\wdelez}{\mbox{$\sqrt{\delta_{1,2}}$}}

\newcommand{\unC}{\mbox{$\underline{C\!}\,$}}
\newcommand{\unCt}{\mbox{$\underline{C\!\!\:t\!}\,$}}

\newcommand{\unD}{\mbox{$\underline{D\!}\,$}}
\newcommand{\unDn}{\mbox{$\underline{D\!}\,$}_0}

\newcommand{\unDd}{\mbox{$\underline{D\!}\,$}_3}

\newcommand{\unDv}{\mbox{$\underline{\vec{D}}$}}
\newcommand{\uncD}{\mbox{$\underline{\cal D\!}\,$}}

\newcommand{\uneins}{\mbox{$\underline{1\hspace{-1.0mm} {\rm l}}$}}

\newcommand{\unF}{{\underline{F}}}

\newcommand{\unFz}{{\underline{F\!}\,}_2}
\newcommand{\unFd}{{\underline{F\!}\,}_3}

\newcommand{\unGa}{\mbox{$\underline{\Gamma}$}}

\newcommand{\unHhh}{\mbox{$\doublehat{{\underline{H\!}\,}}$}}
\newcommand{\unHtil}{\mbox{$\widetilde{{\underline{H\!}\,}}$}}

\newcommand{\unmuv}{\mbox{$\underline{\muv}$}}

\newcommand{\unM}{\mbox{$\underline{\cal M\!}\,$}}
\newcommand{\uncM}{\mbox{$\underline{\cal M\!}\,$}}
\newcommand{\unMn}{\mbox{$\underline{\cal M\!}\,_{\sssty 0}$}}

\newcommand{\unMc}{\mbox{$\underline{M\!}\,_{c}$}}

\newcommand{\Pro}{{ \mbox{\,P$\!\!\!\!\!${\rm I}$\,\,\,$} }}
\newcommand{\Protil}{{ {\widetilde{\mbox{\,P$\!\!\!\!\!${\rm I}$\,\,\,$}}} }}

\newcommand{\unPro}{{\mbox{\underline{\,P$\!\!\!\!\!${\rm I}$\,\,$}$\,$}}}

\newcommand{\QPro}{\mbox{Q$\!\!\!\!\!\:${\small \sf  
l\normalsize}$\,\,\,$}}

\newcommand{\unR}{\mbox{$\underline{\cal R\!\!\:}\,$}}
\newcommand{\uncR}{\mbox{$\underline{\cal R\!\!\:}\,$}}

\newcommand{\unS}{\mbox{$\underline{S\!}\,$}}

\newcommand{\uncT}{\mbox{$\underline{\cal T\!}\,$}}
\newcommand{\unrT}{\mbox{$\underline{T\!}\,$}}
\newcommand{\unT}{\mbox{$\underline{\cal T\!}\,$}}

\newcommand{\unTg}{\mbox{$\underline{T\!g\!}\,$}}

\newcommand{\unU}{\mbox{$\underline{U\!}\,$}}

\newcommand{\unV}{\mbox{$\underline{V\!}\,$}}

\newcommand{\lhat}{\hat l}

\newcommand{\Hh}{\hat H}
\newcommand{\Kh}{\hat K}
\newcommand{\Oh}{\hat O}

\newcommand{\doublehat}[1]{
  \hat{\hat{\mbox{\phantom{$#1$}}}}
  \!\!\!\!\!\!\!\: #1
  }

\newcommand{\Hhh}{\doublehat{H}}

\newcommand{\brak}[1]{\mbox{$\la#1\,|$}}
\newcommand{\ket}[1]{\mbox{$ |\,#1\ra$}}

\newcommand{\EINT}{E^{\sssty INT}}
\newcommand{\EEXT}{E^{\sssty EXT}}

\newcommand{\EKcEvdedz}{E(\cEv,\dele,\delz)}

\newcommand{\EiKaFdEevd}{E^{\sssty INT}(\alpha,F_3,\cE\evd)}

\newcommand{\Atil}{\widetilde A}

\newcommand{\ctil}{\tilde c}
\newcommand{\Ctil}{\widetilde C}
\newcommand{\Etil}{\widetilde E}
\newcommand{\etatil}{\tilde\eta}

\newcommand{\Htil}{\widetilde H}
\newcommand{\cHtil}{\widetilde \cH}
\newcommand{\ntil}{\tilde n}
\newcommand{\cTtil}{\widetilde \cT}
\newcommand{\cRtil}{\widetilde \cR}

\newcommand{\AkT}{A_{\mbox{\tiny T}}}
\newcommand{\AkR}{A_{\mbox{\tiny R}}}
\newcommand{\AvT}{\vec{A}_{\mbox{\tiny T}}}
\newcommand{\AvR}{\vec{A}_{\mbox{\tiny R}}}

\newcommand{\av}{\vec{a}}
\newcommand{\Av}{\vec{A}}

\newcommand{\Bv}{\vec{B}}

\newcommand{\Dv}{\vec{D}}

\newcommand{\eve}{\vec{e}_1}
\newcommand{\evz}{\vec{e}_2}
\newcommand{\evd}{\vec{e}_3}

\newcommand{\cEv}{\vecgr{\cal E}}
\newcommand{\cEvP}{\cEv_{\mbox{\tiny P}}}
\newcommand{\cEvR}{\cEv_{\mbox{\tiny R}}}

\newcommand{\cEsvs}{\cEv'^{\sssty (\sigma)}}

\newcommand{\cER}{{\cal E}_{\mbox{\tiny R}}}

\newcommand{\etav}{\vecgr{\eta}}
\newcommand{\etavb}{{\bar{\etav}}}
\newcommand{\etavR}{\etav_{\mbox{\tiny R}}}

\newcommand{\Iv}{\vec{I}}

\newcommand{\muv}{\vecgr{\mu}}
\newcommand{\nv}{\vec{n}}

\newcommand{\pv}{\vec{p}}

\newcommand{\Pv}{\vec{P}}

\newcommand{\Rv}{\vec{R}}

\newcommand{\Rvdot}{\dot{\vec{R}}}

\newcommand{\sigv}{\vecgr{\sigma}}

\newcommand{\xv}{\vec{x}}

\newcommand{\xvA}{\xv_{\sssty \!A}}
\newcommand{\xvl}{\xv_{\ell}}

\newcommand{\Uzepp}{\unU^{\sssty (2,1)++}}
\newcommand{\Uzemm}{\unU^{\sssty (2,1)--}}
\newcommand{\Uzepm}{\unU^{\sssty (2,1)+-}}
\newcommand{\Uzemp}{\unU^{\sssty (2,1)-+}}
\newcommand{\Uezpp}{\unU^{\sssty (1,2)++}}
\newcommand{\Uezmm}{\unU^{\sssty (1,2)--}}
\newcommand{\Uezpm}{\unU^{\sssty (1,2)+-}}
\newcommand{\Uezmp}{\unU^{\sssty (1,2)-+}}

\newcommand{\Ustpmpm}{\unU^{\sssty (\sig,\tau)\pm\pm}}

\newcommand{\Ustpmmp}{\unU^{\sssty (\sig,\tau)\pm\mp}}

\newcommand{\Ustrs}{\unU^{\sssty (\sig,\tau)r\,s}}
\newcommand{\Utsmsmr}{\unU^{\sssty (\tau,\sig)(-s)(-r)}}
\newcommand{\Utssr}{\unU^{\sssty (\tau,\sig)s\,r}}

\newcommand{\iKkap}{{\sssty (\kappa )}}
\newcommand{\iKrho}{{\sssty (\rho )}}
\newcommand{\iKsig}{{\sssty (\sigma )}}
\newcommand{\iKtau}{{\sssty (\tau )}}
\newcommand{\iKi}{{\sssty (i)}}

\newcommand{\iKen}{{\sssty (n)}}
\newcommand{\iKn}{{\sssty (0)}}
\newcommand{\iKe}{{\sssty (1)}}
\newcommand{\iKz}{{\sssty (2)}}
\newcommand{\iKd}{{\sssty (3)}}
\newcommand{\iKv}{{\sssty (4)}}
\newcommand{\iKvar}[1]{{\sssty (#1)}}
\newcommand{\isKvar}[1]{{'\sssty (#1)}}
\newcommand{\ikvar}[1]{\mbox{\raisebox{0.2ex}{$\ssty (#1)$}}}

\newcommand{\ivar}[1]{{\sssty #1}}

\newcommand{\vzHep}{\left.^4_2\mbox{He}^+\right.}
\newcommand{\eeH}{\left.^1_1\mbox{H}\right.}

\newcommand{\lbra}[1]{\mbox{$ ( \widetilde{#1}$}}
\newcommand{\lbrak}[1]{\mbox{$ ( \widetilde{#1}|$}}
\newcommand{\rket}[1]{\mbox{$ | #1 )$}}

\newcommand{\evrzza}{\mbox{$ | {2,a})$}}
\newcommand{\evrzzas}{\mbox{$ | {2,a'})$}}

\newcommand{\evrbet}{\mbox{$ | {\beta})$}}

\newcommand{\levlE}{\mbox{$ ( \widetilde{\alpha,F_3 ,{\cal E}\evd}|$}}
\newcommand{\revrE}{\mbox{$ | \alpha,F_3,{\cal E}\evd )$}}

\newcommand{\levlRvar}[3]{\mbox{$
                       ( \widetilde{#1,#2,{\cal E}\evd},#3 |$}}
\newcommand{\revrRvar}[3]{\mbox{$ | #1,#2,{\cal E}\evd ,#3 )$}}

\newcommand{\revrnE}{\mbox{$ | \alpha,F_3,{\cal E}\evd,\vec{n})$}}

\newcommand{\WWA}{Wigner-Weisskopf-approximation\ }

\date{}
\begin{document}
{\sloppy
\begin{titlepage}
%

\title{
{\normalsize 
\hfill HD--THEP--97--36
}
\vspace{1cm}
\\
{\LARGE\bf\sf Parity Violating\\
Energy Shifts and Berry Phases in Atoms, I\thanks{
Work supported by Deutsche Forschungsgemeinschaft, Project No.\ Na 296/1--1} \stepcounter{footnote}}
}

\author{
{\sc 
D. Bru\ss,\thanks{Present address: ISI, Villa Gualino, Viale Settimio Severo 65, 10133 Torino, Italy.}\ \
T. Gasenzer,\thanks{Supported by Cusanuswerk}\ \ and 
O. Nachtmann}
}

\date{\small\sl 
Institut  f\"ur Theoretische Physik, Universit\"at Heidelberg\\
Philosophenweg 16, D-69120 Heidelberg, Germany
}

\maketitle

\begin{center}
\parbox[t]{\textwidth}{\small
We present a study of parity (P) violating contributions to the eigenenergies of stationary systems containing atoms in spatially inhomogeneous external electric fields. In this context the subtle interplay of P-violation and time reversal (T) invariance plays an important role. If the entire field configuration is chosen to exhibit chirality the energies are in general shifted by pseudoscalar contributions which change sign under a planar reflection of the field. 
In part I we consider sudden variations of the fields and calculate P-violating energy shifts using perturbation theory.
In part II the adiabatic case will be treated and the connection to geometrical (Berry-) phases will be elucidated.
To calculate the effects we use the standard model of elementary particle physics where the P-odd interaction arises through the exchange of Z-bosons between the quarks in the nucleus and the atomic electrons. We consider in detail hydrogen-like systems in unstable levels of principal quantum number $n=2$. We study atoms with vanishing nuclear spin like $\vzHep$ and with nuclear spin $I=1/2$ like $\eeH$. The nominal order of P-violating effects is $10^{-5}\!\!\dots 10^{-9}\,$Hz which is determined by the mixing of the $2S_{1/2}$ and $2P_{1/2}$ states. However we point out that with certain configurations of the external fields, it is possible to enhance the P-violating energy shifts dramatically!
Instead of energy shifts linear in the P-violation parameters we get then shifts proportional to the square root of these parameters. Numerically we find such energy shifts which only appear for unstable states to be of order $10^{-5}...\,1\,\,$Hz. Under a reversal of the handedness of the external field configuration these P-violating shifts get multiplied by a phase factor $i$, i.e.\ the shifts in the real and imaginary part of the complex eigenenergies are exchanged. 
Application of our technique to hydrogen-like atoms with a nucleus of spin $I=1/2$ yields P-violating energy shifts which are very sensitive to the nuclear spin dependent P-odd force, which receives a rather large contribution from the polarized strange quark density in polarized nuclei. 
Thus, a measurement of these energy shifts could provide an important tool to elucidate nuclear properties connected to the so called ``spin crisis''.  
We also present a method for treating degenerate perturbation theory which combines advantages of both, Kato's and Bloch's methods.
} 
\end{center}

\end{titlepage}

%
%
\section{Introduction}
\label{secIntro}
\setcounter{equation}{0}
In this paper we will discuss various possibilities to obtain parity (P)  
violating energy shifts for atomic systems due to effects of the weak neutral  
current (NC) interactions. Energy shifts, i.e.\ frequencies can be measured in  
atomic physics with very high accuracy. Thus P-violating energy shifts could  
open up a new window for measurements of the tiny NC effects in atoms. There  
are some ``no go'' theorems for such energy shifts which are related to time  
reversal (T) invariance. However, each theorem needs some assumptions, and in  
the following we will show how the ``no go'' theorems can be avoided by  
considering situations where these assumptions are not fulfilled.\\

In our paper we will stay in the framework of the Standard Model (SM) of  
elementary particle physics \cite{GWS67} which led to detailed predictions of  
P-violating NC effects in atoms \cite{Bouchiat74,Moskalev76}. By now such  
effects are well studied theoretically and have been verified experimentally  
for heavy atoms containing many electrons (for reviews cf.\  
\cite{Khriplovich91,Bouchiat97,Budker98}). But P-violating effects have not been measured yet for  
hydrogen-like atoms, where theoretical calculations can be pushed to very  
high accuracy \cite{Sapirstein90}. Proposals for P-violation experiments with  
hydrogenic systems have been made e.g.\ in \cite{Dunford78}-\cite{BBN95}.
In \cite{BBN95}, a variety of P-violating polarization rotations was studied  
theoretically. It was also shown that, in principle, P-violating  
contributions to the energy levels of (unstable) hydrogen-like atoms in  
suitable external electric fields should exist. In the present work we will  
address that question in detail.

In the days of high precision tests of the SM at high
energies, e.g.\ at LEP and SLC (cf.\ \cite{Carter92} for a review), we should  
list some motivations for our study:

(1) In order to check the ``running'' of the SM couplings as it is predicted  
by the renormalization theory, atomic systems should be useful, providing low  
energy information complementary to high energy physics measurements. In this  
way a very sensitive test of higher order effects in the SM
and a probing for new physics beyond the SM is possible.

(2) P-violating effects in atoms can give interesting information on  
properties of atoms and nuclei. The experimental study of anapole moments of nuclei has recently become possible in this way \cite{Wood97,Budker98}.
P-violation in atoms may also shed light on the so-called
``spin crisis'' for the nucleons \cite{Ashman89,Adams94}.

(3) In part II we will show how P-violating energy shifts in atoms can
emerge as a topological effect in quantum mechanics, i.e.\ as P-violating  
geometrical (Berry-) phases. To study such effects should be interesting in  
itself.

(4) There are many recent advances in experimental techniques of atomic physics,
like the production and storage of ``cold'' heavy ion beams in storage rings
(cf.\ \cite{Habs93} and references therein) and the observation of interference
effects with atomic beams (cf.\ \cite{Mlynek92} for a review). Recent  
progress has been made on techniques of atomic mirrors \cite{Roach95Hughes97}, microtraps for ultracold atoms \cite{MagneticMicrotraps} and of semiconductor nano-structures, 
which could be used to realize experimental configurations as they will be  
discribed below. One could also imagine applications of the recently observed carbon nanotubes \cite{Ebbesen96} or of the football-shaped fullerenes \cite{Kroto85Kraetschmer90}. In a natural way they represent cavities with a diameter of a few $\mathrm{\AA}$, which are required for the purpose of measuring our P-violation effects for atoms being stored inside them.\\

This paper continues the work done in \cite{BN83,BBN95} and much of the  
formalism will be taken over from there. We are studying excited  
hydrogen-like atomic systems within external electric and magnetic field configurations.  
In the following we will briefly recall the methods including the \WWA  
\cite{Weisskopf30} for unstable states.
We then examine P-violating energy shifts of atoms in external electric  
fields for two basically different experimental setups with either sudden or  
adiabatically smooth local variations of the electric field vector. Part I will be restricted to the sudden approximation case, while the adiabatic situation will be studied in part II which is in preparation. 
In Section \ref{sec2} we consider the case of an atom inside a box with two segments only and show that there are no P-violating energy shifts due to T-invariance.
In Section \ref{sec3} we present results from perturbation theory for boxes with three segments. 
Our conclusions are drawn in Section \ref{secConcl}, where we also discuss the relation of the P-violating energy shifts presented in our paper to those in chiral molecules. 

A short account of some of the results presented in detail here has been published in \cite{BGNLetter97}.

From the methodological point of view we deal in Section \ref{sec3} with degenerate perturbation theory, which has been treated in the classical papers by Kato and Bloch \cite{Kato49,Bloch58} and which is masterly explained in \cite{Messiah}. We introduce a variant of degenerate perturbation theory which we find very convenient four our problems and which should be interesting for many other applications. Our ansatz is analogous to the introduction of normal coordinates in the problem of small oscillations in mechanics (cf.\ e.g.\ \cite{Goldstein}).

In part II, which is in preparation, we will describe P-violating effects in smoothly varying fields using a  
WKB-approximation. We relate the energy shifts of stationary systems via  
boundary conditions back to geometrical phases which the atomic wave functions  
acquire in the external field, whose topological properties in a suitable  
space are crucial. 

All our formulae are written in natural units, $\hbar = c = 1$, if other  
units are not explicitly indicated.

%
\subsection{P-violation in hydrogen-like atomic systems in external fields}
\label{sec1.1}
We consider a hydrogen-like system, i.e.\ a nucleus with a single electron or  
muon around it. It is crucial for our purpose to consider the coupling of the  
internal motion of the atom, in general
in an unstable state, with the c.m.\ motion of the atom as
a whole. Thus we will first discuss the formalism which we will
use for describing this situation. For simplicity we will assume throughout the paper, except in Sect.\ \ref{sec1.3},
that for both, the c.m.\ motion and for the internal atomic
motion, we can make the non-relativistic approximation.

Our system consists of two charged particles, the lepton $\ell$
and the nucleus $A$ with $Z$ protons and $N$ neutrons, where
$\ell$ and $A$ are bound together and experience an external
electromagnetic field described by a 4-vector potential $(A^0(x),\Av(x))$.
As basic dynamic variables we have the c.m.\ coordinate $\Rv$ and the
relative coordinate $\xv$:
\bea
   \Rv &=& \frac{\ml\xvl+\mA\xvA}{\ml+\mA},\nonumber\\
   \xv &=& \xvl - \xvA,
\eea
where $\xvl,\xvA$ are the position vectors of $\ell,A$, respectively. A simple
exercise shows that in the non-relativistic case\footnote{A 
detailed discussion of the c.m.\ and relative coordinates for relativistic systems including spin can be found in \cite{Osborn68}.}
the Hamiltonian of the system can be written
as follows:
\bea
    H &=& \frac{1}{2m}\lk(\Pv-q\Av(\Rv,t)\rk)^2 + qA^0(\Rv,t)\nonumber\\
      & & +\,\HINT - \Dv\cdot\cEv(\Rv,t) - \muv\cdot\Bv(\Rv,t)\nonumber\\
      & & +\,H_0^{(\gam)} + H_{rad}
\label{eq1.2}
\eea
Here $m$ and $q$ are the total mass and charge of the atom:
\bea
    m &=& \mA + \ml\nonumber\\
    q &=& (Z-1)\,e.
\eea
We denote by $e$ $(e>0)$ the proton charge.
The components of the momentum operator for the c.m.\ motion are
\be
   P_j = \frac{1}{i}\ptdnd{R_j},\quad (j=1,2,3).
\ee
The Hamiltonian for the internal motion in the c.m.\ system
in the absence of external fields is denoted by $\HINT$. The
electric and magnetic dipole moment operators are
$\Dv$ and $\muv$, respectively, and $\cEv(\Rv,t)$, $\Bv(\Rv,t)$ are the  
electric and magnetic fields at the c.m.\ position:
\bea
    \cEv(\Rv,t) &=& -\nab A^0(\Rv,t)-\ptdnd{t}\Av(\Rv,t),\nonumber\\
     \Bv(\Rv,t) &=& \nab\cross\Av(\Rv,t).
\eea
Finally $H_0^{(\gam)}$ is the Hamiltonian of the free photons and $H_{rad}$
describes the coupling of the atom to the photons.

In (\ref{eq1.2}) we have neglected the variation of the electric and magnetic
fields over the atomic dimensions. We have also neglected terms
proportional to $\Rvdot\cross\cEv$ and $\Rvdot\cross\Bv$ coming from the fact  
that the
electric and magnetic fields felt by the atom are those in
its actual rest system. Such terms give rise to the motional
Stark and Zeeman effects. Thus for our purpose the motion of the
atom should be very slow. Since we are interested here in the
principle of P-violating energy shifts, we want to study
the simplest possible cases. We only have to make sure
that our approximations do not introduce any spurious sources
of P-violation.

Let us now have a closer look at the Hamiltonian of the
internal atomic motion which we write as
\be
   \HINT = \HINT_0 + \hpv,
\ee
where $\HINT_0$ is the P-conserving piece and $\hpv$ the P-violating
one.

In the framework of the SM the relevant effective P-violating Hamiltonian  
$\hpv$ is --- to leading order --- due to $Z$ exchange between 
the atomic lepton  
and the quarks in the nucleus. In the non-relativistic limit one finds (cf.\  
e.g.\ \cite{Khriplovich91}; we use the notation as in Section 2 of  
\cite{BBN95}):
\bea
   \hpv  &=& \hpve + \hpvz, \\
   \hpve &=& \frac{G}{4\sqrt{2}} \frac{1}{m_\ell } \qwe  (Z,N) \cdot
                  \left\{ \del^3(\xv) (\sigv \cdot \pv) +
                 (\sigv \cdot
              \pv)  \del^3(\xv)   \right\}, \\
   \hpvz &=& \frac{G}{4\sqrt{2}} \frac{1}{m_\ell } \qwz (Z,N) \cdot
                 \left\{ \del^3(\xv) (\Iv \cdot \sigv)
                 (\sigv \cdot \pv) + (\sigv \cdot \pv)
                  (\Iv \cdot \sigv)
                   \del^3(\xv)   \right\},
\eea
where $\hpvz$ depends on the nucleus' spin $\Iv$ and $\hpve$ does not. $G$ is  
Fermi's constant, furthermore, $\pv$ and $\sigv$ are the momentum and spin  
operators of the lepton $\ell$ and $\qwez$ are the weak charges of the nucleus
\bea
\label{eq1.4}
   \qwe (Z,N) & = & Z\,(1-4 \,\sint )-N  ,\\
   \qwz (Z,N) & = & -\frac{1}{I}\, (1-4 \,\sint )\left[ \Delta u(Z,N) -
                  \Delta d(Z,N) -\Delta s(Z,N) \right],
\label{eq1.5}
\eea
where $\theW$ is the weak angle and the $\Delta q$ are defined as the ``total  
polarizations'' carried by the quark species $q=u,d,s$. It is assumed in  
(\ref{eq1.5}) that possible contributions of the $c,b$ and $t$ quarks in the  
nucleus can be neglected. We refer the reader to \cite{BBN95} and to the  
references therein for more details.

The energy levels of the hydrogen-like atom in the Coulomb approximation
are $1S$, $2S$, $2P$ etc., where $2S$ and $2P$
are degenerate. These levels are split by the fine structure, the Lamb shift and
further QED effects as well as by the hyperfine structure if the nuclear spin  
$I\not=0$. The resulting near degeneracy of the opposite parity states $2S$  
and $2P$ makes the $n=2$ levels in principle ideal candidates for looking for  
P-violating effects. Also, the $2S$ state is long lived. Thus we will  
concentrate on the atoms in $n=2$ states in the following.

The relevant dimensionless parameters measuring the ``wrong''
parity admixture in the $n=2$ states are $\deli(Z,N),\,i=1,2$, defined in  
(3.17) of \cite{BBN95},
which are proportional to $\qwi(Z,N)$. For electronic atoms we have
approximately (cf.\ (3.21) of \cite{BBN95}):
\be
\label{eq1.5a}
  \deli(Z,N) \simeq -\qwi(Z,N)\cdot 6.14\times 10^{-12}.
\ee

Let us discuss next the Hilbert space of state vectors
for our system. We will consider atoms in $n=2$ states, denoted
$|2,a),\,a=1,2,...$ (cf.\ Sect.\ 3.1 of \cite{BBN95}), and at arbitrary c.m.\  
position $\Rv$, and their
decay products, i.e.\ atoms in $n=1$ states plus photons.
A basis in our Hilbert space is then provided by the vectors
\bea
   &&|\Rv;\,2,a\ra,\nonumber\\
   &&|\Rv;\,1,b;\,\gam,k\ra,
\label{eq1.5.2}
\eea
where we label by $k=1,2,...$ a convenient basis set of photon
states and by $b=1,2,...$, the internal $n=1$ atomic
states. We assume these states to be normalized to
%
\bea
    \la\Rv';\,2,a'\,|\,\Rv;\,2,a\ra
    &=& \deld(\Rv'-\Rv)\,\del_{a'a}\nonumber\\
    \la\Rv';\,1,b';\,\gam,k'\,|\,\Rv;\,1,b;\,\gam,k\ra
    &=& \deld(\Rv'-\Rv)\,\del_{b'b}\,\del_{k'k},
\eea
with all other scalar products vanishing.

Consider now a general state vector $|\,t\ra$ at some time $t$. The part
of the state vector describing the undecayed $n=2$ atoms
is (cf.\ (3.6), (3.7) of \cite{BBN95}):
\be
   \Psi(\Rv,t) = \sum_a\,\psi_a(\Rv,t)\,|\,2,a\,),
\ee
where
%
\be
   \psi_a(\Rv,t) = \la\Rv;\,2,a\,|\,t\ra.
\ee
The wave function $\Psi$ can be thought of as a
Schr\"odinger wave function for the c.m.\ motion, but
with a number of components equal to the number $N$
of linearly independent $n=2$ states labeled by $a$. Thus
we can also write $\Psi$ as an $N$-component wave function:
\be
  \Psi(\Rv,t)\quad\longleftrightarrow\quad
  \lk(\ba{c} \psi_1(\Rv,t)\\ \psi_2(\Rv,t)\\ \vdots\\ \psi_N(\Rv,t)\ea\rk).
\label{eq1.5.6}
\ee
Similarly, the $n=1$ states plus photons are described by multi-component  
wave functions (cf.\ (3.8) of \cite{BBN95})
\be
   \Phi_k(\Rv,t) = \Big(\la\Rv;\,1,b;\,\gam,k\,|\,t\ra\Big).
\ee
The general methods \cite{Weisskopf30} for obtaining the non-hermitian mass
matrix can now be applied (cf.\ Sect.\ 3.2 of \cite{BBN95}). Starting from  
the Hamiltonian (\ref{eq1.2}) this gives us an effective Schr\"odinger  
equation for the state vector $\Psi(\Rv,t)$ as follows:
\be
   i\ptdnd{t}\Psi(\Rv,t) = (\heff\Psi)(\Rv,t),
\label{eq1.5.8}
\ee
where $\heff$ is a non-hermitian operator:
\bea
   \heff       &=& \frac{1}{2m}\lk(\Pv-q\Av(\Rv,t)\rk)^2+q\,A^0(\Rv,t)
                   +\unM(\Rv,t),
\label{eq1.5.9}\\
   \unM(\Rv,t) &=& \unMn - \unDv\cdot\cEv(\Rv,t)-\unmuv\cdot\Bv(\Rv,t),\\
   \unMn       &=& \Big((\,2,a\,|\,\HINT_0+\hpv\,|\,2,b\,)\Big)
                   -\frac{i}{2}\,\unGa,\nonumber\\
   \unDv       &=& \Big((\,2,a\,|\,\Dv\,|\,2,b\,)\Big),\nonumber\\
   \unmuv      &=& \Big((\,2,a\,|\,\muv\,|\,2,b\,)\Big).
\label{eq1.5.11}
\eea
As in \cite{BBN95}, quantities with a bar underneath denote matrices in
the space of $n=2$ states. Thus $\unDv$ and $\unmuv$ are the corresponding
matrices of the electric and magnetic dipole operators,
respectively. In (\ref{eq1.5.11}) the matrix $\unGa$ is the decay rate matrix  
for the $n=2$ states as defined in (3.13) of \cite{BBN95}. Here we have  
assumed that the decay rates from specific states $|\,2,a\,)$ do {\it not}
depend on their c.m.\ motional state. This is an approximation
which neglects e.g.\ relativistic time dilatation effects. But
again, no spurious P-violation is introduced in this way.

In the following we will be concerned with finding solutions
to (\ref{eq1.5.8}) for various field configurations. In particular, we will
be interested in electric and magnetic field configurations which are
constant in time and consider the eigenstates $\Psi(\Rv)$ of $\heff$ for
this situation:
\be
   \heff\Psi(\Rv) = E\,\Psi(\Rv).
\label{eq1.5.12}
\ee
In general the eigenvalue $E$ is, of course, complex. The wave
functions $\Psi(\Rv)$ satisfying (\ref{eq1.5.12}) correspond to states with  
single exponential decay. We note that a non-hermitian operator $\cO$ does not need to be diagonalizable, i.e.\ in general there will be no linear operator $V$ satisfying $V^{-1}\cO V=$ diagonal operator. A finite $N$-dimensional non-hermitian matrix is always diagonalizable if it has $N$ different eigenvalues. In the case of degeneracies of eigenvalues diagonalization is in general not possible even for finite $N$, and the right (left) eigenstates form no longer a basis (cf.\ e.g.\ \cite{Smirnov49}). In the following we will in the general analysis always {\it assume} that all non-hermitian matrices/operators occurring are diagonalizable. In the specific examples discussed we check explicitly that this assumption is valid.

The basic question which we study in this paper is as
follows: under what circumstances can the complex energy
eigenvalue $E$ have {\it linear terms} in the P-violation
parameters $\delez(Z,N)$. Since these parameters are very small, we will
neglect terms quadratic in $\delez$ in our calculations below. How
could experimentalists look for terms linear in $\delez$? The typical
strategy which we have in mind is as follows. Consider an electric
field arrangement $\cEv(\xv)$ and the parity-transformed one:
\be
  \cEvP(\xv) = -\cEv(-\xv).
\ee
Let $\EKcEvdedz$ be the energy eigenvalue of some level in the field  
$\cEv(\xv)$ and $E(\cEvP,\dele,\delz)$ the eigenvalue of the corresponding  
level in the field $\cEvP(\xv)$.
The basic observable to study experimentally is the {\it energy difference}:
\be
   \Del E = \EKcEvdedz - E(\cEvP,\dele,\delz).
\label{eq1.5.14}
\ee
Clearly, in order for $\Del E$ to have a chance to be nonzero,
the field configurations $\cEv(\xv)$ and $\cEvP(\xv)$ must be essentially  
different, i.e.\ they should not be transformable into each other by
a proper rotation. In other words: the field configuration
$\cEv(\xv)$ must have {\it handedness} or {\it chirality}.\\

Now we note that from considering a P-transformation we get:
\be
   \EKcEvdedz = E(\cEvP,-\dele,-\delz).
\ee
This leads to
\be
   E(\cEv,0,0) = E(\cEvP,0,0).
\ee
If we first exclude the case where there is a degeneracy
of levels for $\delez=0$, the eigenvalues $E$ are differentiable in
$\delez$ at this point (cf.\ appendix C of \cite{BBN95}) and we get:
\be
   \Del E = 2\sum_{i=1,2}\,\deli\,\lk.\ptdxndy{\EKcEvdedz}{\deli}
              \rk|_{\delez=0}
            + \cO(\dele^2,\dele\delz,\delz^2).
\ee
Thus, the measurable asymmetry (\ref{eq1.5.14}) is directly related to the
terms in the energy eigenvalues $\EKcEvdedz$ which are linear in $\delez$.
In practice degeneracies of levels for $\delez=0$ {\it do}
occur and have to be dealt with carefully. As we will show, such degeneracies can be used to obtain P-violating energy differences (\ref{eq1.5.14}) which are proportional to $\sqrt{\deli}$. Thus we get huge enhancements of P-violating energy shifts of order $1/\sqrt{\deli}\simeq 10^6$ for special external field arrangements. Varying the fields around these special values one should observe sharp resonance type phenomena. 

Of course, instead of the asymmetry (\ref{eq1.5.14}) we could consider
the asymmetry of the energies in the field $\cEv(\xv)$ and a field
$\cEvR(\xv)$ obtained from $\cEv$ by reflection on some plane. Maybe it
is also possible to make a continuous change from $\cEv(\xv)$ to $\cEvP(\xv)$
or $\cEvR(\xv)$ and observe the corresponding continuous change of
the energy eigenvalues.


\subsection{Atoms in a constant electric field}
\label{sec1.2}
Let us consider a simple case to start with: an atom in
a constant electric field $\cEv$. We assume the atom to be
confined inside a region $\cB$ which can be of
arbitrary shape. This confinement of the atom in $\cB$
could, for instance, be realized by having the walls
of a suitable trap containing the atom built as ``mirrors''
(cf.\ \cite{Roach95Hughes97}). We represent the ``vessel'' or ``trap'' containing the atom by a potential $V_b(\Rv)$ which is zero inside the vessel
and infinite outside. This leads to the following boundary
condition for the wave function $\Psi(\Rv)$:
\be
   \Psi(\Rv) = 0 \quad\mbox{for}\quad\Rv\in\ptd\cB
\label{eq1.5.18}
\ee
where $\ptd \cB$ denotes the boundary of $\cB$.
In the following (\ref{eq1.5.18}) is always understood to be imposed.
The effective Hamiltonian (\ref{eq1.5.9}) reads now
\be
   \heff = \frac{\Pv^2}{2m}-q\,\Rv\cdot\cEv+\unMn-\unDv\cdot\cEv.
\label{eq1.5.19}
\ee
The eigenvalue equation (\ref{eq1.5.12})
%
can easily be discussed for this case. Clearly, we can diagonalize
the operators governing the c.m.\ motion,
\be
   \HEXT = \frac{\Pv^2}{2m}-q\,\Rv\cdot\cEv,
\ee
and the internal motion,
\be
   \unM = \unMn-\unDv\cdot\cEv,
\ee
separately, form the tensor product of the corresponding
eigenvectors and add the corresponding eigenvalues. Indeed, let
$
%
   \phi_n(\Rv),\,n=1,2,...,
$
be a complete set of orthonormal eigenvectors of $\HEXT$:
\be
   \HEXT\,\phi_n(\Rv) = \EEXT_n\,\phi_n(\Rv),
\ee
where $\EEXT_n$ are the eigenvalues which are, of course, real
since $\HEXT$ is a hermitian operator. Let $\evrbet$
be the right eigenvectors of $\unM$ to eigenvalues $\EINT(\beta)$, which
can be complex:
\be
   \unM\,\evrbet = \EINT(\beta)\,\evrbet,\quad(\beta=1,2,...,N).
\ee
Then the eigenvectors $\Psi$ of $\heff$ (\ref{eq1.5.19}), (\ref{eq1.5.12}) are
\be
   \Psi_{\beta,n}(\Rv) = \phi_n(\Rv)\,\evrbet \equiv |\,\beta,n\,),
\ee
with eigenvalues
\be
   E(\beta,n) = \EEXT_n + \EINT(\beta),
\label{eq1.5.27}
\ee
i.e.\ we have
\be
   \heff\,\Psi_{\beta,n}(\Rv) = E(\beta,n)\,\Psi_{\beta,n}(\Rv).
\ee

Here we can assume without loss of generality that
the electric field points in 3-direction:
\be
   \cEv=\cE\evd.
\label{eq1.5.29}
\ee
The eigenvalues and eigenvectors of the corresponding mass matrix
\be
   \unM = \unMn - \unDd\cE
\label{eq1.5.30}
\ee
were discussed extensively in \cite{BBN95}. For later use and to fix the
notations we recall the main features here.

For an electric field (\ref{eq1.5.29}) the mass matrix $\unM$  
(\ref{eq1.5.30}) commutes
with the 3-component $\unFd$ of the  total angular momentum operator:
\be
   [\unM , \unF_3] =0.
\label{eq1.9}
\ee
We can diagonalize $\unM$ and $\unFd$ simultaneously. The right eigenvectors  
$\revrE$ satisfy:
\bea
    \unFd \, \revrE & = &  F_3 \, \revrE  ,\nonumber \\
    \unM \, \revrE & = &  \EiKaFdEevd \, \revrE;
\label{eq1.10}
\eea
the left eigenvectors $\levlE$ satisfy:
\bea
   \levlE\unFd &=& \levlE F_3   ,\nonumber \\
   \levlE\unM &=& \levlE\EiKaFdEevd.
\label{eq1.11}
\eea
Here we label the eigenvectors by $(\alp,F_3)$ instead of $\beta$ as above,  
with $\alp=1,2,...$ numbering the eigenvectors to the same $F_3$. We will  
assume
that there is no degeneracy of the complex energy levels for fixed $F_3$:
\be
\label{eq1.12}
   \EiKaFdEevd\not = \EINT(\alp',F_3,\cE\evd)\quad {\rm for}\quad
                     \alp\not=\alp'.
\ee
This
is true for all systems in external fields which we will consider.
The condition (\ref{eq1.12})  guarantees that the
right eigenvectors (\ref{eq1.10}) (left eigenvectors (\ref{eq1.11})) form a  
basis (the dual basis) in the $n=2$ subspace where the normalization can be  
chosen such that
\be
   \levlE\alp',F_3',\cE\evd) = \del_{\alp,\alp'} \, \del_{F_3,F_3'}.
\label{eq1.13}
\ee
We introduce quasiprojectors (cf.\ (3.28) of \cite{BBN95}):
\be
   \unPro (\alp,F_3,\cE\evd) = \revrE \levlE
\label{eq1.15}
\ee
which satisfy
\bea
   \unPro(\alp,F_3,\cE\evd)  \unPro(\alp',F_3',\cE\evd) &=&
       \del_{\alp,\alp'} \,
       \del_{F_3,F_3'}  \unPro(\alp,F_3,\cE\evd)  ,\nonumber \\
   \Tr \,\unPro(\alp,F_3,\cE\evd) & = & 1,
\label{eq1.16}
\eea
but are not necessarily hermitian.
We have
\bea
   \uneins & = & \sum_{\alp,F_3} \unPro(\alp,F_3,\cE\evd) ,\\
\label{eq1.17}
   \unM    & = & \sum_{\alp,F_3} \EiKaFdEevd\,\unPro(\alp,F_3,\cE\evd) ,\\
\label{eq1.18}
   \unFd   & = & \sum_{\alp,F_3} F_3 \, \unPro(\alp,F_3,\cE\evd) .
\label{eq1.19}
\eea
The total energies (\ref{eq1.5.27}) are
\be
   E(\alp,F_3,\cE\evd,n) = \EEXT_n+\EINT(\alp,F_3,\cE\evd).
\label{eq1.19.1}
\ee

It was shown explicitly in Sect.\ 3.4 of \cite{BBN95} that the complex
energy eigenvalues $\EINT(\alp,F_3,\cE\evd)$ have no contributions linear in  
the P-violation parameters $\delez$. This was derived by considering a
reflection R on the 1-3-plane for the atom at rest. Now we
have the atom in a box which need not be R-invariant. But
for the constant electric field we have no coupling between the c.m.\
and the internal motions in our approximations. Thus even if
the external field arrangement is not R-symmetric here,
we still get no terms linear in $\delez$ in the total
energy eigenvalues $E(\alp,F_3,\cE\evd,n)$ (\ref{eq1.19.1})
for the constant external field.

In the following we will investigate situations where
a non-trivial coupling between c.m.\ and internal motions
is present. We will see that indeed for suitable external
field arrangements having no reflection symmetry the total
energy eigenvalues can get terms linear in the P-violation
parameters $\delez$.

  
\subsection{General consequences of T-invariance}
\label{sec1.3}
In this Section we will work out some general
consequences of T-invariance for stable and unstable
atomic systems in a static external electromagnetic field,
described by the 4-potential $(A^\mu(\xv))=(\Av(\xv),A^0(\xv))$.
Let us start by considering the atomic system coupled to the
radiation field at the level of quantum field theory (QFT). The
Hamiltonian describing this system is
\be
\label{eq1.49}
   H[A]=H_0(t)+H'[t,A],
\ee
where $H_0$, which includes the P-violating term
$\hpv$, is the Hamiltonian for $A^\mu=0$,
and $H'$ is the interaction term with the external field. In terms of the fundamental lepton and quark fields the latter reads
\be
\label{eq1.50}
   H'[t,A]=\int d^3\!x\,j_\mu(\xv,t)A^\mu(\xv),
\ee
with $j_\mu(\xv,t)$ the electromagnetic current operator.
For static fields the total Hamiltonian is time-independent,
as we know from Noether's theorems. In the following we will
work in the Sch\"odinger picture of quantum mechanics, identifying
the Hamiltonians there with those of (\ref{eq1.49}) at $t=0$. Thus
we set $H_0\equiv H_0(0)$ and $H'[A]\equiv H'[0,A]$.
We will now assume that another splitting of $H$ can be made:
\be
\label{eq1.51}
  H[A]=\Htil_0[A]+H_{rad},
\ee
where $\Htil_0[A]$ describes stable atomic
systems in the external field $A^\mu(\xv)$ and $H_{rad}$
their coupling to the radiation field. This coupling will
render the above systems except the lowest level unstable.
Let $\ket{\ikvar{0},\alp,A}$ be the eigenstates of $\Htil_0[A]$:
\be
\label{eq1.52}
   \Htil_0[A]\,\ket{\ikvar{0},\alp,A}=E_0(\alp,A)\,\ket{\ikvar{0},\alp,A},\quad
   (\alp=1,2,...),
\ee
where $E_0(\alp,A)$ are real eigenvalues.
In practice it turns out to be convenient to work with some
suitable basis of orthonormal states
\be
\label{eq1.53}
   \ket{a,\ntil}\qquad (a=1,2,...;\ntil=1,2,...)
\ee
for the eigenstates in (\ref{eq1.52}). Below, the index
$a$ will be chosen to be related to the internal state of
motion of the atom, $\ntil$ to the state of the c.m.\ motion.
We denote by $\cHtil$ the Hilbert space spanned by
the states (\ref{eq1.53}).

With the inclusion of $H_{rad}$ of (\ref{eq1.51}) the states
$\ket{a,\ntil}$ become in general unstable. Using the standard
Wigner-Weisskopf procedure (cf.\ Sect.\ 3.2 of \cite{BBN95}) we
find the effective Hamiltonian $\heff$ describing
the time evolution of the undecayed states $|a,\ntil\ra$
through the (approximate) solution of Schr\"odinger's equation
with the initial condition of having some combination of the states
$\ket{a,\ntil}$ at time $t=0$ (cf.\ (3.9) of \cite{BBN95}):
\bea
\label{eq1.53a}
   i\ptdnd{t}\,\ket{t} &=& H[A]\,\ket{t},\nonumber\\
   \ket{t=0}           &=& \sum_{a,\ntil}c_{a,\ntil}\ket{a,\ntil}.
\eea
For $t>0$ but neither too small nor too large times we get
\be
\label{eq1.54}
   \brak{a,\ntil}\,\exp(-iH[A]\cdot t)\,\ket{a',\ntil'}
   \simeq\brak{a,\ntil}\,\exp(-i\heff[A]\cdot t)\,\ket{a',\ntil'}.
\ee
This represents the general definition of the effective
Hamiltonian, which reduces to (\ref{eq1.5.9}) in the non-relativistic
limit.

We consider now time reversal (T) invariance. From QFT we learn that there exists an antiunitary
operator $V({\mathrm T})$ which transforms the Hamiltonians as follows:
\bea
\label{eq1.55}
   \lk(V({\mathrm T})\,H'[A]\,  V^{-1}({\mathrm T})\rk)^\dagger &=& H'[\AkT],\\
\label{eq1.56}
   \lk(V({\mathrm T})\,H[A]\,   V^{-1}({\mathrm T})\rk)^\dagger &=& H[\AkT], \\
\label{eq1.57}
   \lk(V({\mathrm T})\,\Htil_0[A]\,V^{-1}({\mathrm T})\rk)^\dagger &=& \Htil_0[\AkT].
\eea
Here $\AkT$ denotes the T-transformed potential:
\bea
\label{eq1.58}
   \AkT^0(\xv) &=& A^0(\xv),\nonumber\\
   \AvT(\xv) &=& -\Av(\xv).
\eea
From (\ref{eq1.52}) and (\ref{eq1.57}) we find that $V({\mathrm T})$ transforms
the eigenstates $\ket{\ikvar{0},\alp,A}$ of $\Htil_0[A]$
into eigenstates $\ket{\ikvar{0},\alp,\AkT}$ of $\Htil_0[\AkT]$.
Recall that $E_0(\alp,A)$ are the eigenvalues of the
hermitian operator $\Htil_0[A]$ and thus real. We assume
now that the space $\cHtil$ spanned by the basis vectors
$\ket{a,\ntil}$ in (\ref{eq1.53}) is large
enough to contain both the states $\ket{\ikvar{0},\alp,A}$
and $\ket{\ikvar{0},\alp,\AkT}$. The action of $V({\mathrm T})$ on the
states $\ket{a,\ntil}$ with chosen phase convention is then
given by some matrix $\cTtil$ over $\cHtil$:
\be
\label{eq1.58a}
   V({\mathrm T})\,\ket{a,\ntil}=\ket{a',\ntil'}\,
   \cTtil_{a'\ntil',a\ntil}.
\ee
The antiunitarity of $V({\mathrm T})$ implies:
\be
\label{eq1.58b}
   \cTtil^\dagger\cTtil=\eins.
\ee
The effective Hamiltonians $\heff[A]$ (\ref{eq1.54})
and $\heff[\AkT]$ can be
identified with non-hermitian matrices over $\cHtil$
\be
\label{eq1.59}
   \heff[A]\equiv\Big(\brak{a,\ntil}\,\heff[A]\,\ket{a',\ntil'}\Big).
\ee
From (\ref{eq1.54}), (\ref{eq1.56}) and (\ref{eq1.58}) we
now get as a consequence of T-invariance:
\be
\label{eq1.60}
   \lk(\cTtil^\dagger \heff[\AkT]\cTtil\rk)^T=\heff[A].
\ee
This is the generalization of (3.42) of \cite{BBN95}
for atomic systems in arbitrary static external fields.

Let now $\{E(\alp,A)\}$ be the spectrum of eigenvalues
of $\heff[A]$ and $\Protil(\alp,A)$
the quasiprojectors corresponding to $E(\alp,A)$,
$\alp=1,2,...$.
\be
\label{eq1.60a}
   \heff[A]=\sum_\alp E(\alp,A)\,\Protil(\alp,A).
\ee
An argument using the resolvents of $\heff[A]$
and $\heff[\AkT]$ analogous to (3.43) ff.\ of \cite{BBN95}
shows that with a suitable numbering we have
\bea
\label{eq1.61}
   E(\alp,\AkT) &=& E(\alp,A),\\
\label{eq1.62}
   \lk(\cTtil^\dagger\Protil(\alp,\AkT)\cTtil\rk)^T &=& \Protil(\alp,A).
\eea
These equations embody the general consequences of T-invariance for
the complex (real) eigenvalues and quasiprojectors (projectors)
of unstable (stable) atomic states in static external fields.

Let us consider next a general improper rotation R around some
point $\xv^\iKn$, which we choose without loss of generality
as $\xv^\iKn=0$:
\bea
\label{eq1.63}
   \mbox{R}:\quad \xv &\to& R\cdot\xv,\nonumber\\
              \det\,R  &=&  -1.
\eea
For the external field we set:
\bea
\label{eq1.64}
   \AkR^0(\xv) &=& A^0(R^{-1}\xv),\nonumber\\
   \AvR(\xv)   &=& R\cdot\Av(R^{-1}\cdot\xv)
\eea
The QFT-operator $U({\mathrm R})$ corresponding to R transforms
the Hamiltonian (\ref{eq1.49}) as follows:
\be
\label{eq1.65}
    U({\mathrm R})\,H[A,\dele,\delz]\,U^{-1}({\mathrm R})
  = H[\AkR,-\dele,-\delz],
\ee
where we indicate now explicitly the
dependence on the P-violation parameters $\delez$.
Assuming the space $\cHtil$ to be large enough to contain
for a state also its $U({\mathrm R})$-transform we can represent
$U({\mathrm R})$ in $\cHtil$ by a matrix $\cRtil$:
\bea
   U({\mathrm R})\,|a,\ntil) &=& |a',\ntil')\,\cRtil_{a'\ntil',a\ntil},\nonumber\\
   \quad \cRtil^\dagger\cRtil &=& \eins.
\label{eq1.66}
\eea
For the effective Hamiltonian this leads to
\be
\label{eq1.67}
   \cRtil^\dagger\,\heff[\AkR,-\dele,-\delz]\,\cRtil
   = \heff[A,\dele,\delz].
\ee
Considering again the resolvents of $\heff[A,\dele,
\delz]$ and $\heff[\AkR,-\dele,-\delz]$
we find that the numbering of the spectra of complex
energy eigenvalues can be chosen such that
\bea
\label{eq1.68}
   E(\alp,\AkR,-\dele,-\delz) &=& E(\alp,A,\dele,\delz),\\
\label{eq1.69}
   \cRtil^\dagger\,\Protil(\alp,\AkR,-\dele,-\delz)\,\cRtil
   &=& \Protil(\alp,A,\dele,\delz).
\eea

From (\ref{eq1.61}) and (\ref{eq1.68}) we find the following\\
{\sl Theorem:} If the external field $A$
has the property that the T-transformation for it is equivalent
to some improper rotation R
\be
\label{eq1.70}
  \AkT=\AkR,
\ee
then
\be
\label{eq1.71}
   E(\alp,A,\dele,\delz)=E(\alp,A,-\dele,-\delz).
\ee
For the case that the energy level $E(\alp,A,\dele,\delz)$
is {\it not degenerate} for $\dele=\delz=0$, this implies
vanishing linear terms in $\delez$:
\be
\label{eq1.72}
   \lk.\ptdnd{\del_j}E(\alp,A,\dele,\delz)\rk|_{\dele=\delz=0}=0,\quad (j=1,2).
\ee
As an example for the use of this theorem we consider
a hydrogen atom at rest at $\xv=0$ in static and
spatially homogenous electric and magnetic fields
$\cEv, \Bv$. Let $R$ be the reflection on the plane
through $\xv=0$ spanned by $\cEv$ and $\Bv$.
We can write
\bea
\label{eq1.72a}
   A^0(\xv) &=& -\xv\cdot\cEv,\nonumber\\
   \Av(\xv) &=& -\eh\xv\cross\Bv,
\eea
which fulfils
\be
\label{eq1.72b}
   \AkT^\mu(\xv)=\AkR^\mu(\xv).
\ee
Furthermore, for $\cEv\not=0$ and $\Bv\not=0$ there
are in general no degeneracies of levels for $\dele=\delz=0$.
Thus we conclude that the energy eigenvalues of the stable and
unstable levels of the hydrogen atom in a constant electric
plus magnetic field have no terms linear in the P-violation
parameters $\delez$.\\

To illustrate the general T- and R-transformations we will now close 
this Section by discussing these for the non-relativistic effective Hamiltonian (\ref{eq1.5.9}).
The space $\cHtil$ (cf.\ (\ref{eq1.53}) ff.) is then spanned by the
vectors $|\Rv;2,a\ra$ of (\ref{eq1.5.2}). We can choose a
convenient orthonormal basis set of {\it real}
functions $\xi_{\ntil}(\Rv)$ for the c.m.\ motion and set:
\be
\label{eq1.73}
   \ket{a,\ntil} = \xi_{\ntil}(\Rv)\,\evrzza.
\ee
The T-transformation (\ref{eq1.58a}) is represented by
\be
\label{eq1.74}
   V({\mathrm T})\,\ket{a,\ntil} = \xi_{\ntil}(\Rv)\,\evrzzas\,
   \unT_{a'a},
\ee
where $\unT$ is the matrix describing the T-transformation
of the $n=2$ atomic states (cf.\ (3.33)
of \cite{BBN95}). Thus we find here
\be
\label{eq1.75}
   \cTtil_{a'\ntil',a\ntil}=\del_{\ntil'\ntil}\,
   \unT_{a'a}.
\ee
It is now easy to verify (\ref{eq1.60}) explicitly for
$\heff[A]$ of (\ref{eq1.5.9}). Thus, as it should be,
the general results (\ref{eq1.61}), (\ref{eq1.62}) hold in
the non-relativistic limit.

Let us furthermore define the matrix $\unR$
of the R-transformation (\ref{eq1.63}) for the $n=2$ states
\be
\label{eq1.76}
   U({\mathrm R})\,\evrzza=\evrzzas\,\unR_{a'a}.
\ee
and $(\cR_{\ntil'\ntil}')$ by
\be
\label{eq1.77}
   \xi_{\ntil}(R^{-1}\cdot\Rv)=\xi_{\ntil'}(\Rv)
   \cR_{\ntil'\ntil}'.
\ee
Then the matrix $\cRtil$ of (\ref{eq1.66}) is given by
\be
\label{eq1.78}
   \cRtil_{a'\ntil',a\ntil}=\cR_{\ntil'\ntil}'\,
   \unR_{a'a}.
\ee
This is the matrix which is to be used in 
(\ref{eq1.68}) and (\ref{eq1.69}) in the
non-relativistic limit.

In the following we will frequently make use of the general
consequences of the T- and R-transformations spelled
out in this Section.

\section{Energy shifts for piecewise constant electric fields; exact calculation}
\label{sec2}
\setcounter{equation}{0}
We will now consider simple electric field arrangements
which are neither parity- nor reflection-symmetric. We imagine
a rectangular box which we subdivide into several segments, in
each of which the $\cEv$-vector is pointing in a fixed direction,
with its modulus also having in general different values.
Within any of the segments the electric field
is assumed to be homogeneous, which results in a sudden
variation at the interface, where the segments are connected
to each other, and at the box surface which forms the
boundary of the system.  Of course, Maxwell's equations
require then charge distributions on the interfaces.
Clearly it would be difficult to realize this experimentally,
and we choose this situation only since it is simple to
be analysed from the theoretical point of view. A more realistic
experimental situation could be taken into account by
including corrections to the sudden variation by means of
perturbation theory in a small parameter describing e.g. the extension
of the transition regions between the segments in relation to
the total size of the segments.

Let the lengths of the edges of our rectangular box $\cB$ be
$A_{1,2,3}$ in the 1,2,3-direction (Fig.\ 1):
\be
\label{eq2.1}
  {\rm box}\ \cB:\quad 0\leq R_j\leq A_j,\quad
                        (j=1,2,3).
\ee
In 1-direction we subdivide the box into segments numbered
by $\sig$:
\be
\label{eq2.2}
  {\rm segment}\  \cS^\iKsig:\quad
  R^\iKvar{\sig-1}_1\leq R_1\leq R_1^\iKsig,\quad
  0\leq R_j\leq A_j,\ (j=2,3),
\ee
where $\sig=1,...,\sig_{max}$ and
\be
\label{eq2.3}
  R^\iKn\equiv0<R_1^\iKe<...<R_1^\iKvar{\sig_{max}}\equiv A_1.
\ee
This is illustrated in Fig.\ 1 for $\sig_{max}=3$.
The electric field in $\cS^\iKsig$ is denoted by
$\cEv^\iKsig$.

We will consider both neutral atoms and charged ions in such
segmental boxes. The effective Hamiltonian (\ref{eq1.5.19})
reads now
\be
\label{eq2.3a}
   \heff=\frac{1}{2m}\Pv^2+\unMn-(q\Rv+\unDv)\cdot\sum_\sig
   \cEv^\iKsig\,\Tet(R_1^\iKsig-R_1)\,
   \Tet(R_1-R_1^\iKvar{\sig-1}),
\ee
The problem is to solve the eigenvalue equation (\ref{eq1.5.12})
with the boundary condition (\ref{eq1.5.18}).
There is a direct way to solve this problem by making an
ansatz of linear combinations of constant field wave functions
in each segment. Continuity conditions to be imposed on the
wave function and its first derivative at the interfaces lead to
a determinant equation from which the eigenenergies may be
extracted. This technique will be outlined in the next paragraph
and yields the minimum number of different segments, i.e.\
electric fields  required to obtain P-odd effects. We will show
that for boxes with 2 segments no energy shifts
linear in the $\deli\ (i=1,2)$ occur in our theoretical framework. 
However, for boxes with 3 segments
such energy shifts {\it do} occur as we will prove in Sect.\ 
\ref{sec3} using perturbation theory.

%
\subsection{A neutral atom in a box with 2 segments}
\label{sec2.1}
In this Section we consider a neutral atom in internal states
of principal quantum number $n=2$ in a box with
two segments. The effective Hamiltonian (\ref{eq2.3a}) reads now:
\be
\label{eq2.4}
  \heff=\frac{1}{2m}\Pv^2+\unMn-\unDv\cdot\cEv(\Rv),
\ee
where
\be
\label{eq2.5}
   \cEv(\Rv) = \cEv^\iKe\Tet(R_1^\iKe-R_1)\,\Tet(R_1)
   +\cEv^\iKz\Tet(A_1-R_1)\,\Tet(R_1-R_1^\iKe).
\ee
We have to solve (\ref{eq1.5.12}) with the
boundary condition (\ref{eq1.5.18}):
\bea
\label{eq2.5a}
   \heff\Psi(\Rv) &=& E\,\Psi(\Rv),\\
\label{eq2.6}
   \Psi(\Rv) &=& 0\quad \mbox{for\ } \Rv\in\ptd\cB.
\eea

The Hamiltonian (\ref{eq2.4}) commutes with the part describing
the motion in the 2- and 3-directions:
\be
\label{eq2.7}
   H_{2,3}=\frac{1}{2m}\lk(P^2_2+P^2_3\rk).
\ee
Here and in the following the boundary condition (\ref{eq2.6}) is
always understood to be imposed. Thus we can diagonalize $\heff$
and $H_{2,3}$ simultaneously. The normalized ei\-genfunctions
of $H_{2,3}$ are
\be
\label{eq2.8}
   \phi_{n_2,n_3}(R_2,R_3) = \frac{2}{\sqrt{A_2A_3}}
   \sin\frac{n_2\pi R_2}{A_2}
   \cdot\sin\frac{n_3\pi R_3}{A_3},
\ee
$(n_2=1,2,3,...;\,n_3=1,2,3,...)$. We have
\bea
\label{eq2.9}
   H_{2,3}\,\phi_{n_2,n_3}(R_2,R_3) 
   &=& E_{2,3}(n_2,n_3)\,\phi_{n_2,n_3}(R_2,R_3),\nonumber\\
   E_{2,3}(n_2,n_3) 
   &=& \frac{1}{2m}\lk[\lk(
   \frac{n_2\pi}{A_2}\rk)^2+\lk(\frac{n_3\pi}{A_3}\rk)^2\rk].
\eea
We can now make a product ansatz for the eigenfunctions
of $\heff$ (\ref{eq2.4}):
\bea
\label{eq2.10}
  \Psi(\Rv) &=& \phi_{n_2,n_3}(R_2,R_3)\,\phi(R_1),\nonumber\\
  \phi(R_1) &=& \lk\{\ba{ccc}
  \phi^\iKe(R_1) & {\rm for} & 0\leq R_1\leq R_1^\iKe,\\
  \phi^\iKz(R_1) & {\rm for} & R_1^\iKe\leq R_1\leq A_1,\ea\rk.
\eea
where $\phi(R_1)$ has $N$ components (cf.\ (\ref{eq1.5.6})),
corresponding to the number of linearly independent $n=2$ internal
states.

Let us denote by $\rket{\ikvar{\sig},\beta}$ the right eigenvectors of the mass
matrices in the Sectors $\sig=1,2$; corresponding to the internal,
in general complex energy eigenvalues $\EINT(\ikvar{\sig},\beta)$:
\be
\label{eq2.11}
  (\unMn-\unDv\cdot\cEv^\iKvar{\sig})\,\rket{\ikvar{\sig},\beta}
  = \EINT(\ikvar{\sig},\beta)\,\rket{\ikvar{\sig},\beta},\quad
  (\beta=1,...,N).
\ee
The corresponding left eigenvectors are denoted by
$  \lbrak{\ikvar{\sig},\gam},\quad (\gam=1,...,N)$.
We have
\be
\label{eq2.12}
  \lbra{\ikvar{\sig},\gam}\rket{\ikvar{\sig},\beta}=\delta_{\gam\beta}
  \quad{\rm for}\quad\sig=1,2 
\ee
(no summation over $\sig$).
Furthermore we define matrices $\unU^\iKvar{1,2}$ and $\unU^\iKvar{2,1}$
with matrix elements:
\bea
\label{eq2.13}
  U^\iKvar{1,2}_{\gam,\beta}(\dele,\delz)
  &:=& \lbra{\ikvar{1},\gam}\rket{\ikvar{2},\beta},\nonumber\\
  U^\iKvar{2,1}_{\gam,\beta}(\dele,\delz)
  &:=& \lbra{\ikvar{2},\gam}\rket{\ikvar{1},\beta}.
\eea
We indicate explicitly the dependence of these
matrices on the P-violation parameters $\delez$.
Of course, for the case of nuclear spin $I=0$ there is
no dependence on $\delz$.
As we discussed in Sect.\ \ref{sec1.2}, 
the complex energies $\EINT(\ikvar{\sig},\beta)$ 
have no terms linear in $\delez$.
We make now the following ansatz for $\phi^\iKvar{1,2}(R_1)$ in
(\ref{eq2.10}):
\bea
\label{eq2.14}
  \phi^\iKe(R_1) &=& \sum_\beta\zeta_\beta^\iKe\sin\lk[Q_\beta^\iKe R_1\rk]
   \rket{\ikvar{1},\beta},\nonumber\\
  \phi^\iKz(R_1) &=& \sum_\beta\zeta_\beta^\iKz\sin
   \lk[Q_\beta^\iKz\lk(R_1-A_1\rk)\rk]\rket{\ikvar{2},\beta},
\eea
where $Q_\beta^\iKsig$ and $\zeta_\beta^\iKsig$ are
constants to be determined. With (\ref{eq2.10})
and (\ref{eq2.14}) we already satisfy the boundary condition (\ref{eq2.6}).
Let us now insert the ansatz (\ref{eq2.10}), (\ref{eq2.14})
in (\ref{eq2.5a}) and let $E$ be the sought eigenvalue of $\heff$. We find
that the $Q_\beta^\iKvar{\sig}$ in (\ref{eq2.14}) are to be
determined from
\be
\label{eq2.15}
  \frac{1}{2m}\lk(Q_\beta^\iKsig\rk)^2
  = E-E_{2,3}(n_2,n_3)-\EINT(\ikvar{\sig},\beta),\quad
  (\sig=1,2;\ \beta=1,...,N).
\ee
For definiteness, let $Q_\beta^\iKsig$ be
the root of (\ref{eq2.15}) with positive real part and if this is
zero, the one with positive imaginary part. Finally, the
constants $\zeta_\beta^\iKsig$ in (\ref{eq2.14}) are
determined by the transition conditions at the interface
$R_1=R_1^\iKe$. There $\Psi(\Rv)$ and $\ptd\Psi(\Rv)/
\ptd R_1\equiv\ptd_1\Psi(\Rv)$ must be continuous. These
conditions lead to:
\bea
\label{eq2.16}
  \sum_\beta\zeta_\beta^\iKe\sin\lk[Q_\beta^\iKe R^\iKe_1\rk]
    \rket{\ikvar{1},\beta}
    &=& \sum_\beta\zeta_\beta^\iKz\sin\lk[Q_\beta^\iKz\lk(R^\iKe_1-A_1\rk)\rk]
    \rket{\ikvar{2},\beta},\\
\label{eq2.17}
  \sum_\beta\zeta_\beta^\iKe Q_\beta^\iKe\cos\lk[Q_\beta^\iKe R_1^\iKe\rk]
    \rket{\ikvar{1},\beta}
    &=& \sum_\beta\zeta_\beta^\iKz Q_\beta^\iKz\cos\lk[Q_\beta^\iKz
    \lk(R_1^\iKe-A_1\rk)\rk]\rket{\ikvar{2},\beta}.\ 
\eea
Multiplying (\ref{eq2.16}) and (\ref{eq2.17}) with
$\lbrak{\ikvar{2},\gam}$ and $\lbrak{\ikvar{1},\gam}$, respectively,
where $\gam=1,...,N$, we get the following matrix equation:
\be
\label{eq2.18}
  M_c(E,\dele,\delz)\cdot\zeta=0,
\ee
where
\be
\label{eq2.19}
  M_c(E,\dele,\delz) = \lk(\ba{cc}
   \unU^\iKvar{2,1}(\dele,\delz)\cdot\unS^\iKe(E,\dele,\delz)
 &-\unS^\iKz(E,\dele,\delz)\\
   \unC^\iKe(E,\dele,\delz)
 &-\unU^\iKvar{1,2}(\dele,\delz)\cdot
   \unC^\iKz(E,\dele,\delz)\ea\rk),
\ee
\be
\label{eq2.20}
  \zeta=\lk(\zeta_1^\iKe,...,\zeta_N^\iKe,\zeta_1^\iKz,...\zeta_N^\iKz\rk)^T.
\ee
In (\ref{eq2.19})
$\unS^\iKvar{\sig}$ and $\unC^\iKvar{\sig}$ ($\sig=1,2$) are diagonal $N\times N$ matrices,
with the diagonal elements given by
\bea
\label{eq2.21}
  S^\iKe_{\beta,\beta}(E,\dele,\delz)
  &=& \sin\lk[Q_\beta^\iKe R_1^\iKe\rk],\nonumber\\
  S^\iKz_{\beta,\beta}(E,\dele,\delz)
  &=& \sin\lk[Q_\beta^\iKz\lk(R_1^\iKe-A_1\rk)\rk],\nonumber\\
  C^\iKe_{\beta,\beta}(E,\dele,\delz)
  &=& Q_\beta^\iKe\cos\lk[Q_\beta^\iKe R_1^\iKe\rk],\nonumber\\
  C^\iKz_{\beta,\beta}(E,\dele,\delz)
  &=& Q_\beta^\iKz\cos\lk[Q_\beta^\iKz\lk(R_1^\iKe-A_1\rk)\rk].
\eea

Note that in the matrices $\unS^\iKsig,\unC^\iKsig$ 
the dependence on the parameters $\delez$
enters only through the internal energy eigenvalues
$\EINT(\ikvar{\sig},\beta)$ (cf.\ (\ref{eq2.15})), which are {\it even}
under the replacement $\delez\to-\delez$. Therefore
we have
\bea
\label{eq2.22}
  \unS^\iKsig(E,\dele,\delz) &=& \unS^\iKvar{\sig}
  (E,-\dele,-\delz),\nonumber\\
  \unC^\iKsig(E,\dele,\delz) &=& \unC^\iKvar{\sig}
  (E,-\dele,-\delz),\quad
  (\sig=1,2).
\eea
The condition for (\ref{eq2.18}) having a nontrivial solution-vector $\zeta$ is:
\be
\label{eq2.23}
  \det M_c(E,\dele,\delz)=0.
\ee
The roots of this equation give the complex energy
eigenvalues of $\heff$ (\ref{eq2.4}). We want to see if these
roots have terms linear in $\delez$.

%
\subsubsection{Results for neutral atoms with $I=0$}
\label{sec2.1.1}
Let us first consider the case of a neutral hydrogen-like
atom with vanishing nuclear spin. This is not realizable for conventional atoms but an atom consisting of a $\pi^+$ meson and an electron is a physical system to which our considerations apply. For this case the representations of $\unMn$
and $\unDv$ may be found in Appendix B of [12]. They
are 8$\times$8-matrices and thus the continuity conditions for the
wave function and its derivative w.r.t.\ $R_1$ lead to a set of 16 linear
equations (cf.\ (\ref{eq2.18})):
\be
\label{eq2.30}
   M_c(E,\dele)\cdot\zeta=0
\ee
From the determinant condition
\be
\label{eq2.31}
   \det M_c(E,\dele)=0
\ee
we get the energy eigenvalues.
As we show in Appendix \ref{appB} we can simplify the structure of $M_c$ considerably if we choose a suitable set of basis states having simple transformation properties under T (time reversal) and R (reflection on the 1--3-plane). We show in Appendix \ref{appB} that (\ref{eq2.31}) is equivalent to an equation having no linear terms in $\dele$:
\be
\label{eq2.42}
   \det M_c'(E,\dele)=f_0(E)+\dele^2 f_2(E)+\cO(\dele^4)=0
\ee
Now we would like to use the
implicit function theorem to calculate from (\ref{eq2.42}) the
dependence of the solutions $E$ on $\dele$.

However, we have to be careful. It turns out that for $\dele=0$
the energy levels are twofold degenerate, i.e.\ $f_0(E)$ has
double zeroes. These degeneracies are even preserved to
order $\dele$. We show all this in Appendix B, where we
use the T-symmetry to derive that $\det M_c'=0$ is equivalent
to
\be
\label{eq2.43}
   \det M_c^{asym}(E,\dele)=0,
\ee
where $ M_c^{asym}=-( M_c^{asym})^T$ is an antisymmetric matrix.
Now, from a basic theorem of linear algebra (cf.\ Appendix \ref{appB}), the determinant
of an antisymmetric matrix with even number of rows is always
a square:
\be
\label{eq2.44}
   \det M_c^{asym}(E,\dele)=[g(E,\dele)]^2.
\ee
Therefore the solutions $E$ of (\ref{eq2.43}) are twofold
degenerate for any choice of $\cEv^\iKe,\cEv^\iKz$.
Furthermore we find that
\be
\label{eq2.45}
   g(E,\dele)=g_0(E)+\dele^2g_2(E)+\cO(\dele^4)
\ee
with $g_0(E)$ having only nondegenerate zeroes. The implicit
function theorem implies then that the zeroes of $g(E,\dele)$
have no terms linear in $\dele$:
\be
\label{eq2.45.1}
  \lk.\ptdxndy{E(\dele)}{\dele}\rk|_{\dele=0} = 0.
\ee
To conclude: Using the R and T symmetry relations
we find that, for a box with two electric fields $\cEv^\iKe,
\cEv^\iKz$,  the complex eigenenergies of the atom have
no terms linear in $\dele$.

%
\subsubsection{Results for neutral atoms with $I=1/2$}
\label{sec2.1.2}
In this subsection we will consider the case of a neutral atom with $I=1/2$, as
e.g.\ $\eeH$. The eigenenergies are again obtained from the
condition
\be
\label{eq2.46}
   \det M_c(E,\dele,\delz)=0
\ee
for the continuity matrix $M_c$ given in (\ref{eq2.19}). 
Now $M_c$ is a 32$\times$32 matrix. 
Using the $\pm$ eigenstates
as defined in Appendix \ref{appA} and their T and R symmetry relations we show in Appendix \ref{appB} that (\ref{eq2.46}) is equivalent to 
\be
\label{eq2.56}
   \det M_c'(E,\dele,\delz)=f_0(E)
   +\sum_{i,j}\deli\delta_jf_{2,ij}(E)+\ {\rm terms\ of\ 4th\
   order\ in}\ \delez.
\ee
Here $f_0(E)$ has (in general) no double zeroes and thus we
can immediately conclude from the implicit function theorem
that the energy eigenvalues, i.e.\ the solutions of (\ref{eq2.56}),
have no terms linear in $\delez$:
\be
\label{eq2.57}
   \lk.\frac{\partial E(\dele,\delz)}{\partial\delta_j}
   \rk|_{\dele=\delz=0} = 0,
\ee
Thus, within our theoretical framework, we find that, for a
$\eeH$-atom in a box with 2 segments only there are no
energy shifts linear in $\delez$.

%
\subsection{A charged ion with nuclear spin $I=0$ in a box with
two segments}
\label{sec2.2}
In the real world conventional one-electron atoms with vanishing nuclear spin
are always electrically charged. Therefore we investigate now
the modifications of the results of Section \ref{sec2.1.1} due to the
c.m.\  motion being influenced by the external electric field.

Let $q=e(Z-1)$ be the charge of the ion. The effective Hamiltonian
is given in (\ref{eq2.3a}) where we choose 2 segments, i.e.\
$\cEv(\Rv)$ as in (\ref{eq2.5}). In  Sect.\ \ref{sec2.1} the c.m.\
part of the wave functions was built up by sines and cosines.
Now, for a charged ion, we need the wave functions for a potential
linear in the coordinate.

Let us consider first
the general one-dimensional Schr\"odinger equation with a
potential linear in the coordinate,
\be
\label{eq2.58}
   \lk[-\frac{\partial^2}{\partial R^2}-2\,m(q\cE R+E)\rk]
   \psi(R)=0.
\ee
The solutions of (\ref{eq2.58}) are obtained with the help of the 
Airy functions defined by
\bea
\label{eq2.59a}
   {\mathrm Ai}(z) 
   &=& \ed\sqrt{z}\lk(I_{-1/3}(\zeta)-I_{+1/3}(\zeta)\rk),
   \nonumber\\
   {\mathrm Bi}(z) 
   &=& \sqrt{\mbox{$\frac{z}{3}$}}\lk(I_{-1/3}(\zeta)+I_{+1/3}(\zeta)\rk)
\eea
for $z>0$ and 
\bea
\label{eq2.59b}
   {\mathrm Ai}(z) 
   &=& \ed\sqrt{-z}\lk(J_{-1/3}(\zeta)+J_{+1/3}(\zeta)\rk),
   \nonumber\\
   {\mathrm Bi}(z) 
   &=& \sqrt{\mbox{$-\frac{z}{3}$}}\lk(J_{-1/3}(\zeta)-J_{+1/3}(\zeta)\rk)
\eea
for $z<0$,
where $J_p$ ($I_p$) denote the (modified) Bessel functions,
\be
\label{eq2.60}
   \zeta:=\mbox{$\frac{2c}{3}$}|z|^{3/2},\quad
   c^2=2mq\cE. 
\ee
The general solution of (\ref{eq2.58}) is then
\be
\label{eq2.61}
   \psi(R)=\xi_+\psi_+(R,E,\cE)-\xi_-\psi_-(R,E,\cE),
\ee
where
\bea
\label{eq2.61a}
   \psi_+(R,E,\cE) &=& {\mathrm Ai}(R+2mE/c^2),\nonumber\\
   \psi_-(R,E,\cE) &=& {\mathrm Bi}(R+2mE/c^2),
\eea
and $\xi_\pm$ are constants.
If we impose as boundary condition the vanishing of the wave
function at $R=0$ and $R=A$,
\be
\label{eq2.62}
   \psi(0)=\psi(A)=0,
\ee
we get corresponding eigenenergies $E$ by solving the equation
\be
\label{eq2.63}
   \det\lk(\ba{cc}
   \psi_+(0,E,\cE)&\psi_-(0,E,\cE)\\
   \psi_+(A,E,\cE)&\psi_-(A,E,\cE)\ea\rk)=0.
\ee
Let us number these eigenenergies by $E_m(A,\cE)$ and the
corresponding normalized eigenfunctions (\ref{eq2.61})
by $\phi_m(R,\cE)$, where $m=1,2,...\,$. We can choose the functions
$\phi_m(R,A,\cE)$ to be real and normalized:
\be
\label{eq2.64}
   (m,A,\cE|m',A,\cE)=\int^A_0dR\,\phi_m(R,A,\cE)\,\phi_{m'}(R,A,\cE)=
   \delta_{m'm}.
\ee

Now we come to our problem of finding the eigenvalues of $\heff$  
(\ref{eq2.3a}) for a box with 2 segments. We make the following
ansatz for the complete wave function $\Psi(\Rv)$:
\be
\label{eq2.64a}
   \Psi(\Rv)=\lk\{\ba{ccc}
   \Psi^\iKe(\Rv)&{\rm for}& 0\leq R_1\leq R_1^\iKe,\\
   \Psi^\iKz(\Rv)&{\rm for}& R_1^\iKe\leq R_1\leq A_1,
   \ea\rk.
\ee
\bea
\label{eq2.65}
   \Psi^\iKsig(\Rv)  &=&  \sum_{\beta,m_2,m_3}
       \zeta^\iKsig_{\beta,m_2,m_3} 
       \phi_{m_2}(R_2,A_2,\cE^\iKsig_2)
       \phi_{m_3}(R_3,A_3,\cE_3^\iKsig)\nonumber\\
   &\times& \lk[
       \xi^\iKsig_{+,\beta,m_2,m_3}
       \psi_+(R_1,E_{\beta,m_2,m_3}^\iKsig,\cE_1^\iKsig) 
      -\xi^\iKsig_{-,\beta,m_2,m_3}
       \psi_-(R_1,E_{\beta,m_2,m_3}^\iKsig,\cE_1^\iKsig)
      \rk]
      |\ikvar{\sig},\beta),\nonumber\\
   && (\sig=1,2),
\eea
where we use the same notation as in  Sect.\ \ref{sec2.1}, and we have with $E$ the sought energy eigenvalue of $\heff$:
\be
\label{eq2.66}
   E^\iKsig_{\beta,m_2,m_3}=E-E_{m_2}(A_2,\cE_2^\iKsig)
   -E_{m_3}(A_3,\cE_3^\iKsig)-\EINT(\ikvar{\sig},\beta).
\ee
Furthermore, we set
\bea
\label{eq2.67}
   \xi_{\pm,\beta,m_2,m_3}^\iKe
   &=& \psi_\mp(0,E^\iKe_{\beta,m_2,m_3},\cE_1^\iKe),\nonumber\\
   \xi_{\pm,\beta,m_2,m_3}^\iKz
   &=& \psi_\mp(A_1,E^\iKz_{\beta,m_2,m_3},\cE_1^\iKz).
\eea
This ansatz fulfils already the boundary conditions at the
complete surface of the box. We still have to impose the
continuity of $\Psi(\Rv)$ and $\partial\Psi(\Rv)/\partial R_1$ at
the interface, i.e.\ at $R_1=R_1^\iKe$. This gives
conditions analogous to (\ref{eq2.16}), (\ref{eq2.17}), which we can
write as linear equation for the now infinite component vector $\zeta$:
\be
\label{eq2.68}
   M_c(E,\dele)\cdot\zeta=0,
\ee
where
\bea
\label{eq2.69}
   &\zeta=\lk(\zeta^\iKe_{\beta,m_2,m_3},\ 
              \zeta^\iKz_{\beta',m_2',m_3'}\rk)^T,&\nonumber\\
   &\beta,\beta'=1,...,N;\ m_2,m_2'=1,2,...; m_3,m_3'=1,2,...\,.&
\eea
Explicitly we have:
\be
\label{eq2.70}
   M_c(E,\dele)=\lk(\ba{cc}
   \unU^\ivar{(2,1)}(\dele)\unS^\iKe(E,\dele) & -\unS^\iKz(E,\dele)\\
   \unC^\iKe(E,\dele) & -\unU^\ivar{(1,2)}(\dele)\unC^\iKz(E,\dele)
   \ea\rk),
\ee
\be
\label{eq2.71}
   \unU^\iKvar{\sig,\tau}(\dele)
   =\lk((\widetilde{\ikvar{\sig},\alp}|\ikvar{\tau},\beta)
        (m_2,A_2,\cE^\iKsig_2|m_2',A_2,\cE_2^\iKvar{\tau})
        (m_3,A_3,\cE_3^\iKsig|m_3',A_3,\cE_3^\iKvar{\tau})\rk),
\ee
\bea
\label{eq2.72}
   \unS^\iKsig(E,\dele) 
   &=& \Big(\delta_{\alp\beta}
       \Big[\xi^\iKsig_{+,\beta,m_2,m_3}
        \psi_+(R_1^\iKe,E_{\beta,m_2,m_3}^\iKsig,\cE_1^\iKsig)\nonumber\\
   & & \quad -\xi^\iKsig_{-,\beta,m_2,m_3}
        \psi_-(R_1^\iKe,E_{\beta,m_2,m_3}^\iKsig,\cE_1^\iKsig)\Big]
        \delta_{m_2,m_2'}\delta_{m_3,m_3'}\Big),\\
\label{eq2.73}
   \unC^\iKsig(E,\dele)
   &=& \Big(\delta_{\alp\beta}\ptdnd{R_1}
       \Big[\xi^\iKsig_{+,\beta,m_2,m_3}
       \psi_+(R_1,E_{\beta,m_2,m_3}^\iKsig,\cE_1^\iKsig)\nonumber\\
   & & \quad -\xi^\iKsig_{-,\beta,m_2,m_3}
       \psi_-(R_1,E_{\beta,m_2,m_3}^\iKsig,\cE_1^\iKsig)\lk.\Big]
       \rk|_{R_1=R_1^\iKe}
       \delta_{m_2,m_2'}\delta_{m_3,m_3'}\Big).
\eea
Now we choose as in Sect.\ \ref{sec2.1.1} and Appendix \ref{appB} the $\pm$ basis states (\ref{eq2.32})
for the internal motion of the ion. The matrices $\unU^\iKvar{\sig,
\tau}$ can then be split into submatrices $\Ustrs$ $(r,s\in\{+,-\})$ as in (\ref{eq2.33}), (\ref{eq2.34}). Of course,
all these matrices are now infinite dimensional. It is easy to check that
the matrices $\Ustrs$ satisfy again the relations
(\ref{eq2.35a})--(\ref{eq2.35c}). Thus for any finite truncation
in the transverse modes, say
\bea
\label{eq2.74}
   &&m_2,m_2'\leq K_2,\nonumber\\
   &&m_3,m_3'\leq K_3,
\eea
we can immediately take over the methods of Sect.\ \ref{sec2.1.1} and conclude
that there are no shifts in the eigenenergies $E$ which are linear
in $\dele$.

Of course, we would now like to take the limit $K_{2,3}\to\infty$.
We will not give a rigorous mathematical treatment of this limit
but discuss it in physical terms. Suppose we have an eigenenergy
$E$. Then we see from (\ref{eq2.66}) that the energies $E^\iKsig
_{\beta,m_2,m_3}$ will be more and more negative for higher
and higher transverse energies $E_{m_2},E_{m_3}$, i.e.\ for
$m_2,m_3\to\infty$.
But for large negative $E^\iKsig_{\beta,m_2,m_3}$ the Airy functions
(\ref{eq2.59a}), (\ref{eq2.59b}) will be exponentially decreasing or increasing.
Clearly, the corresponding contribution in $\Psi^\iKsig(\Rv)$
(\ref{eq2.65}) must in essence be exponentially decreasing
when going away from the interface or surface of the box, and this decrease
will be the faster the more negative $E^\iKsig_{\beta,m_2,m_3}$ is. We
can write the energy eigenvalue $E$ as
\be
\label{eq2.75}
   E=\frac{\la\Psi|\heff|\Psi\ra}{\la\Psi|\Psi\ra}.
\ee
Choosing now some large enough $K_2,K_3$, we see that the contribution
from the high transverse modes with $m_2>K_2,m_3>K_3$ will in essence come from the interface and/or the surfaces and can
be made arbitrarily small in the integrals (\ref{eq2.75}).
Here it is important that $\heff$ contains no parts with high
frequency in any direction. Thus we estimate that the limit
$K_{2,3}\to\infty$ should be harmless.

In conclusion: In this Section we found that, within
the framework of our calculations, there are no energy shifts linear in the P-violation parameter $\dele$ also for a
charged ion with nuclear spin $I=0$ in a box with two segments.

\section{Energy shifts from perturbation theory for atoms in a box with three segments}
\label{sec3}
\setcounter{equation}{0}
In Sections \ref{sec2.1} and \ref{sec2.2} we considered boxes with two
segments, but did not find any P-violating energy shifts.
This is somewhat surprising, since the setup with two electric
field vectors together with the fixed direction marked
by the external geometry {\it does} form a chiral
situation. But apparently, this is not sufficient to have
energy shifts revealing parity violation.

The above results require another segment to be added
and to consider a situation where one can form a nontrivial
trihedral out of the electric field vectors alone.
For the simplest case of neutral atoms with $I=0$, this
adds another 16 rows and columns to $\unMc$ (\ref{eq2.19}),  
(\ref{eq2.36}), as now continuity conditions have to be
imposed at two interfaces and the wave function $\phi_c^\iKz$
of the middle segment consists of sine {\it and} cosine terms. The result
is a determinant equation for a nontrivial 32$\times$32-matrix,
which we have not analysed, since it is rather complicated. Instead,
we turn in this Section to simpler calculations based
on Rayleigh-Schr\"odinger perturbation theory.\\

Let us consider a box with three segments (Fig.\ 1) where the
electric fields are
\be
\label{eq2.76}
   \cEv^\iKsig = \cE\evd+\cEv^{'\iKsig},
               \quad(\sig=1,2,3).
\ee
We assume that the electric field in 3-direction, $\cE\evd$
is dominant:
\be
\label{eq2.77}
   |\cEv^{'\iKsig}|\ll|\cE|,
   \quad(\sig=1,2,3).
\ee
We will then write down a perturbation expansion
with respect to the terms induced in $\heff$ (\ref{eq2.3a})
by the fields $\cEv^{'\iKsig}$. Indeed, let us write $\heff$
(\ref{eq2.3a}) as follows:
\be
\label{eq2.78}
   \heff = \heffn + \heffe,
\ee
where
\be
\label{eq2.79}
   \heffn = \frac{1}{2m}\Pv^2+\unMn-(q R_3+\unD_3)\cE,
\ee
\be
\label{eq2.80}
   \heffe = -(q\Rv+\unDv)\cdot\sum^3_{\sig=1}\cEv^{'\iKsig}
   \,\Theta(R_1^\iKsig-R_1)\,\Theta(R_1-R_1^\iKvar{\sig-1}).
\ee
Denoting by R the reflection on the 1--3-plane $x_2=A_2/2$, we
see that $\heffn$ is R symmetric whereas
$\heffe$ in general is not. Thus we expect from the
results of Sect.\ \ref{sec1.3} that the eigenenergies of $\heffn$ will
not have linear terms in $\delez$ but that such
terms will be generated upon inclusion of $\heffe$
with a suitable arrangement of fields $\cEv^{'\iKsig}$.

The idea is now to calculate $\deli$-linear contributions
to the eigenenergies $E$ of $\heff$ in perturbation theory
based on the separation (\ref{eq2.78}).
The eigenvectors as well as the eigenvalues $E^\iKn$ of $\heffn$
may be determined numerically in a simple way and the series of higher  
corrections to $E^\iKn$ contains just matrix elements of $\heffe$
w.r.t.\ these eigenvectors.

%
\subsection{Results for a neutral atom with $I=0$}
\label{sec3.1}
Let us again consider first a neutral atom with nuclear spin
$I=0$. The eigenvectors of $\heffn$ are then easily written
down, using the results of Sect.\ \ref{sec1.2}:
\bea
\label{eq2.81}
   \revrnE         &:=& \phi_{\nv}(\Rv)|\alp,F_3,\cE\evd),\nonumber\\
   \phi_{\nv}(\Rv) &= & \prod^3_{i=1}\sqrt{\frac{2}{A_i}}
   \sin\lk(\frac{n_i\pi R_i}{A_i}\rk).
\eea
The corresponding eigenvalues $E^\iKn(\alp,F_3,\cE\evd,\nv=(n_1,n_2,n_3))$
satisfy
\bea
\label{eq2.83}
   \heffn|\alp,F_3,\cE\evd,\nv)
     &=& E^\iKn(\alp, F_3,\cE\evd,\nv) |\alp,F_3,\cE\evd,\nv),\\
   E^\iKn(\alp, F_3,\cE\evd,\nv)
     &=& \frac{1}{2m}\sum_{i=1}^3 \lk(\frac{n_i\pi}{A_i}\rk)^2
         + E^\ivar{INT}(\alp,F_3,\cE\evd).
\eea

We know from the results of Sections 3.3 and 3.4 of \cite{BBN95} that
$E^\ivar{INT}(\alp, F_3, \cE\evd)$ has no terms linear in
$\dele$ and that these energy eigenvalues for $F_3$ and
$-F_3$ are degenerate. These statements are then also true
for $E^\iKn(\alp,F_3,\cE\evd,\nv)$, i.e.\ we have
\be
\label{eq2.84}
   E^\iKn(\alp,F_3,\cE\evd,{\nv})=E^\iKn(\alp,-F_3,\cE\evd,\nv),
\ee
and, neglecting terms of order $\dele^2, E^\iKn
(\alp, F_3,\cE\evd,\nv)$ is independent of $\dele$.

Now we go from $\heffn$ to $\heff$ (\ref{eq2.78}) considering
$\heffe$ as a perturbation. The twofold degeneracy (\ref{eq2.84})
of the zero order energies forces us to use degenerate perturbation
theory (cf.\ \cite{Messiah}, \cite{Kato49}, \cite{Bloch58}). Also we have
to remember that our effective Hamiltonian is non-hermitian. As
stated in Section \ref{sec1.1}, we will always assume in the general discussion that all non-hermitian matrices
occurring are diagonalizable (For our specific numerical examples this assumption holds, as we see from the explicit calculations). Then the standard degenerate
perturbation theory treatment for hermitian matrices of
\cite{Messiah}, \cite{Kato49}, \cite{Bloch58} can be taken over for our case, the
only difference being that projectors have to be replaced
by quasiprojectors.

Consider now the perturbation of the degenerate energy level
characterized by $(\alp,|F_3|,\cE\evd,\nv)\equiv\balp$,
\be
\label{eq2.85}
   E^\iKn_{\balp}\equiv E^\iKn(\alp,|F_3|,
   \cE\evd,\nv)=E^\iKn(\alp,-|F_3|,\cE\evd,\nv).
\ee
The corresponding right eigenvectors of $\heffn$
\be
\label{eq2.86}
   |\balp,\pm)\equiv|\alp,\pm|F_3|,\cE\evd,\nv)
\ee
span a space $\cR^\iKn_{\balp}$. According to
the methods of Kato and Bloch (cf.\ \cite{Messiah}-\cite{Bloch58})
the perturbed eigenvalues can be obtained by solving an
eigenvalue problem in $\cR^\iKn_{\balp}$.
Using first Kato's method (cf.\ \cite{Messiah}) we define
the following matrices:
\bea
\label{eq2.87}
   \Pro^\iKn_{\balp} 
   &=& \sum_{m=\pm}|\balp,m)(\widetilde{\balp,m}|,\\
\label{eq2.88}
   {\QPro}^\iKn_{\balp}
   &=& \eins-\Pro^\iKn_{\balp},\\
\label{eq2.89}
   \frac{\QPro^\iKn_{\balp}}{\eps^k}
   &=& \sum_{\bbet\not=\balp}
       \frac{\Pro^\iKn_{\bbet}}{(E^\iKn_{\balp}-E^\iKn_{\bbet})^k},
       \quad(k=1,2,...);\nonumber\\
\label{eq2.90}
   S_{\balp,0} &=& -\Pro^\iKn_{\balp},\nonumber\\
   S_{\balp,k} &=& \frac{\QPro^\iKn_{\balp}}{\eps^k},
   \quad {\rm for}\quad k=1,2,....
\eea
Here the $\Pro_{\bbet}^\iKn,\bbet\not=\balp$,
denote the quasiprojectors on all the other degenerate
eigen-subspaces of $\heffn$, corresponding to the
eigenvalues $E^\iKn_{\bbet}$. We define furthermore
the operators:
\be
\label{eq2.91}
   (k_1,...,k_{n+1}):=
   S_{\balp,k_1}\heffe S_{\balp,k_2}\cdot\cdot\cdot\heffe
   S_{\balp,k_{n+1}},
\ee
where the $k_j$ are non-negative integers.
We note the relation
\be
\label{eq2.91a}
   (k_1,...,k_n,k_{n+1})\cdot(l_1,l_2,...,l_{m+1})=
   [1-\delta_{0,k_{n+1}}-\delta_{0,l_1}]
   (k_1,...,k_n,k_{n+1}+l_1,l_2,...,l_{m+1}).
\ee

With the help of the operators (\ref{eq2.91}) we define two further
matrices over $\cR^\iKn_{\balp}$
as follows:
\bea
\label{eq2.92}
   \Kh_{\balp} &=& \Pro^\iKn_{\balp}+\sum^\infty_{n=2}{A'}^\iKen,\nonumber\\
   \Hh_{\balp} &=& E_{\balp}^\iKn\Kh_{\balp}+\sum^\infty_{n=1}{B'}^\iKen,
\eea
where
\bea
\label{eq2.93}
   {A'}^\iKen &=& -{\sum}_\iKen(0,k_1,...,k_{n-1},0),\nonumber\\
   {B'}^\iKen &=& {\sum}_\iKvar{n-1}(0,k_1,...,k_{n-1},0),
\eea
and
\be
\label{eq2.94}
   {\sum}_\iKvar{p}(0,k_1,...,k_{n-1},0)
\ee
denotes the sum over all non-negative integers
$k_1,...,k_{n-1}$ satisfying
\be
\label{eq2.95}
   k_1+...+k_{n-1} = p.
\ee
In Kato's method the eigenvalues $E_{\balp}$ of
$\heff=\heffn+\heffe$ are obtained as solutions
of the following eigenvalue problem in $\cR^\iKn_{\balp}$:
\be
\label{eq2.96}
   (\Hh_{\balp}-E_{\balp}\Kh_{\balp})\,|\,\,\ra = 0,
\ee
where $|\,\,\ra \in \cR^\iKn_{\balp}$.

We show now that both, $\Kh_{\balp}$ and
$\Hh_{\balp}$ are matrices over $\cR^\iKn_{\balp}$
which are multiples of the 2$\times$2 unit matrix there.
Indeed, let us first consider the matrix elements
of $\Kh_{\balp}$:
\bea
\label{eq2.97}
   \Kh_{\balp}|\balp,m) &=& |\balp,m')(\Kh_{\balp})_{m'm},\nonumber\\
   (\Kh_{\balp})_{m'm}  &=& (\widetilde{\balp,m'}|\Kh_{\balp}|\balp,m).
\eea
From T-invariance we get, using (\ref{eqA.2}), (\ref{eqA.3})
and (\ref{eqA.11}) of Appendix \ref{appA} and that $\phi_{\nv}(\Rv)$
in (\ref{eq2.81}) is a real function:
\bea
\label{eq2.98}
   \lk(\uncT^\dagger(0,k_1,...,k_{n-1},0)\uncT\rk)^T
   &=& (0,k_{n-1},...,k_1,0),\\
\label{eq2.99}
   \lk(\unT^\dagger\Kh_{\balp}\unT\rk)^T
   &=& \Kh_{\balp},\\
\label{eq2.100}
   (\widetilde{\balp,m'}|\Kh_{\balp}|\balp,m)
   &=& (\widetilde{\balp,m'}|
       \unT(\unT^\dagger\Kh_{\balp}\unT)\unT^\dagger
       |\balp, m)^T\nonumber\\
   &=& (-1)^{2F+(m+m')|F_3|}
   (\widetilde{\balp,-m|}\Kh_{\balp}|\balp,-m')\nonumber\\
   & & (m,m'\in\{+1,-1\}).
\eea
Note that the matrix $\unT$ acts on the internal atomic
state vectors only, but the transposition has to be taken
including the c.m.\ part of the state vectors. In (\ref{eq2.100})
$F$ and $|F_3|=-F,...,F$ are half-integers. Therefore we get:
\bea
\label{eq2.101}
   (\widetilde{\balp,+}|\Kh_{\balp}|\balp,+) 
   &=& (\widetilde{\balp,-}|\Kh_{\balp}|\balp,-),\nonumber\\
   (\widetilde{\balp,+}|\Kh_{\balp}|\balp,-) 
   &=& 0,
\eea
i.e.\ $\Kh_{\balp}$ is proportional to the unit matrix
in $\cR_{\balp}^\iKn$. The proof that also
$\Hh_{\balp}$ is a multiple of the unit matrix in $\cR_{\balp}^\iKn$ 
is analogous. As a result we see that the twofold
degeneracy of the eigenvalue (\ref{eq2.85}) is maintained to
{\it all orders} in the perturbation theory.

The actual calculation of the perturbed eigenvalues is best done
with the help of Bloch's method \cite{Bloch58} which gives a smaller
number of terms which have to be calculated in each order.
Let us define the matrix $\Hhh_{\balp}$ by
\bea
\label{eq2.102}
   \Hhh_{\balp}\Kh_{\balp}       &=& \Hh_\alp,\nonumber\\
   \Hhh_{\balp}\Pro^\iKn_{\balp} &=& \Hhh_{\balp}.
\eea
One finds the following expansion for $\Hhh_{\balp}$:
\be
\label{eq2.103}
   \Hhh_{\balp} = E_\balp^\iKn\Pro^\iKn_{\balp}+\sum^\infty_{n=1}{C'}^\iKen,
\ee
where
\bea
\label{eq2.104}
   {C'}^\iKe  &=& (0,0),\nonumber\\
   {C'}^\iKen &=& {\sum}_\iKvar{n-1}'(0,k_1,...,k_{n-1},0),\quad(n\geq2).
\eea
Here $\sum_\iKvar{n-1}'$ denotes the sum as in
(\ref{eq2.94}), (\ref{eq2.95}), with the additional requirement
\be
\label{eq2.105}
   k_1+...+k_m\geq m\quad {\rm for}\quad m=1,...,n-1.
\ee
Clearly, the eigenvalue problem (\ref{eq2.96})
is equivalent to
\be
\label{eq2.106}
   (\,\Hhh_{\balp}-E_{\balp})|\,\,\ra = 0,
\ee
with $|\,\,\ra  \in \cR^\iKn_{\balp}$.
Furthermore $\Hhh_{\balp}$ is a multiple of the unit matrix in  
$\cR^\iKn_{\balp}$, as we see from (\ref{eq2.102}). Thus we find
\bea
\label{eq2.107}
   E_{\balp} &=& \eh \Tr\,\Hhh_{\balp}
            =  E^\iKn_{\balp}+\sum^\infty_{n=1}E^\iKen_{\balp},
   \nonumber\\
   E^\iKen_{\balp} &=& {\sum}_\iKvar{n-1}'\eh \Tr(0,k_1,...,k_{n-1},0),
   \quad (n=1,2,\dots).
\eea
The first three correction terms read:
\bea
\label{eq2.108}
   E^\iKe_{\balp} &=& \eh \Tr(0,0),\nonumber\\
   E^\iKz_{\balp} &=& \eh \Tr(0,1,0),\nonumber\\
   E^\iKd_{\balp} &=& \eh \Tr\{(0,1,1,0)+(0,2,0,0)\}.
\eea
Using (\ref{eq2.91a}) and arguments analogous to (\ref{eq2.97}) ff.
we can write $E^\iKd_{\balp}$ as follows:
\be
\label{eq2.108a}
   E^\iKd_{\balp}=\eh\Tr(0,1,1,0)-\lk[\eh\Tr(0,2,0)\rk]\lk[\eh\Tr(0,0)\rk].
\ee

Now we consider a reflection R on the 1--3-plane
$R_2=A_2/2$ (cf.\ Fig.\ 1):
\be
\label{eq2.108b}
   \mbox{R}: \lk(\ba{c} x_1\\ x_2    \\ x_3\ea\rk)
          \to\lk(\ba{c} x_1\\ A_2-x_2\\ x_3\ea\rk).
\ee
The reflected field $\cEvR^\iKsig$ has its 1- and 3-components 
unchanged, the 2-component changes sign (cf.\ (\ref{eq1.64})). 
According to Sect.\ \ref{sec1.1} we want to calculate
the following energy difference (cf.\ (\ref{eq1.5.14}) with
$\cEvP\to \cEvR)$
\be
\label{eq2.109}
   \Delta E_{\balp}=E_{\balp}(\cEv)-E_{\balp}(\cEvR)
\ee
which is a parity-odd observable. From (\ref{eq2.79}),
(\ref{eq2.80}) we see that for our neutral atom $(q=0)$ we have
\bea
\label{eq2.110}
   \heffn(\cEvR,\dele) &=& \heffn(\cEv,\dele),\nonumber\\
   \heffn(\cEvR,\dele) &=& \cRtil\,\heffn(\cEv,-\dele)\,\cRtil^\dagger,
\eea
\bea
\label{eq2.111}
   \heffe(\cEv')  &=& -\sum^3_{\sig=1}{\cEv'}^\iKsig\unDv
   \cdot\Theta(R_1^\iKsig-R_1)\,\Theta(R_1-R_1^\iKvar{\sig-1}),\nonumber\\
   \heffe(\cEvR') &=& \cRtil\,\heffe(\cEv')\,\cRtil^\dagger.
\eea
Here $\cRtil$ is the operator representing the reflection
R as defined in (\ref{eq1.66}). From the R-transformation of
the states (\ref{eq2.81}) as given in (\ref{eqA.12}) we get the following:
\bea
\label{eq2.111b}
   (0,k_1,...,k_{n-1},0)|_{\cEvR,\dele}
   &=& \cRtil(0,k_1,...,k_{n-1},0)|_{\cEv,-\dele}\cRtil^\dagger,\\
\label{eq2.112}
   \Hhh_{\balp}(\cEvR,\dele)
   &=& \cRtil\,\Hhh_{\balp}(\cEv,-\dele)\,\cRtil^\dagger,
\eea
\bea
\label{eq2.113}
   \Delta E_{\balp}
   &=& \eh\Tr\lk[\,\Hhh_{\balp}(\cEv,\dele)-\Hhh_{\balp}(\cEvR,\dele)\rk]
   \nonumber\\
   &=& \eh\Tr\lk[\,\Hhh_{\balp}(\cEv,\dele)-\Hhh_{\balp}(\cEv,-\dele)\rk].
\eea

Let us now have a closer look at the traces occurring in
(\ref{eq2.107}). We define the spherical components of the
dipole operator $\unDv$ and the fields $\cEv^{'\iKsig}$
as usual:
\bea
\label{eq2.114}
   \unD_\pm &=& \mp\mbox{$\frac{1}{\sqrt2}$}(\unD_1\pm i\unD_2),\nonumber\\
   \unDn    &=& \unDd
\eea
and analogously $\cE_m'^\iKsig\ (m=\pm,0; \sig=1,2,3)$.
Under the reflection R (\ref{eq2.108b}) we have
\be
\label{eq2.115}
   \unR^\dagger\unD_m\unR = (-1)^m\unD_{-m}.
\ee
From (\ref{eq2.91}) and (\ref{eq2.111}) we get
\bea
\label{eq2.116}
   &&\eh\Tr(0,k_1,...,k_{n-1},0)\\
   &&\quad=\ \sum_{\sig,m}\lkDk 
      D^\iKvar{\sig_1}_{m_1},k_1,D^\iKvar{\sig_2}_{m_2},...,
      k_{n-1},D^\iKvar{\sig_n}_{m_n}\rkDk_{\dele}
      \cdot(-1)^{m_1+...+m_n}
      \cdot\cE^\isKvar{\sig_1}_{-m_1}
      \cdot\cdot\cdot\cE^\isKvar{\sig_n}_{-m_n},\nonumber
\eea
where we set
\bea
\label{eq2.117}
   &&\lkDk D^\iKvar{\sig_1}_{m_1},k_1,D^\iKvar{\sig_2}_{m_2},k_2,...,k_{n-1},
           D^\iKvar{\sig_n}_{m_n}\rkDk_{\dele}\\
   &&\quad =(-1)^n\eh\Tr\Big\{S_{\balp,0}\unD_{m_1}
            \Theta(R_1^\iKvar{\sig_1}-R_1)\,\Theta(R_1-R_1^\iKvar{\sig_1-1})
           \nonumber\\
   &&\qquad\qquad\times S_{\balp,k_1}\cdot\cdot\cdot S_{\balp,k_{n-1}}
           \unD_{m_n}
           \Theta(R_1^\iKvar{\sig_n}-R_1)\,\Theta(R_1-R_1^\iKvar{\sig_n-1})
           S_{\balp,0}\Big\}.\nonumber
\eea

The point is now to realize that these
expressions are invariant under rotations around the 3-axis
performed in the internal space only. This is clear
from the factorization of the states (\ref{eq2.81})
into the internal and the c.m.\  motion parts and the
definitions (\ref{eq2.90}), (\ref{eq2.117}) and leads to
the following results:
\bea
\label{eq2.118}
   \lkDk D_{m_1}^\iKvar{\sig_1},k_1,D_{m_2}^\iKvar{\sig_2},k_2,...,k_{n-1},
   D_{m_n}^\iKvar{\sig_n}\rkDk_{\dele}
   &=& 0\nonumber\\
   {\rm for}\quad \sum_{j=1}^nm_j
   &\not=& 0.
\eea

The time reversal transformation T gives (cf.\ (3.38)
of \cite{BBN95} and (\ref{eq2.98})):
\be
\label{eq2.119}
   \lkDk D^\iKvar{\sig_1}_{m_1},k_1,...,k_{n-1},D^\iKvar{\sig_n}_{m_n}
   \rkDk_{\dele} =
   \lkDk D^\iKvar{\sig_n}_{m_n},k_{n-1},...,k_1,D^\iKvar{\sig_1}_{m_1}
   \rkDk_{\dele}.
\ee
For $n=2$ we get a further relation from T. Due to
the factorization of the states (\ref{eq2.81}) into the
internal and c.m.\ parts we have
\be
\label{eq2.119a}
   \lkDk D^\iKvar{\sig_1}_{m_1},\ k_1,\ D^\iKvar{\sig_2}_{m_2}\rkDk_{\dele}
 = \lkDk D^\iKvar{\sig_1}_{m_2},\ k_1,\ D^\iKvar{\sig_2}_{m_1}\rkDk_{\dele},
\ee
as is proved easily using (3.38) and (3.47)
of \cite{BBN95} and (\ref{eq2.90}), (\ref{eq2.117}).\\

The reflection (\ref{eq2.108b}) gives
\bea
\label{eq2.120}
   {\cER'}^\iKsig_{m}
   &=& (-1)^m\,{\cE'}^\iKsig_{-m},\\
   \lkDk D^\iKvar{\sig_1}_{m_1},k_1,...,k_{n-1},
         D^\iKvar{\sig_n}_{m_n}\rkDk_{\dele}
   &=& (-1)^{\sum^n_{j=1}m_j}\,
   \lkDk D^\iKvar{\sig_1}_{-m_1},k_1,...,k_{n-1},
         D^\iKvar{\sig_n}_{-m_n}\rkDk_{-\dele}.\nonumber\\
\label{eq2.121}
\eea

With these relations we get for
$\Delta E_{\balp}$ (\ref{eq2.109}), (\ref{eq2.113}):
\bea
\label{eq2.122}
   \Delta E_{\balp}
   &=& \sum^\infty_{n=1}\Delta E^{(n)}_{\balp},
   \nonumber\\
   \Delta E^\iKen_{\balp}
   &=& {\sum}_\iKvar{n-1}'\sum_{\sig,m}
       \eh\lk[\lkDk D^\iKvar{\sig_1}_{m_1},k_1,...,k_{n-1},
                    D^\iKvar{\sig_n}_{m_n}\rkDk_{\dele}
             -\lkDk D^\iKvar{\sig_1}_{m_1},k_1,...,k_{n-1},
                    D^\iKvar{\sig_n}_{m_n}\rkDk_{-\dele}\rk]\nonumber\\
   & &\qquad\times\lk[(-1)^{\sum^n_{j=1}m_j}\,
      \cE^\isKvar{\sig_1}_{-m_1}\cdot\cdot\cdot\cE^\isKvar{\sig_n}_{-m_n}
     -\cE^\isKvar{\sig_1}_{m_1} \cdot\cdot\cdot\cE^\isKvar{\sig_n}_{m_n}
      \rk],
\eea
where only the terms with $\Sigma m_j=0$ are non-zero (cf.\ (\ref{eq2.118})).
In (\ref{eq2.122}) the energy shifts $\Delta E^{(n)}_{\balp}$
are written in a way to exhibit explicitly that they are
odd functions of $\dele$ and proportional to parity-odd
combinations of the electric field components ${\cE'}_m^\iKsig$.

Explicitly we get from (\ref{eq2.122}) for the lowest
$n$-values:
\bea
\label{eq2.123}
   \Delta E^\iKe_{\balp}
   &=& \eh\Tr(0,0)_{\dele}-\eh\Tr(0,0)_{-\dele}\nonumber\\
   &=& \sum_\sig\eh \lk[\lkDk D^\iKsig_0\rkDk_{\dele}
                       -\lkDk D^\iKsig_0\rkDk_{-\dele}\rk]
       \lk[\cE^\isKvar{\sig}_0-\cE^\isKvar{\sig}_0\rk]\nonumber\\
   &=& 0,
\eea
\bea
\label{eq2.124}
   \Delta E_{\balp}^\iKz
   &=& \eh\Tr(0,1,0)_{\dele}-\eh\Tr(0,1,0)_{-\dele}\nonumber\\
   &=& \eh\sum_\sig\Big[
       \lkDk D^\iKvar{\sig_1}_+,1,D^\iKvar{\sig_2}_-\rkDk_{\dele}
      -\lkDk D^\iKvar{\sig_1}_+,1,D^\iKvar{\sig_2}_-\rkDk_{-\dele}\nonumber\\
   & &\qquad
      -\lkDk D^\iKvar{\sig_1}_-,1,D^\iKvar{\sig_2}_+\rkDk_{\dele}
      +\lkDk D^\iKvar{\sig_1}_-,1,D^\iKvar{\sig_2}_+\rkDk_{-\dele}\Big]
      \lk[\cE^\isKvar{\sig_1}_-\cE^\isKvar{\sig_2}_+-
          \cE^\isKvar{\sig_1}_+\cE^\isKvar{\sig_2}_-\rk]\nonumber\\
   &=& 0,
\eea
where we used (\ref{eq2.119a}). In a similar way we find
\be
\label{eq2.124.1}
   \Tr(0,2,0)_{\dele}-\Tr(0,2,0)_{-\dele} = 0.
\ee
The first non-vanishing result is the term with $n=3$. We get
from (\ref{eq2.108a}):
\bea
\label{eq2.125}
   \Delta E^\iKd_{\balp}
   &=& \eh\Tr(0,1,1,0)_{\dele}-\eh\Tr(0,1,1,0)_{-\dele}\nonumber\\
   & &\ -\ \eh \Tr(0,2,0)_{\dele}\eh \Tr(0,0)_{\dele}
        +\eh \Tr(0,2,0)_{-\dele}\eh \Tr(0,0)_{-\dele}.
\eea
From (\ref{eq2.123}) and (\ref{eq2.124.1})
we find that the 3rd and 4th terms
on the r.h.s. cancel. Thus we finally obtain:
\bea
\label{eq2.126}
   \Delta E_{\balp}^\iKd
   &=& \sum_{\sig,m}\eh\lk[
       \lkDk D^\iKvar{\sig_1}_{m_1},1,
             D^\iKvar{\sig_2}_{m_2},1,
             D^\iKvar{\sig_3}_{m_3}\rkDk_{\dele}
      -\lkDk D^\iKvar{\sig_1}_{m_1},1,
             D^\iKvar{\sig_2}_{m_2},1,
             D^\iKvar{\sig_3}_{m_3}\rkDk_{-\dele}\rk]\nonumber\\
   & &\qquad\times\lk[(-1)^{m_1+m_2+m_3}
   \cE^\isKvar{\sig_1}_{-m_1}
   \cE^\isKvar{\sig_2}_{-m_2}
   \cE^\isKvar{\sig_3}_{-m_3}
  -\cE^\isKvar{\sig_1}_{m_1}
   \cE^\isKvar{\sig_2}_{m_2}
   \cE^\isKvar{\sig_3}_{m_3}\rk].
\eea

To show that this term is actually non-zero, let us
consider the example where the perturbation is absent
in Section $\sig=2$:
\be
\label{eq2.127}
   {\cEv'}^\iKz=0.
\ee
We then get from (\ref{eq2.126}):
\be
\label{eq2.128}
   \Delta E^\iKd_{\balp}
   =\dele\lk[c^\iKe_{\balp}\evd\cdot{\cEv'}^\iKe
            -c^\iKd_{\balp}\evd\cdot{\cEv'}^\iKd\rk]
    \frac{1}{\cE^3}\lk(\evd\cdot\lk({\cEv'}^\iKe\times{\cEv'}^\iKd\rk)\rk),
\ee
where we set for $\sig=1,3$:
\bea
\label{eq2.129}
   \dele c^\iKsig_{\balp}
   &=& -2i\,\cE^3\cdot\lk\{
       \lkDk D^\iKsig_+,1,D^\iKsig_0,1,D^\iKtau_-\rkDk_{\dele}
      +\lkDk D^\iKsig_0,1,D^\iKsig_+,1,D^\iKtau_-\rkDk_{\dele}\rk.\nonumber\\
   & & \qquad\lk.
      +\lkDk D^\iKsig_+,1,D^\iKtau_-,1,D^\iKsig_0\rkDk_{\dele}
      -(\dele\to-\dele)\rk\},
\eea
($\tau=3$ for $\sig=1$; $\tau=1$ for $\sig=3$).\\

To give some numbers we consider a box (Fig.\ 1) of dimension
\be
\label{eq2.130}
   A_1 = A_2 = A_3 = 8 \mbox{nm}
\ee
and segment lengths (cf.\ (\ref{eq2.2})):
\be
\label{eq2.131}
   R_1^\iKe = 0.4\cdot A_1,\qquad
   R_1^\iKz = 0.6\cdot A_1.
\ee
In Fig.\ 2a we plot the real parts of the corresponding unperturbed
\nopagebreak
eigenenergies $E^\iKn_{\balp}$ (\ref{eq2.85}) for our
``hydrogen'' atom with $I=0$ as function of the electric field
$\cE$. We note that the transverse parts of the wave functions,
characterized by the integers $n_2,n_3$ (cf.\ (\ref{eq2.81})) do
not mix under perturbations by $\heffe$ (\ref{eq2.80}) for
our case here $(q=0)$. Thus we consider only the levels with
the lowest transverse energies corresponding to $n_2=n_3=1$. For
the Lamb shift $L$ and the $2P_{3/2}-2P_{1/2}$ energy difference
$\Delta$ as well as the widths $\Gamma_S,\Gamma_P$ we take the
values from real hydrogen (cf.\ Table III of \cite{BBN95}). The
zero of the energy scale is put at the energy of the $2P_{1/2}$
level in vacuo.

Now we introduce the perturbation in the electric field
\be
\label{eq2.132}
   {\cEv'}^\iKe = \lk(\ba{c} \cE'\\ 0       \\ \cE'\ea\rk),\quad
   {\cEv'}^\iKz = 0,\quad
   {\cEv'}^\iKd = \lk(\ba{c} 0   \\ \eta\cE'\\ 0   \ea\rk).
\ee
This leads to an energy shift $E_{\balp}^\iKn\to E_{\balp}$
as calculated in (\ref{eq2.107}) and to a P-violating energy
difference $\Delta E_{\balp}$. This is given to lowest
nontrivial order in (\ref{eq2.128}), from which we see that
$\Delta E^\iKd_{\balp}$ is proportional to $\eta{\cE'}^3$
and thus changes sign if we change the chirality of the field
arrangement $(\eta\to -\eta)$:
\be
\label{eq2.133}
   \Delta E^\iKd_{\balp}=\dele c^\iKe_{\balp}
   \eta\lk(\frac{\cE'}{\cE}\rk)^3.
\ee
We plot the real and imaginary parts of the coefficient $c_{\balp}^\iKe$
for the state $\balp=(2S_{1/2},|F_3|=1/2,
n_1=1,n_2=1,n_3=1)$ in Fig.\ 3. Of course, we would like to
make $|\Delta E^\iKd_{\balp}|$ as big as possible
by increasing $\cE'$. However, there are limits to this in our
{\it calculation}. The perturbation treatment is only reliable
as long as the energy shifts induced by the perturbation
are small compared to the spacing of the unperturbed levels.
Thus, we can at best trust our calculation for values of
$|\cE'|\leq\cE'_{max}$ which
we estimate from the requirement that the first plus
second order energy shift is equal to the minimum of the
unperturbed energy differences:
\be
\label{eq2.134}
   |E_{\balp}^\iKe(\pm\cE_{max}')|+|E_{\balp}^\iKz(\pm\cE_{max}')|
   = \min_{\bbet\not=\balp} |E_{\bbet}^\iKn-E_{\balp}^\iKn|.
\ee
Here we choose the plus or minus sign according to which
gives the lower value for $\cE_{max}'$.
We plot $\cE_{max}'$ in Fig.\ 4 for the level
$\balp=(2S_{1/2},|F_3|=1/2,n_1=1,n_2=1,n_3=1)$.
The P-violating energy and lifetime shifts for $\cE'=\cE_{max}'$
are then shown in Fig.\ 5. From this calculation we can thus establish
theoretically for our ``hydrogen'' atom with $I=0$ a P-violating
energy shift of the order of
\be
\label{eq2.135}
   \Re\Delta E_{\balp}^\iKd/h\simeq\dele\cdot 5\cdot10^3\,\mbox{Hz}
   = -2.46\cdot10^{-9}\,\mbox{Hz}.
\ee

Since $\dele$ is of the order $10^{-11}...10^{-12}$ (cf.\ (3.20),
(3.21) of \cite{BBN95}), this is an extremely small energy
shift. Fortunately there is --- at least in theory --- a way
to enhance the effect dramatically. In fact, we will show now
how to obtain energy shifts not of order $\dele$ but
of order $\sqrt{\dele}$.

Consider Figure 2a. There is a crossing between the levels $\balp=(2S_{1/2},|F_3|=1/2;\nv=(1,1,1))$ and $\bbet = (2P_{1/2},|F_3|=1/2;\nv=(2,1,1))$ in the real parts at an electric field strength $\cE\simeq 500$ V/cm. Correspondingly, the maximally allowed perturbation field strength $\cE'_{max}$ is at a minimum at $\cE\simeq 500$ V/cm as can be seen in Fig.\ 4. This can be easily understood from (\ref{eq2.134}) and the fact that $|\Im E^\iKn_{\balp}/h|\simeq 16\,$MHz and $|\Im E^\iKn_{\bbet}/h|\simeq 38\,$MHz are by two orders of magnitude smaller than $|\Re E^\iKn_{\balp,\bbet}/h|\simeq 1.6\,$GHz (cf.\ Figs.\ 2a, 2b). Then, the r.h.s.\ of (\ref{eq2.134}) becomes very small at the zero of the real part $\Re(E^\iKn_{\bbet}-E^\iKn_{\balp})$ and so does $\cE'_{max}$. However, this only means that our calculation as done so far is limited. In fact,
the P-violating energy and lifetime shifts at such a crossing
of levels may nevertheless have a maximum. We will show this in
the following using again degenerate perturbation theory
treating all the crossing levels together.

We consider a configuration with a crossing of levels in
both the real and the imaginary parts at the same field
strength:
\bea
\label{eq2.136a} 
   \Re\lk(E^\iKn_{\bbet}(\cE\evd)-E^\iKn_{\balp}(\cE\evd)\rk)=0,\\
\label{eq2.136b} 
   \Im\lk(E^\iKn_{\bbet}(\cE\evd)-E^\iKn_{\balp}(\cE\evd)\rk)=0.
\eea
We find (cf.\ Fig.\ 2b) that (\ref{eq2.136b}) is fulfilled
at
\be
\label{eq2.136c}
   \cE_{res}\simeq 1024\ \mathrm{V/cm}
\ee
for $\alpha=2S_{1/2},\ \beta=2P_{1/2}$. If one now
adjusts the box lengths to
\be
\label{eq2.136d}
   A_1 = 5.84\ {\mathrm nm},\quad 
   A_2 = A_3 = 8\ {\mathrm nm},
\ee
then (\ref{eq2.136a}) is satisfied for $\balp=(2S_{1/2},|F_3|=1/2;n_1=1),
\bbet=(2P_{1/2},|F_3|=1/2;n_1=2)$ and so is (\ref{eq2.136b}) as there
are no contributions to the imaginary part of the energy from the c.m.\ motion. Note that (\ref{eq2.136a}) could also be fulfilled by adjusting
$A_2,A_3$, but the differences between energy levels
with different c.m.\ quantum numbers $n_2,n_3$ will not
appear for neutral atoms as energy denominators in the
perturbation series, since the perturbation fields
$\cEsvs$ are chosen not to vary along the 2- and 3-directions.
Then, due to orthogonality, only matrix elements
\[
  \lbrak{\alpha, F_3;(n_1,n_2,n_3)}\heffe|\beta,F_3';(n_1',n_2,n_3))
\]
diagonal in $n_2,n_3$ are nonzero.

In order to calculate a reliable value for
the P-violating energy shifts, we use degenerate
perturbation theory in the 4-dimensional subspace
spanned by the states
\be
\label{eq2.136e}
  |\balp,\pm):=
  |2S\eh,\pm|F_3|,\cE_{res}\evd;(1,1,1)),\quad
  |\bbet,\pm):=
  |2P\eh,\pm|F_3|,\cE_{res}\evd;(2,1,1)),
\ee
where $|F_3|=1/2$.
In analogy to (\ref{eq2.87})--(\ref{eq2.89}) we define
the matrices:
\bea
\label{eq2.137}
   \Pro^\iKn_{(\balp\bbet)}
   &:=& \sum_{m=\pm}\lk[|\balp,m)\lbrak{\balp,m}
                       +|\bbet,m)\lbrak{\bbet,m}\rk],\\
\label{eq2.138}
   \QPro^\iKn_{(\balp\bbet)}
   &:=& 1-\Pro^\iKn_{(\balp\bbet)},\\
\label{eq2.139}
   \frac{\QPro^\iKn_{(\balp\bbet)}}{\eps^k}
   &:=& \sum_{\bar\gam\not=\balp\atop\bar\gam\not=\bbet}
       \frac{\Pro^\iKn_{\bgam}}
       {(E^\iKn_{\balp}-E^\iKn_{\bar\gam})^k},\quad(k=1,2,...)\\
\label{eq2.140}
   S_{(\balp\bbet),0}
   &:=& -\Pro^\iKn_{(\balp\bbet)},\nonumber\\
   S_{(\balp\bbet),k}
   &:=& \frac{\QPro^\iKn_{(\balp\bbet)}}{\eps^k},\quad {\rm for}\quad k=1,2,...
\eea
(Remember that in (\ref{eq2.139}) $E^\iKn_{\balp}=E^\iKn_{\bbet}\not=E^\iKn_{\bgam}$).
Using these matrices we define, similarly to (\ref{eq2.91}),
the operators
\be
\label{eq2.141}
   (k_1,...,k_{n+1}) 
   := S_{(\balp\bbet),k_1}\heffe S_{(\balp\bbet),k_2}
      \cdot\cdot\cdot\heffe S_{(\balp\bbet),k_{n+1}}.
\ee

In the following we will show that all energy levels
remain at least twofold degenerate to all orders of perturbative
theory also in this case. We define
\bea
\label{eq2.142}
   \Kh_{(\balp\bbet)}
   &:=& \Pro^\iKn_{(\balp\bbet)}+\sum^\infty_{n=2}{A'}^\iKen,\\
\label{eq2.143}
   \Hh_{(\balp\bbet)}
   &:=& E^\iKn_{\balp}\Kh_{(\balp\bbet)}+\sum^\infty_{n=1}{B'}^\iKen,
\eea
with ${A'}^\iKen,{B'}^\iKen$ defined as in (\ref{eq2.93}),
but in terms of the operators (\ref{eq2.141}). 

Instead of discussing now Kato's eigenvalue problem (\ref{eq2.96})
we will now introduce and use a modified version of Bloch's method \cite{Bloch58}. We define a matrix $\Htil_{(\balp\bbet)}$ by the conditions
(see Appendix \ref{appC}):
\bea
\label{eq2.147a}
   (\Kh_{(\balp\bbet)})^{1/2}\Htil_{(\balp\bbet)}(\Kh_{(\balp\bbet)})^{1/2} 
   &=& \Hh_{(\balp\bbet)},\nonumber\\
   \Htil_{(\balp\bbet)}\Pro^\iKn_{(\balp\bbet)} 
   &=& \Htil_{(\balp\bbet)},\nonumber\\
   \Pro^\iKn_{(\balp\bbet)}\Htil_{(\balp\bbet)} 
   &=& \Htil_{(\balp\bbet)}.
\eea
Bloch's operator $\Hhh_{(\balp\bbet)}$ defined as in (\ref{eq2.102}) with $\balp\rightarrow(\balp\bbet)$ is in general non-hermitian even if $\Hh_{(\balp\bbet)}$ and $\Kh_{(\balp\bbet)}$ are hermitian. In contrast, 
$\Htil_{(\balp\bbet)}$ is hermitian if  $\Hh_{(\balp\bbet)}$ and $\Kh_{(\balp\bbet)}$ are hermitian, which would be the case for stable states, i.e.\ $\Gam_S=\Gam_P=0$. We will need this property of $\Htil_{(\balp\bbet)}$ later when we want to check, that our effect proportional to $\sqrt{\dele}$ vanishes for the case of stable states.
We assume that the perturbation expansion of $\Htil_{(\balp\bbet)}$ is well defined and converges (cf.\ Appendix \ref{appC} for the conditions for this to be the case). The eigenvalue problem (\ref{eq2.96}) is then equivalent to the following standard diagonalization problem:
\be
\label{eq2.147b}
   (\,\Htil_{(\balp\bbet)} - E\,)\,|\,\,\ra = 0,
\ee
where $|\,\,\ra \in \cR^\iKn_{(\balp\bbet)}$.

Now we will show, that the eigenvalue problem factorizes and leaves a twofold degeneracy of the eigenenergies. We define the 2$\times$2 matrices:
\be
\label{eq2.148}
   \unHtil_{\bgam'\bgam}
   := \lk(\lbrak{\bgam',m'}\Htil_{(\balp\bbet)}|\bgam,m)\rk),
\ee
where $\bgam',\bgam\in\{\balp,\bbet\};m',m\in\{+1,-1\}$.
Now the secular matrix reads:
\be
\label{eq2.149}
   \unS(E):=\lk(\ba{cc}
   \unHtil_{\balp\balp}-E &\unHtil_{\balp\bbet}\\
   \unHtil_{\bbet\balp}   &\unHtil_{\bbet\bbet}-E\ea\rk).
\ee
By simple manipulations we obtain that
\be
\label{eq2.150}
   \det\unS(E)=\det\unS'(E)\lk[\det\unHtil_{\balp\bbet}\det
   \unHtil_{\bbet\balp}\rk]^{-1},
\ee
where
\be
\label{eq2.151}
   \unS'(E):=\lk(\ba{cc}
   \unHtil_{\balp\balp}-E
  &\unHtil_{\balp\bbet}   \unHtil_{\bbet\balp}   \\
   \unHtil_{\balp\bbet}   \unHtil_{\bbet\balp}   
  &\unHtil_{\balp\bbet}   (\unHtil_{\bbet\bbet}-E)\unHtil_{\bbet\balp}
   \ea\rk).
\ee
Using T-invariance we show in Appendix \ref{appC} that $\unHtil_{\balp\balp}$, $\unHtil_{\bbet\bbet}$ and $\unHtil_{\balp\bbet}\unHtil_{\bbet\balp}$ are all multiples of the 2$\times$2 unit matrix.

Then it follows from (\ref{eq2.150}), (\ref{eq2.151}) that
\be
\label{eq2.153}
   \det \unS(E) = \Big((a-E)(b-E) - c\Big)^2,
\ee
where
\bea
\label{eq2.152}
   a &:=& \eh\Tr\,\unHtil_{\balp\balp},\nonumber\\
   b &:=& \eh\Tr\,\unHtil_{\bbet\bbet},\nonumber\\
   c &:=& \eh\Tr(\unHtil_{\balp\bbet}\unHtil_{\bbet\balp}).
\eea
Thus $\det\unS(E)$ is a perfect square, all eigenvalues $E$ are twofold degenerate and $\det\unS(E)=0$ if and only if
\be
\label{eq2.154}
   (a-E)(b-E) - c = 0.
\ee
This is a quadratic equation for $E$ which has the solutions
\bea
\label{eq2.155}
   E_{\balp\bbet,\pm} 
         &:=& \eh\lk[a+b\pm\sqrt{(a-b)^2+4c}\,\rk]\nonumber\\
         &=&  \eh\lk[a+b\pm\sqrt{[-i(a-b)+2\sqrt{c}\,]
                                [\,i(a-b)+2\sqrt{c}\,]}\,\rk].
\eea
The complex numbers $a,b,c$ can be expanded in $\dele$:
\bea
\label{eq2.156}
   a &=& a_0 + \dele a_1 +...\nonumber\\
   b &=& b_0 + \dele b_1 +...\nonumber\\
   c &=& c_0 + \dele c_1 +...
\eea
and thus we get for the eigenenergies (\ref{eq2.155}):
\bea
\label{eq2.157}
   E_{\balp\bbet,\pm} 
   &=& \eh\Big[a_0+b_0+\dele(a_1+b_1)+\cO(\dele^2)\nonumber\\
   & & \quad\quad\pm\Big\{[-i(a_0-b_0)+2\sqrt{c_0}\,]
                 [\,i(a_0-b_0)+2\sqrt{c_0}\,]\nonumber\\
   & & \quad\quad+\dele[2(a_0-b_0)(a_1-b_1)+4c_1]+\cO(\dele^2)\Big\}^{1/2}\Big]
\eea
Now it is clear that in the case 
\be
\label{eq2.158}
  (a_0-b_0)^2+4c_0 \not= 0
\ee
we will get at best P-violating energy shifts linear in $\dele$. If, however, 
\be
\label{eq2.159}
   i(a_0-b_0)+2\sqrt{c_0} = 0
\ee
or 
\be
\label{eq2.159.1}
  -i(a_0-b_0)+2\sqrt{c_0} = 0
\ee
and together with (\ref{eq2.159}) or (\ref{eq2.159.1}):
\be
\label{eq2.160}
  2(a_0-b_0)(a_1-b_1)+4c_1 \not= 0,
\ee
we will obtain 
\bea
\label{eq2.161}
   E_{\balp\bbet,\pm} 
   &=& \eh\Big[a_0+b_0
       \pm\sqrt{\dele}\Big\{2(a_0-b_0)(a_1-b_1)+4c_1\Big\}^{1/2}\nonumber\\
   & & \ \ +\cO(\dele)\Big].
\eea
This is a P-violating splitting which {\it can} be
distinguished from a P-conserving one as it changes its phase by a
factor of $i$ under a change of sign of $\dele$ or a reversal of the handedness of the external field arrangement which changes the signs of $a_1,b_1,c_1$.
In other words: we have to search for a degeneracy of two complex energy levels (\ref{eq2.155}) in an external electric field with non-zero handedness in the absence of P-violation, i.e.\ for $\dele=0$. Then the small P-violation effects proportional to $\dele$ in the matrix elements can produce an energy shift of order $\sqrt{\dele}$.

Note, that this is clearly an effect which results from the non-hermiticity of $\Htil_{(\balp\bbet)}$, i.e., which is only possible for unstable states. For the case of stable states, $\Htil_{(\balp\bbet)}$ is hermitian (cf.\ (\ref{eq2.147a})) and therefore we get for the submatrices (\ref{eq2.148})
\be
\label{eq2.161a}
   \unHtil_{\balp\balp} = \unHtil_{\balp\balp}^\dagger,\qquad
   \unHtil_{\bbet\bbet} = \unHtil_{\bbet\bbet}^\dagger,\qquad
   \unHtil_{\balp\bbet} = \unHtil_{\bbet\balp}^\dagger.
\ee
From (\ref{eq2.152}) we find then that $a,b,c$ and likewise $a_0,b_0,c_0$, and $a_1,b_1,c_1$ are real and
\bea
\label{eq2.161b}
   c\ge0,\quad c_0\ge0;\nonumber\\
   c_1=0\quad\mbox{if}\quad c_0=0.
\eea
Now the condition (\ref{eq2.159}) is equivalent to (\ref{eq2.159.1}) and leads to $a_0=b_0$ and $c_0=0$ which implies through (\ref{eq2.161b}) also $c_1=0$.
Therefore, (\ref{eq2.160}) cannot be fulfilled simultaneously with (\ref{eq2.159}) or, equivalently, (\ref{eq2.159.1}) for the case of stable states, and there will be no energy shifts proportional to $\sqrt{\dele}$ for them.\\

On the contrary, if $a,b$ and $c$ are complex numbers, which is the case for unstable states, one can in general fulfill (\ref{eq2.159}) or (\ref{eq2.159.1}) together with (\ref{eq2.160}). We will show by the results of a numerical investigation, that this is indeed the case. We choose the following general configuration of three perturbation fields:
\bea
\label{eq2.162}
   \cEsvs &=& \etav^\iKsig\cE';\quad(\sig=1,2,3).
\eea
In Appendix \ref{appD} we give a proof that for sufficiently small $\cE'$ there
exist indeed ``resonance'' values for the vectors $\etav^\iKsig$ where energy shifts proportional to $\wdele$ occur. As an example we choose $\cE'=3\,$V/cm. Then the calculation to first order in $\cE'$ gives the following resonance values:
\be
\label{eq2.168.9}
   \etav^\iKe_{res,-} = \lk( \ba{c} 1   \\ 
                                    0   \\
                                    -61.31 \ea\rk),\quad
   \etav^\iKz_{res,-} = \lk( \ba{c} 0   \\ 
                                    0   \\
                                    574.93 \ea\rk),\quad
   \etav^\iKd_{res,-} = \lk( \ba{c} 0   \\ 
                                    1   \\
                                    -63.31 \ea\rk),
\ee
where (\ref{eq2.159.1}) and (\ref{eq2.160}) are fulfilled. Higher orders in $\cE'$ can only shift these values slightly.\\

In Figs.\ 6\,a and b we plot the decadic logarithms of the moduli of the real and imaginary parts of the P-violating energy shift (cf.\ (\ref{eq2.161}))
\be
\label{eq2.168.13}
  E_{\balp\bbet,+}(\{\cEv^{'\iKsig}\})-E_{\balp\bbet,+}(\{\cEvR^{'\iKsig}\})
\ee 
for our ``hydrogen'' in a field configuration given by (\ref{eq2.131}), (\ref{eq2.136c}) and (\ref{eq2.168.9}) vs.\ the deviation $\eta_3^\iKz-\eta_{3,res,-}^\iKz$ of $\eta_3^\iKz$ from the resonance value $\eta_{3,res,-}^\iKz=574.93$ for which the P-even splitting of levels is removed, revealing an enhancement of the P-violating splitting of the order of $1/\sqrt{\dele}$. The width of the peak is proportional to $\sqrt{\dele}$ and found to be $\Del\eta^\iKz_3 \simeq 0.4\cdot10^{-4}$, corresponding to $\cE'\Del\eta^\iKz_3=1.2\cdot10^{-4}\,$V/cm. 
Note, however, that in Fig.\ 6a the differences of the real parts of the energies of the levels
are at maximum of the order of $10^{-2}\,$Hz whereas the lifetime of the levels is calculated to be $0.29\cdot10^{-8}\,\mathrm{s}$. This corresponds to a line width $\Delta E/h = 5.44\cdot10^7\,$Hz. Thus it is certainly not an easy task to detect frequency shifts of order $10^{-2}\,$Hz in such broad lines.
But our calculation is only meant to provide an illustration how energy shifts proportional to $\sqrt{\dele}$ can be obtained. Certainly, for a realistic experimental situation one would try to optimize such an effect.

%
\subsection{Results for an ion with $I=0$}
\label{sec3.2}
In this section we will discuss the modifications one has to include
for the physically relevant case of ions without nuclear spin,
as $\vzHep$.
As we have already  discussed in Section \ref{sec2.2} one has to introduce
Airy functions instead of sines and cosines for the c.m.\ wave
functions. We write the (right) eigenstates of $\heffn$ as
\bea
\label{eq2.169}
   |\alpha,F_3,\cE\evd,\nv) 
   &:=& \revrE\cdot\phi_{\nv}(\Rv),\nonumber\\
   \phi_{\nv}(\Rv)
   &:=& \cN_{n_3}\lk[\psi_+(R_3,E_{n_3},\cE)-\xi_{n_3}
        \psi_-(R_3,E_{n_3},\cE)\rk]\nonumber\\
   & & \times\prod^2_{i=1}\sqrt{\frac{2}{A_i}}\sin\lk(\frac{n_i\pi R_i}{A_i}
   \rk),
\eea
where the Airy functions $\psi_\pm$ are defined as  in
(\ref{eq2.59a}), (\ref{eq2.59b}), (\ref{eq2.61}) f. The eigenenergies $E_{n_3}$
corresponding to the Airy  modes, which are numbered by $n_3$, are
obtained by solving a determinant equation analogous to
(\ref{eq2.63}), the coefficients $\xi_{n_3}$  and the normalization
$\cN_{n_3}$ follow from (\ref{eq2.61}) and (\ref{eq2.64}), respectively.
Using the eigenstates (\ref{eq2.169}) one computes the perturbation series
in an analogous manner as described in (\ref{eq2.87})--(\ref{eq2.95}) for Kato's and (\ref{eq2.102})--(\ref{eq2.105}) for Bloch's method.
Note that the perturbation $\heffe$ (\ref{eq2.80}) to the effective Hamiltonian $\heffn$ now
contains the additional term
\be
\label{eq2.171}
   -q\,\Rv\sum^3_{\sig=1}{\cEv'}^\iKsig
   \Theta(R_1^\iKsig- R_1               )
   \Theta(R_1       - R_1^\iKvar{\sig-1}),
\ee
which was absent in the $q=0$ case.
Therefore, the terms of order $n=1,2,...$ of the series (\ref{eq2.92}), (\ref{eq2.93}), (\ref{eq2.104}) have the following structure (symbolically)
\bea
\label{eq2.171a}
   {A'}^\iKen &\prop& \lk( (q\Rv + \unDv)^n \rk)\nonumber\\
            &\prop& \lk( (\unDv)^n + n(\unDv)^{n-1}(q\Rv) + ...\rk)
\eea
and similarly for ${B'}^\iKen$, ${C'}^\iKen$.

Since matrix elements of (\ref{eq2.171}) w.r.t.\ the states (\ref{eq2.169}) are diagonal in the internal
quantum numbers $(\alpha, F_3)$, $\dele$-linear
P-violating contributions to the series can only arise
from matrix elements of $\unDv$. Therefore, up to the third
order, $n\leq3$, there will be no P-violating contributions
to the energy shifts $\Delta E^\iKen_{\balp}$ (cf.\ (\ref{eq2.123})\,ff.) involving matrix elements of (\ref{eq2.171}). This follows
from our result that such contributions appear only for
$n\geq3$ in the $q=0$ case. Hence, the expressions (\ref{eq2.123}),
(\ref{eq2.124}), (\ref{eq2.128})\,f. for the P-violating energy shifts $\Delta E_{\balp}$ up to $n=3$ remain
valid for ions with $I=0$.

However, when one calculates the maximally allowed perturbation field
strength $\cE'_{max}$ using (\ref{eq2.134}), the contributions
from the matrix elements of (\ref{eq2.171})
to the P-conserving energy shifts $E^\iKe_{\balp},E^\iKz_{\balp}$ have
to be taken into consideration.\\

We will now present results for $\vzHep$-ions in a box of lenghts: 
\be
\label{eq2.172}
   A_1=A_2=A_3 =1 \mathrm{nm}.
\ee
Note that compared to the case of our ``hydrogen'' (cf.\ (\ref{eq2.130})) we rescaled the $A_i$ by a factor
\be
\label{eq2.173}
   Z^{-2}(Z+N)^{-1/2}=1/8
\ee
for $\vzHep$ ($Z=N=2$) in order to obtain an approximate scaling factor of
$Z^4$ for the c.m.\ energies. For $\cE'=0$ the contribution of the c.m.-mode in $i$-direction reads 
\be
\label{eq2.174}
   E_{n_i}
   =      \frac{n_i^2\pi^2}{2\,m\,A_i^2}.
\ee
As the mass of the nucleus of a light atom may be approximated by $(Z+N)$ times the mass of the proton, we obtain the desired scaling factor $Z^4$ for the c.m.\ energies (\ref{eq2.174}), if we multiply the widths $A_i$ of the box used for hydrogen with (\ref{eq2.173}): 
\be
\label{eq2.174.1}
   \frac{n_i^2\pi^2 Z^4 [Z+N]}{2\,(Z+N)\,m\,A_i^2}
   = Z^4E_{n_i}\quad (i=1,2,3;\ n_i=1,2,...).
\ee
As it was mentioned in section 3.2 of \cite{BBN95}, one obtains ``reduced $Z=1$'' problems from the general $Z$ case if one multiplies all electric field strengths of the $Z=1$ case by $Z^5$. In the following we will choose such rescaled fields in addition to the rescaled box widths and observe, that all energies will approximately be larger by a factor of $Z^4=16$ compared to the results for ``hydrogen''. 

In Fig.\ 7 we plot the lowest levels of
the c.m.\ motion in 3-direction vs.\ the electric field strength
$\cE$, which are the solutions to (\ref{eq2.63}), the
eigenenergies of the Airy states. As the zero of the energy scale we choose the ``minimum'' of the energy in 3-direction
\be
\label{eq2.171.1}
  E_{3,min}:= \lk\{\ba{lll}
              0            & \mbox{for} & \cE\le0,\\
              -q\,A_3\,\cE & \mbox{for} & \cE>0.  \ea\rk.
\ee
For $\cE=0$ the quadratic
dependence on $n_3$ corresponding to the sine-/cosine modes
is recovered.

The Lamb shift $L$ and the energy difference $\Delta$ as well
as the widths $\Gam_S, \Gam_P$ for $\vzHep$ are
given in Table II of \cite{BBN95}.
In the following we again restrict ourselves to levels with the lowest c.m.\ energies corresponding to $n_i=1,\ i=1,...,3$. 

We plot $\cE'_{max}$ in Fig.\ 8 for the level
$\balp=(2S_{1/2},|F_3|=\eh,\nv=(1,1,1))$. In Figs.\ 9\,a, b we show the P-violating (a) energy shifts and (b)
shifts of the decay width for $\cE'=\cE'_{max}$. We have
chosen the same configuration of the perturbation fields as
in (\ref{eq2.131}), (\ref{eq2.132}) with $\eta=1$.

From our calculation we obtain a theoretical energy shift of
the order of
\be
\label{eq2.175} 
   \Re\Delta E^\iKd_{\balp}/h\simeq\dele\cdot 1.2\cdot 10^5\ {\mathrm Hz}
   = 1.36\cdot10^{-6}\ {\mathrm Hz},
\ee
which is larger by a factor of $550\gsim Z^4\dele(2,2)/\dele(1,0)$
compared to the result (\ref{eq2.135}) for the ``hydrogen atom'' with
$I=0$.

As one can expect there is the same possibility as in the $q=0$ case
to obtain P-violating energy shifts of the order of $\sqrt{\dele}$.

The calculation is the same as described in Section \ref{sec3.1},
(\ref{eq2.136a})--(\ref{eq2.162}), and Appendix \ref{appD}, (\ref{eq2.163})--(\ref{eq2.168}) and (\ref{eq2.168.4})\,f., except that it now
involves the states (\ref{eq2.169}) and the
additional piece (\ref{eq2.171}) to $\heffe$.
The secular matrix may be block-diagonalized in the same
way as in (\ref{eq2.148})--(\ref{eq2.153}) because (\ref{eq2.171}) is diagonal  in the internal quantum
numbers $(\gam,m)$.

We have chosen the crossing of the levels
$\balp=(2S_{1/2},\nv=(1,1,1))$,
$\bbet=(2P_{1/2},\nv=(2,1,1))$, (\ref{eq2.136a}), (\ref{eq2.136b}), 
which occurs at:
\bea
\label{eq2.176}
   &\cE_{res} = Z^5\cdot 30.11\ \mathrm{kV/cm}&\\
\label{eq2.177}
   {\rm for}\qquad
        &A_1 = 0.76\ {\mathrm nm},
         A_2 = A_3 = 1\ {\mathrm nm}&.
\eea
For the case of a $\vzHep$ ion in a box described by (\ref{eq2.131}), (\ref{eq2.176}), (\ref{eq2.177}), we choose again perturbing fields as in (\ref{eq2.162}). We show in Appendix \ref{appD} that 
for
\be
\label{eq2.178.6}
  \cE' = 96\,\mathrm{V/cm}.
\ee
the resonance values for $\etav^\iKsig$ to first order in $\cE'$ are
\be
\label{eq2.178.5}
   \etav^\iKe_{res,-} = \lk( \ba{c} 1   \\ 
                                    0   \\
                                    -47.37 \ea\rk),\quad
   \etav^\iKz_{res,-} = \lk( \ba{c} 0   \\ 
                                    0   \\
                                    356.97 \ea\rk),\quad
   \etav^\iKd_{res,-} = \lk( \ba{c} 0   \\ 
                                    1   \\
                                    -49.73 \ea\rk).
\ee
Here again the analogues of (\ref{eq2.159.1}) and (\ref{eq2.160}) are satisfied.\\

In Fig.\ 10a and b we plot the decadic logarithms of the moduli of the real and twice the imaginary parts of the P-violating energy shift
\be
\label{eq2.178.7}
  E_{\balp\bbet,+}(\{\cEv^{'\iKsig}\})-E_{\balp\bbet,+}(\{\cEvR^{'\iKsig}\})
\ee 
for $\vzHep$ in a field configuration given by (\ref{eq2.176}) and (\ref{eq2.178.5})\,f.\ vs.\ the deviation $\eta_3^\iKz-\eta_{3,res,-}^\iKz$ of $\eta_3^\iKz$ from the resonance value $\eta_{3,res,-}^\iKz=356.97$ for which the P-even splitting of levels is removed, revealing an enhancement of the P-violating splitting of the order of $1/\sqrt{\dele}$. The width of the peak is proportional to $\sqrt{\dele}$ and found to be $\Del\eta^\iKz_3=10^{-3}$ ($\cE'\Del\eta_3^\iKz=9.6\cdot10^{-2}\,$V/cm). 
In Fig.\ 10\,a the differences of the real parts of the energies of the levels
are now at maximum of the order of $1\,$Hz. The lifetime of the levels is calculated to be $0.19\cdot10^{-9}\,\mathrm{s}$, which corresponds to a line width $\Delta E/h = 8.60\cdot10^8\,$Hz. As before it will not be easy to detect frequency shifts of order $1\,$Hz in such broad lines.
Also this calculation is only meant to show that energy shifts proportional to $\sqrt{\dele}$ can in principle be obtained also for ions and it illustrates how $q\not=0$ affects the numerical solutions.

%
\subsection{Results for a neutral atom with $I=1/2$}
\label{sec3.3}
Let us now turn to the case of a neutral atom with $I=1/2$, i.e.\
$\eeH$. As discussed in Appendix \ref{appA}, the eigenstates
of $\heffn(\cE\evd)$ (\ref{eq2.79}) are twofold degenerate in
$F_3=\pm|F_3|$ for fixed $l,j,F$, if $F_3\not=0$:
\be
\label{eq2.182}
   E^\iKn(\alpha,F_3,\cE\evd,\nv)=E^\iKn(\alpha,-F_3,\cE\evd,\nv),
\ee
where $\alpha=(2,l,j,F)$, but there is no
degeneracy between the levels of different $F$, but equal
$|F_3|$.
Hence, for the calculation of perturbative corrections to levels
with $|F_3|\not=0$, the same method involving degenerate
perturbation theory applies as it is described for the $I=0$ case in
Section \ref{sec3.1}. As far as the $|F_3|=0$ levels are concerned, simple
non-degenerate perturbation theory will be sufficient.

Let us concentrate first on the question, if the $F_3$
degeneracy of the $|F_3|\not=0$ states is in general broken
by the perturbation fields ${\cEv'}^\iKsig$. Our results of Chapter
\ref{sec2.1.1} indicate that this will be the case.

Consider the degenerate energy levels
characterized by $\balp=(\alpha,|F_3|,\cE\evd,\nv),\,|F_3|\not=0$. The corresponding right eigenstates of $\heffn$ ((\ref{eq2.79}) with $q=0$),
\be
\label{eq2.183}
   \rket{\balp,\pm}:=\rket{\alpha,\pm|F_3|,\cE\evd,\nv},
\ee
span a two-dimensional space $\cR_{\balp}^\iKn$.
Now we turn to $\heffe$ ((\ref{eq2.80}) with $q=0$) and use the same notation for the perturbation theory treatment
as in (\ref{eq2.87})--(\ref{eq2.96}).

From T-invariance we obtain again the relation (\ref{eq2.100}).
However, now $F$ and $|F_3|$ are integers and we get
\be
\label{eq2.185}
   \lbrak{\balp,+}\Kh_\alp|\balp,+)=\lbrak{\balp,-}\Kh_{\balp}|\balp,-),
\ee
and self-identities for the off-diagonal elements, which therefore
need not to be zero.
The analogous relations apply to the matrix elements of $\Hh_{\balp}$.

For both, the general analysis and for actual calculations we use our modified Bloch method defining $\Htil_\balp$ in analogy to (\ref{eq2.147a}) (cf.\ Appendix \ref{appC}). Then, in order to solve the eigenvalue
problem analogous to (\ref{eq2.147b}), one has to solve the secular equation
\be
\label{eq2.192}
   \det\lk(\unHtil_{\balp}-E_{\balp}\rk) 
   = 0,
\ee
where
\be
\label{eq2.193}
   \unHtil_{\balp}
   = \lk(\lbrak{\balp,m'}\,\Htil_{\balp}|\balp,m)\rk)
   \equiv \lk(\Htil_{\balp,m'm}\rk).
\ee
From T-invariance we find, similarly to (\ref{eq2.185}):
\be
\label{eq2.194}
  \Htil_{\balp,++} = \Htil_{\balp,-}.
\ee
The solutions of (\ref{eq2.192}) are:
\be
\label{eq2.196}
  E_{\balp,\pm} 
  = \unHtil_{\balp,++}\pm\lk[\unHtil_{\balp,+-}\,\unHtil_{\balp,-+}\rk]^{1/2}.
\ee
Thus, in general the twofold degeneracy of the levels of $\heffn$ for $|F_3|\not=0$ is broken by the perturbation.\\

Let us now consider the perturbation expansion of $\unHtil_{\balp}$ according to (\ref{eqC.9}) of Appendix \ref{appC}. 
\be
\label{eq2.196a}
   \unHtil_{\balp} = E_\balp^\iKn\unPro^\iKn_\balp 
                   + \sum_{n=1}^\infty \Ctil^\iKen.
\ee
For the diagonal elements $\Htil_{\balp,mm}$
the same arguments apply which are given
in Section \ref{sec3.1} ((\ref{eq2.114})--(\ref{eq2.126})) for the
trace of $\unHhh_{\balp}$. In this way we see that
there will be no $\delez$-linear contributions
to $\Htil_{\balp,mm}$ below the third order.

As far as the off-diagonal elements $\Htil_{\balp,m'm}$,
$m'\not=m$ are concerned, they are non-zero only above the
first order.
Indeed, for the first order term in (\ref{eq2.196a}) we have:
\be
\label{eq2.196.1}
  \lbrak{\balp,m'}\,{\Ctil}^\iKe\,|\balp,m) 
  = \lbrak{\balp,m'}\,\heffe\,|\balp,m).
\ee
For $m'\not=m$ this matrix element vanishes due to
the same argument which led to (\ref{eq2.118})
and the fact that the difference $\Del F_3$ in $F_3$ between the degenerate
unperturbed states (\ref{eq2.183}) satisfies $|\Del F_3|=|m-m'||F_3|\ge 2$ and that the dipole
operator (\ref{eq2.114}) cannot change $F_3$ by more
than one unit.

We consider now in detail the following configuration of perturbing fields:
\be
\label{eq2.197}
   {\cEv'}^\iKsig:=\etav^\iKsig\cdot\cE',\quad(\sig=1,2,3),
\ee
where the dimensionless vectors $\etav^\iKsig$ are left arbitrary for the
moment. For the expansion of $\Htil_{\balp,m'm}$ we have then from (\ref{eq2.196a}):
\be
\label{eq2.197.1}
   \Htil_{\balp,m'm} 
   = E_{\balp}^\iKn\del_{m'm} 
   + \sum_{n=1}^\infty\,\lbrak{\balp,m'}\,{\Ctil}^\iKen\,|\balp,m),
\ee
with
\bea
\label{eq2.198}
   \lbrak{\balp,m'}\,{\Ctil}^\iKen\,|\balp,m)
  &=& \lk(c^\iKvar{n,0}_{\balp,m'm}
      +\sum^2_{i=1}\deli c^\iKvar{n,i}_{\balp,m'm}\rk){\cE'}^n+\cO(\delez^2)
      \nonumber\\
  & & \qquad m',m\in\{+1,-1\}, n\ge 1.
\eea
From the discussion above we find
\bea
\label{eq2.199}
   c^\iKvar{n,i}_{\balp,m'm} = 0
   &\mbox{for}& n\le 2;\,i=1,2,\ \mbox{if}\ m'=m;\nonumber\\
   c^\iKvar{1,i}_{\balp,m'm} = 0
   &\mbox{for}& i=0,1,2,\ \mbox{if}\ m'\not=m.
\eea
Using these relations we can easily convince ourselves that a consistent evaluation of $E_{\balp,\pm}$ (\ref{eq2.196}) up to order ${\cE'}^2$ requires only the knowledge of the coefficients $c^\iKvar{n,i}_{\balp,m'm}$ for $n\le 2$. We get then:
\bea
\label{eq2.200}
  E_{\balp,\pm}
  &=& E_{\balp}^\iKn 
      +\sum^2_{n=1}c^\iKvar{n,0}_{\balp,++}\,{\cE'}^n\nonumber\\
  & & \pm\ \lk(\lk[c_{\balp,+-}^\iKvar{2,0}+\sum^2_{i=1}
      \deli c_{\balp,+-}^\iKvar{2,i}\rk]
      \lk[c_{\balp,-+}^\iKvar{2,0}
      +\sum^2_{i=1}\deli c_{\balp,-+}^\iKvar{2,i}\rk]\rk)^{\eh}{\cE'}^2
      + \cO({\cE'}^3).
\eea
In the general case where all coefficients $c_{\balp,m'm}^\iKvar{n,i}$
appearing in (\ref{eq2.200}) differ from zero, one therefore gets for the level shifts:
\bea
\label{eq2.201}
   E_{\balp,\pm}
   &=& E_{\balp}^\iKn+E^\iKvar{1,0}_{\balp}+E^\iKvar{2,0}_{\balp}\nonumber\\
   & & \pm\ \eh\lk(\Delta E_{\balp}^\iKvar{2,0}+\sum^2_{i=1}\delta_i
       \Del E^{\iKvar{2,i}}_{\balp}+\cO(\delez^2)\rk)
       + \cO({\cE'}^3).
\eea
In $E_{\balp}^\iKvar{n,i}, \Delta E^\iKvar{n,i}_{\balp}$ $n$ denotes
the order in $\cE'$, $i=0$ the order zero in $\delez$ and
$i=1,2$ the order one in $\dele,\delz$ respectively.
We see that here linear terms in $\delez$ occur already at order ${\cE'}^2$.

To give some numbers we choose an electric field configuration (\ref{eq2.197})
with
\be
\label{eq2.202}
   \eta_3^\iKsig=0;\quad {\rm for}\ \sig=1,3\quad {\rm and}\quad \etav^\iKz=0.
\ee
Then the first order level shift vanishes, $E^\iKvar{1,0}_{\balp}=0$
and we have
\bea
\label{eq2.203}
  E_{\balp,\pm}
  &=& E_{\balp}^\iKn+\Bigg(\sum_{\sig,\tau}\lk\{\ctil^{\,\sig\tau}_{\balp,1}
      (\eta_1^\iKsig\eta_1^\iKtau+\eta_2^\iKsig\eta_2^\iKtau)
      \rk\}\nonumber\\
  & & \pm\Bigg[\sum_{\sig,\tau\atop\rho,\kappa}\Bigg\{
      \ctil^{\,\sig\tau\rho\kappa}_{\balp,2}\,
      (\eta_1^\iKsig\eta_1^\iKtau-\eta_2^\iKsig\eta_2^\iKtau)\,
      (\eta_1^\iKrho\eta_1^\iKkap-\eta_2^\iKrho\eta_2^\iKkap)\nonumber\\
  & & \qquad+\ \ctil^{\,\sig\tau\rho\kappa}_{\balp,3}
      (\eta_1^\iKsig\eta_2^\iKtau+\eta_2^\iKsig\eta_1^\iKtau)
      (\eta_1^\iKrho\eta_2^\iKkap+\eta_2^\iKrho\eta_1^\iKkap)\nonumber\\
  & & \qquad+
      \sum_{i=1}^2\delta_i
      {\ctil}^{'\ivar{\,\sig\tau\rho\kappa}}_{\balp,i}
      (\eta_1^\iKsig\eta_1^\iKtau-\eta_2^\iKsig\eta_2^\iKtau)\,
      (\eta_1^\iKrho\eta_2^\iKkap+\eta_2^\iKrho\eta_1^\iKkap)
      \Bigg\}\Bigg]^{\eh}\Bigg){\cE'}^2\nonumber\\
  & & \qquad+\cO({\cE'}^3),
\eea
where the $\ctil_{\balp,i}, \ctil_{\balp,i}'$ are complex
numbers which depend on $\cE$. The sums over $\sig,\tau,\rho,\kappa$
run over the segment numbers 1,3. In order to compute the P-violating
energy shift for the levels $\balp=(\alpha,|F_3|=1,\nv=(1,1,1))$ for
$\alpha=nljF\in\{2S\eh1,2P\eh1, 2P\frac{3}{2}1,2P\frac{3}
{2}2\}$, we take the values for $L,\Delta,\Gam_S,\Gam_P$ and
the hyperfine splittings $\Atil_1,\Atil_2,\Atil_3$ from Table III of \cite{BBN95}
and chose the following configuration:\footnote{In this paper
we use the notation $A_i$, $i=1,2,3$ for the cartesian lengths
of the box, which are not to be confused with the hfs-splittings
which we denote here by $\Atil_i$, $i=1,2,3$.}
\be
\label{eq2.204}
   A_1=A_2=A_3=8\ \mathrm{nm},
\ee
\be
\label{eq2.205}
   R_1^\iKe/A_1=0.45,\quad R_1^\iKz/A_1=0.55.
\ee
As one can see from (\ref{eq2.203}), the $\delez$-linear contributions
vanish if $\eta_1^\iKsig=0$ or $\eta_2^\iKsig=0;\ \sig=1,3$;
or if $\eta^\iKsig_1=\eta_2^\iKvar{4-\sig}=0,\ \sig
=1,3$ and $\eta_2^\iKsig=\pm\eta_1^\iKvar{4-\sig}$.
We choose
\be
\label{eq2.206}
   \etav^\iKe = \lk(\ba{c}1\\ 0  \\ 0\ea\rk),\quad
   \etav^\iKz = 0,\quad
   \etav^\iKd = \lk(\ba{c}0\\ 0.5\\ 0\ea\rk)
\ee
in order to maximize the P-violating effect. 
In Table I we give numbers for the P-conserving as well as the
P-violating contributions to $E_{\balp,\pm}$ for different levels
$\alpha$ at different electric fields $\cE$.
From the analogue of (\ref{eq2.134}), we found
the maximally allowed perturbation field strengths and used these to compute
the results in Table I.

We establish thus theoretically a P-violating energy shift of the order of
\bea
\label{eq2.207}
   \Re\lk(E(\cE,\{{\cEv'}^\iKsig\})-E(\cE,\{{\cEvR'}^\iKsig\})\rk)/h
   &=& \lk(\ \mbox{1.2}\,\dele - \mbox{20.2}\,\delz 
           + \cO([\cE'/\cE]^3)\rk)\ {\rm kHz}
   \nonumber\\
   &=& -\mbox{2.56}\cdot10^{-8}\ {\rm Hz},
   \nonumber\\
   2\Im\lk(E(\cE,\{{\cEv'}^\iKsig\})-E(\cE,\{{\cEvR'}^\iKsig\})\rk)/\hbar
   &=& \lk(-\mbox{1.1}\,\dele + \mbox{16.3}\,\delz 
          + \cO([\cE'/\cE]^3)\rk)\ 10^4 {\rm s^{-1}}
   \nonumber\\
   &=& \mbox{2.08}\cdot10^{-7}\ {\rm s^{-1}}
\eea
for $\alpha=2S\eh1$ where the numerical values for $\delez$ are taken from Table III of \cite{BBN95}. We observe that the nuclear
spin dependent term $\prop\delz$ of the energy shift is enhanced by a factor
of $\sim 17$ w.r.t.\ the spin independent one. As was already noted in
\cite{BBN95}, 
an enhancement of the terms proportional to $\delz$ over the ones proportional to $\dele$ is expected to be a general feature of P-violating effects for $\eeH$ (cf.\ (6.39), (6.40) of \cite{BBN95}).
Numerically an energy shift (\ref{eq2.207}) is, of course, very small and presumably not easily measured.

Let us now consider the possibility of P-violating
energy shifts of the order of $\sqrt{\delez}$.
From (\ref{eq2.196}) we can easily see how this can be achieved. Let us expand $\Htil_{\balp,m'm}$ in $\delez$:
\bea
\label{eq2.207.2}
  \Htil_{\balp,m'm}
  &=& \Htil^\iKn_{\balp,m'm}
     +\dele\Htil^\iKe_{\balp,m'm}+\delz\Htil^\iKz_{\balp,m'm}\nonumber\\
  & &\ +\ \cO(\dele^2,\dele\delz,\delz^2).
\eea
For {\it unstable} states as we are considering here $\unHtil_\balp$ is a {\it non-hermitian} matrix and we can have
\bea
\label{eq2.207.3}
  \Htil^\iKn_{\balp,+-} &=& 0,\\
\label{eq2.207.4}
  \Htil^\iKi_{\balp,+-} &\not=& 0,\quad i=1,2,\\
\label{eq2.207.5}
  \Htil^\iKn_{\balp,-+} &\not=& 0.
\eea
In this situation the level shifts (\ref{eq2.196}) read:
\bea
\label{eq2.207.6}
  \Etil_{\balp,\pm} 
  &=& \Htil^\iKn_{\balp,++}\nonumber\\
  & &\ \pm\ \lk[\lk(\dele\Htil^\iKe_{\balp,+-}+\delz\Htil^\iKz_{\balp,+-}\rk) 
                  \Htil^\iKn_{\balp,-+} \rk]^{1/2}\nonumber\\
  & &\ +\ \cO(\dele,\delz).
\eea
Without P-violation, i.e. for $\dele=\delz=0$, the levels would be degenerate. 

Note that we need {\it not} to have a crossing
of the energies and lifetimes of the unperturbed states in
order to achieve a degeneracy of levels for nonzero handedness of the external fields.
As the degeneracy of the $\pm|F_3|$ levels gets broken by electric field
perturbations anyway, it suffices to suppress this breaking
by a convenient choice of $\etav^\iKsig,\ \sig=1,2,3$ at
{\it any} field $\cE$.  Hence, we can choose a convenient zero order field strength of e.g.
\be
\label{eq2.209}
   \cE=200\ {\rm V/cm}.
\ee
The condition (\ref{eq2.207.3}) reads now (cf.\ (\ref{eq2.197.1})--(\ref{eq2.199})):
\bea
\label{eq2.209.1}
   \Htil^\iKn_{\balp,+-} 
   &=& \sum_{n=2}^\infty\,c_{\balp,+-}^\iKvar{n,0}{\cE'}^n\nonumber\\
   &=& c_{\balp,+-}^\iKvar{2,0}{\cE'}^2 + c_{\balp,+-}^\iKvar{3,0}{\cE'}^3+...
       \nonumber\\
   &=& 0.
\eea
To order ${\cE'}^2$ we can easily solve this equation. We found a solution to
\be
\label{eq2.209.2}
   c_{\balp,+-}^\iKvar{2,0} = 0
\ee
to be
\bea
\label{eq2.210}
  \etav^\iKe&=&\lk(\ba{c}
  3.91\\ 0.0764\\ 0\ea\rk),
  \ \etav^\iKz=0,\ \etav^\iKd=\lk(\ba{c}
  3.91\\ -0.328\\ 0\ea\rk),
\eea
where we chose the box geometry as in (\ref{eq2.204}), (\ref{eq2.205}).
Note that to order ${\cE'}^2$ the magic condition (\ref{eq2.209.2}) does not depend on $\cE'$ and we get P-violating energy shifts proportional to $\sqrt{\delez}$ for all $\cE'$. However, the requirement of applicability of perturbation theory (cf. (\ref{eq2.134})) gives here as maximally allowed value  \be
\label{eq2.211}
  \cE'_{max} = 8.3\ {\rm V/cm}.
\ee
The reason for this relatively small number is that the r.h.s. of (\ref{eq2.134}) is now of the order of the hfs splitting,
which, at $\cE=200$ V/cm, is by two orders of magnitude smaller than
the differences between levels for nuclear spin $I=0$.
Thus, choosing for instance $\cE'= 5\,{\rm V/cm} < \cE'_{max}$ we are safely in the region where we can apply perturbation theory. 

Using now the implicit function theorem in a way analogous to (\ref{eq2.166.2})\,ff.\ of Appendix \ref{appD} we establish that a solution to (\ref{eq2.209.1}) exists to all orders in $\cE'$ for sufficiently small $\cE'$.
In other words: For such small enough field strength $\cE'$ higher order terms in the perturbation expansion (\ref{eq2.209.1}) will only displace the ``resonance'' position slightly but cannot change the qualitative features. Thus we establish theoretically the existence of P-violating energy shifts proportional to $\sqrt{\delez}$. 

The signal to look for in an experiment is the difference of the eigenenergies $E_{\balp,\pm}$ (\ref{eq2.207.6}) for the perturbation fields ${\cEv'}^\iKsig$ and the reflected ones ${\cEvR'}^\iKsig$,
\be
\label{eq2.212}
  {\cEvR'}^\iKsig = \etavR^\iKsig\cdot\cE',
\ee
where $\etavR^\iKsig$ are the same vectors as in (\ref{eq2.210}) but with the opposite sign of the 2-components.

In Figs.\ 11\,a, b we plot the real and imaginary parts of these energy differences \be
\label{eq2.213}
  E_{\balp,\pm}(\{{\cEv'}^\iKsig\})-E_{\balp,\pm}(\{{\cEvR'}^\iKsig\})
\ee
for $\cE'=5\,$V/cm versus the deviation of $\cE_2^{'\iKd}$ from the resonance value $\cE_{2,res}^{'\iKd}=-1.64\,$ V/cm. Numbers for other values of $\cE'$ are easily obtained from Figs.\ 11\,a, b, since (in our approximation) these energy differences are proportional to $\cE'^2$.
The width of the peak is proportional to $\sqrt{\deli}$ and found to be $2.6\cdot10^{-12}\,$V/cm. The differences of the real parts of the energies  shown in Figs.\ 11a,\,b are at maximum of the order of $10^{-5}\,$Hz whereas the lifetime of the levels is found to be $4.13\cdot10^{-8}\,\mathrm{s}$. This corresponds to a line width $\Delta E/h = 3.86\cdot10^6\,$Hz. Again we note that detecting frequency shifts of order $10^{-5}\,$Hz in such broad lines is certainly not easy.
However, our example shows, how energy shifts proportional to $\sqrt{\delez}$ can be obtained for atoms with $I=1/2$. In Figs.\ 11\,a, b we observe the rise of the signal by a factor $\sim1/\sqrt{\delz}\sim10^6$ at the resonance value.
We are sure that for a realistic experimental situation one can optimize such effects. But it remains to be seen what values of $\Del E$ one can then obtain.

\newpage
\section{Conclusions}
\label{secConcl}
\setcounter{equation}{0}
In this article we have investigated parity violation in hydrogen-like atoms due to Z-boson exchange between the lepton and the quarks in the nucleus. In particular we have considered the consequences of this P-violating neutral current interaction for the eigenenergies of the atoms which are moving in inhomogeneous external electric and magnetic fields in finite regions of space. 

In this first part we have discussed the general formalism to treat a non-relativistic atom in an external field arrangement. We derived the consequences of T-invariance for the eigenenergies of the atoms. In general these eigenenergies arise from an interplay of the internal and the c.m.\ motion of the atom. We have shown that for an atom confined to a finite region of space where we have a homogeneous electric field the eigenenergies get no P-violating contribution. Then we studied configurations of an external electric field within a rectangular box, where the field is homogeneous throughout different segments of the box and shows sudden variations at the interfaces of these segments. In part II we will study configurations where the field varies adiabatically along a certain direction in space.\\

We studied hydrogen like atoms in states of principal quantum number $n=2$ in detail. 
First we considered boxes with two segments only, where the two electric field vectors together with the vector which is normal to the interface form a chirally non-symmetric configuration which could in principle be sufficient to exhibit P-violating effects. However, we showed, that for this geometry there are no P-violating energy shifts due to the T-symmetry at least in our scheme of approximations. Thus we found that a box with a minimum of three segments is needed if the electric field varies along one direction of the rectangular box only.

We studied then explicitly hydrogen-like atoms with vanishing  nuclear spin $I=0$ like $\vzHep$ and with nuclear spin $I=1/2$ like $\eeH$ in a box with three segments using the methods of Rayleigh Schr\"odinger perturbation theory. We indeed found P-violating contributions to the eigenenergies of the systems for field configurations carrying nonzero chirality.
This opens the possibility of determining the parameters of the neutral current P-violation in atoms by frequency measurements.\\

We found energy shifts linear in the P-violation parameters $\deli$ ($i=1,2$),
where $\deli\simeq10^{-12}$ are the mixing amplitudes between $S$ and $P$ levels due to the nuclear spin independent and dependent P-violating contributions to the Hamiltonian divided by the Lamb shift determining the order of magnitude of P-odd effects. Such energy shifts were found to be of the order of $10^{-5}...10^{-9}\,$Hz for boxes of $\sim 5$-$10\,$nm size. To measure these extremely small effects would require an accuracy which only in the context of the hydrogen 21 cm-line-measurements, cf.\ \cite{Sapirstein90}, has been reached so far.\\

There is some similarity of the energy shifts linear in the P-violation parameters for the systems we considered with the P-violating energy difference between the two enantiomers of chiral molecules. Such energy differences are studied theoretically e.g.\ in \cite{ChiralMoleculesTheoryI,ChiralMoleculesTheoryII}. But we emphasize that our physical systems are very different from chiral molecules and also the physics of our P-odd energy shifts is quite different from that in chiral molecules. In chiral molecules the P-odd energy difference occurs with no external field due to a P-violating distortion of the electronic wave functions in the chiral arrangement of the nuclei. The c.m.\ motion of the molecule is irrelevant for this. In our case we consider simple atomic systems where questions of ``spontaneous'' occurence of handed molecules, maybe due to a spontaneous symmetry breaking \cite{Pfeifer} do not occur. We get P-odd energy shifts only in the presence of external fields. The interplay of the c.m.\ and internal motions of the atom is crucial. Since we have only one nucleus and one electron around it our energy shifts certainly do not occur due to the movement of the electron in a chiral lattice of nuclei. As will be seen even more clearly in part II of this work our energy shifts arise in essence through P-violating rotations of the atom as a whole in the external field (cf.\ also \cite{BBN95}). Through the boundary conditions for stationary states such rotations lead to P-odd energy shifts. However, we can make a somewhat loose connection between P-odd energy shifts in chiral molecules and our results in the following way. 
We start from our situation of an atom in a box with a chiral electric field and --- in a ``Gedankenexperiment'' --- imagine to make the box smaller and smaller until it has atomic dimensions, and the field stronger, of the order of atomic fields. Then it will no longer make sense to talk of the motion of an atom in an external field. Instead we have an electron and a nucleus interacting and moving in a common external field. If, in addition, we fix the position of the nucleus we arrive at a situation similar to the single centre approximation, which is used in calculations of the P-odd energy shifts in chiral molecules \cite{ChiralMoleculesTheoryI}. There one considers single orbitals of electrons being exposed to the potential of {\it one} of the nuclei (single centre) and in addition to a chiral external electric field which summarizes the effects of the other nuclei. Again we see that chiral molecules and the situations we considered represent very different limiting cases in the above ``Gedankenexperiment''.\\

We found that there are possibilities to enhance the P-odd effects in the energy levels dramatically, by many orders of magnitude. At a crossing of two levels of {\it unstable states}, i.e.\ {\it complex energy levels} in an electric field with nonzero handedness the P-violating energy shifts can become proportional to $\sqrt{\deli}, \,i=1,2$. The comparison of these levels with the levels for the field configuration of opposite handedness shows a drastic difference: Under a reversal of the handedness the P-violating shift gets multiplied by a {\it phase factor} $i$. This means, that the shifts in the real and imaginary parts of the complex eigenenergies are exchanged. We emphasize that such energy shifts $\prop\wdelez$ only appear for unstable states.

Numerically we found the following results. For the case of a ``hydrogen'' atom with nuclear spin $I=0$ we obtained P-violating energy shifts proportional to $\wdele$ of $\Del E/h\simeq10^{-2}$\,Hz. For $\vzHep$ we found the same kind of shifts with $\Del E/h\simeq1$\,Hz. Considering the real atom $\eeH$ with nuclear spin $I=1/2$, we found energy shifts proportional to $\wdelez$ of the order of $\Del E/h\simeq10^{-5}$\,Hz. These shifts are considerably smaller than the results of the calculation neglecting the nuclear spin of $\eeH$ because the narrow hfs-splitting restricted the reliability of our perturbative calculation to smaller electric field strengths. An improved calculation should overcome these restrictions and could lead to much larger effects. In any case, our results show, that at the (would be) crossing points of the (P-conserving) complex energy levels, there remains a P-violating splitting and the change of this energy difference with the reversal of the field's handedness is enhanced greatly compared to the shifts away from the crossings.

Finally
we note that for the case of hydrogen discussed here all our
energy shifts are much more sensitive to the parameter $\delz$
than to $\dele$ (cf.\ (\ref{eq1.4})--(\ref{eq1.5a})). The reason for this is the
same as given in Sects.\ 6.1 and 6.2 of \cite{BBN95}. Thus a measurement
of our effects could lead to a determination of the
nuclear spin-dependent weak charge $\qwz$ (\ref{eq1.5}) which
depends on the contribution of $s$-quarks to the proton spin. In
this way parity violation in atoms could give information concerning the
``spin crisis'' of the nucleons.

\ \vspace{1cm}\\
{\bf Acknowledgements:} The authors would like to thank H.\ Abele, D.H.\ Beck, C.\ Bouchiat, S.J.\ Brodsky, R.W.\ Dunford, D.\ Gromes, E.\ Henley, J.\ Kluge, H.D.\ Liesen, P.\ Overmann, W.\ Quint, P.\ Sandars, P.\ Schleper, L.M.\ Sehgal, V.\ Telegdi, C.\ Zimmermann and M.\ Zolotorov for useful discussions and suggestions.

\newpage
\begin{appendix}
\section*{Appendix}
%
\renewcommand{\thesection}{\Alph{section}}
\setcounter{section}{0}
%
%
\section{Representations of n$=$2 matrices and states}
\label{appA}
\renewcommand{\theequation}{\Alph{section}.\arabic{equation}}
\setcounter{equation}{0}
In this appendix we discuss
the non-hermitian mass matrix $\unM$ (\ref{eq1.5.30})
and the dipole operator $\unDv$
as well as the matrix representations of the T and R
transformations for atoms with nuclei of $I=0$ and 1/2 transforming
like the $\alpha$-particle and the nucleon, respectively.
In addition we present the T- and R-transformation properties for the eigenvectors of the mass matrix.
Here T is the time reversal operation and R denotes the
reflection on the 1--3-plane. R can be considered as a product of a parity
operation followed by a rotation by $\pi$ around the 2-axis:
\bea
\label{eqA.1}
   &&\mbox{R}:\lk(\ba{c}x_1\\ x_2\\ x_3\ea\rk) \longrightarrow
     \lk(\ba{r}x_1\\ -x_2\\ x_3\ea\rk),\nonumber\\
   &&\mbox{R}=e^{i\pi F_2}\cdot \mbox{P}.
\eea
All our conventions for the states etc. are as in \cite{BBN95}. The
matrices representing T and R are denoted by $\unT$ and $\unR$,
respectively (cf.\ (3.33) and (3.58) of \cite{BBN95}.)

The T- and R-transformation properties of $\uncM$,
of the quasiprojectors $\Pro$ and the eigenvalues $E$ (cf.\ 
Sect.\ \ref{sec1.2}; we drop the index {\sc\sl INT} for internal here)
are given in Sects.\ 3.3, 3.4, and Appendix B of \cite{BBN95}.
We have from T
\bea
\label{eqA.2}
   E(\alp,F_3,\cE\evd,\dele,\delz)
   &=& E(\alpha,-F_3,\cE\evd,\dele,\delz),\\
\label{eqA.3}
   \lk( \unT^{\dagger}\,\unPro(\alpha,F_3,\cE \evd,\dele,\delz)\,\unT\rk)^{T}
   &=& \unPro(\alpha,-F_3,\cE \evd,\dele,\delz)
\eea
and from R:
\bea
\label{eqA.4}
   E(\alpha,F_3,\cE\evd,\dele,\delz)
   &=& E(\alpha,-F_3,\cE\evd,-\dele,-\delz),\\
\label{eqA.5}
   \unR^\dagger{\unPro}(\alpha,F_3,\cE\evd,\dele,\delz)\unR
   &=& \unPro(\alpha,-F_3,\cE\evd,-\dele,-\delz).
\eea
Since $\unT=-\unR$ for $I=0$ and $\unT=\unR$ for $I=1/2$, we
get from (\ref{eqA.2}) to (\ref{eqA.5}),
combining the T- and R-transformations:
\bea
\label{eqA.6}
   E(\alpha,F_3,\cE\evd,\dele,\delz)
   &=& E(\alpha,F_3,\cE\evd,-\dele,-\delz),\\
\label{eqA.6a}
   \Big(\unPro(\alpha,F_3,\cE\evd,\dele,\delz)\Big)^T
   &=& {\unPro}(\alpha,F_3,\cE\evd,-\dele,-\delz).
\eea

From the above relations for the quasiprojectors we turn
now to the T- and R-transformation properties for the right
and left eigenvectors of $\uncM$. Consider first the case of
vanishing electric field, $\cE=0$. The corresponding eigenstates
of $\unMn$ are given up to order $\delta_{1,2}$ in
App.\ B, Tables {\small B.IV} and {\small B.VIII} of \cite{BBN95}. For their T-transformation
we have
\bea
\label{eqA.7}
   \lk({\uncT}^\dagger\rket{2\hat l,j,F,F_3,0\evd,\dele,\delz}\rk)^T
   &=& (-1)^{F+F_3+l}\lbrak{2\hat l,j,F,-F_3,0\evd,\dele,\delz},
\nonumber\\
   \lk(\lbrak{2\hat l,j,F,F_3,0\evd,\dele,\delz}\uncT\rk)^T
   &=& (-1)^{F+F_3+l}\rket{2\hat l,j,F,-F_3,0\evd,\dele,\delz}.
\eea
Here $l=0,1,2, ...$ or in spectroscopic notation $S,P,$ ... denotes
the orbital angular momentum quantum number and the hat indicates
the perturbation by the P-violating Hamiltonian. Similarly
we find for their R transformation
\bea
\label{eqA.8}
   \uncR\rket{2\hat l,j,F,F_3,0\evd,\dele,\delz}
   &=& (-1)^{F+F_3+l}\rket{2\hat l,j,F,-F_3,0\evd,-\dele,-\delz},\nonumber\\
   \lbrak{2\hat l,j,F,F_3,0\evd,\dele,\delz}\uncR^\dagger
   &=& (-1)^{F+F_3+l}\lbrak{2\hat l,j,F,-F_3,0\evd,-\dele,-\delz}
\eea
By an argument similar to the one given below one can see
that with a suitable normalization of the states eqs.\ (\ref{eqA.7})
and (\ref{eqA.8}) are valid not only up to order $\delez$ but to all orders in $\delez$.

We will now define the eigenstates of $\uncM$ for $\cE\not=0$
using the methods explained in App.\ C of \cite{BBN95}. We set
\bea
\label{eqA.9}
  &&\rket{\alpha,F_3,\cE\evd,\dele,\delz}\nonumber\\
  &&\quad =\ \ \Pro(\alpha,F_3,\cE\evd,\dele,\delz)
      \,\rket{\alpha,F_3,0\evd,\dele,\delz}\,
      \,\cN^{-1/2}(\alpha,F_3,\cE\evd,\dele,\delz),\nonumber\\
  &&\lbrak{\alpha,F_3,\cE\evd,\dele,\delz}\nonumber\\
  &&\quad =\ \ \lbrak{\alpha,F_3,0\evd,\dele,\delz}
      \,\Pro(\alpha,F_3,\cE\evd,\dele,\delz)\,
      \,\cN^{-1/2}(\alpha,F_3,\cE\evd,\dele,\delz),
\eea
where
\be
\label{eqA.10}
   \cN(\alpha,F_3,\cE\evd,\dele,\delz)
   =\Tr\lk[\Pro(\alpha,F_3,\cE\evd,\dele,\delz)\,
           \Pro(\alpha,F_3,0\evd,\dele,\delz)\rk]
\ee
and $\alpha$ stands for $(2\hat l,j,F)$. For $\cE\not=0$
these quantum numbers are not conserved and therefore indicate
here the values for the corresponding state at $\cE=0$.
From (\ref{eqA.3}) and (\ref{eqA.7}) we now obtain immediately:
\bea
\label{eqA.11}
  \cN(2\hat l,j,F, F_3,\cE\evd,\dele,\delz)
  &=& \cN(2\hat l,j,F,-F_3,\cE\evd,\dele,\delz),
\nonumber\\
  \lk(\uncT^\dagger|2\hat l,j,F,F_3,\cE\evd,\dele,\delz)\rk)^T
  &=& (-1)^{F+F_3+l}(2\hat l,j,F,\widetilde{-F_3,\cE\evd},\dele,\delz|,
\nonumber\\
  \lk(\lbrak{2\hat l,j,F,F_3,\cE\evd,\dele,\delz}\uncT\rk)^T
  &=& (-1)^{F+F_3+l}|2\hat l,j,F,-F_3,\cE\evd,\dele,\delz).
\eea
Similarly we get from (\ref{eqA.5}) and (\ref{eqA.8}):
\bea
\label{eqA.12}
  \cN(2\hat l,j,F, F_3,\cE\evd,\dele,\delz)
  &=& \cN(2\hat l,j,F,-F_3,\cE\evd,-\dele,-\delz),
\nonumber\\
  \uncR|2\hat l,j,F,F_3,\cE\evd,\dele,\delz)
  &=& (-1)^{F+F_3+l}\rket{2\hat l,j,F,-F_3,\cE\evd,-\dele,-\delz},
\nonumber\\
  \lbrak{2\hat l,j,F,F_3,\cE\evd,\dele,\delz}\uncR^\dagger
  &=& (-1)^{F+F_3+l}\lbrak{2\hat l,j,F,-F_3,\cE\evd,-\dele,-\delz}.
\eea

Up to now we have worked with the basis states (\ref{eq1.10})
which are simultaneous eigenstates of $\unM$ (\ref{eq1.5.30}) and $\unFd$.
Alternatively, we may use the basis set of (left and right)
eigenstates of $\unM$, which, for $|F_3|\not=0$, are defined as follows:
\bea
\label{eq2.29}
  &&|\alpha,|F_3|,\pm,\cE\evd,\dele,\delz)\nonumber\\
  &&\quad:= \edwz\lk[\rket{\alpha,|F_3|,\cE\evd,\dele,\delz}
       \pm(-1)^{1/2-I}\uncR\rket{\alpha,|F_3|,\cE\evd,-\dele,-\delz}\rk],
\nonumber\\
  &&\lbrak{\alpha,|F_3|,\pm,\cE\evd,\dele,\delz}\nonumber\\
  &&\quad:= \edwz\lk[\lbrak{\alpha,|F_3|,\cE\evd,\dele,\delz}
       \pm(-1)^{1/2-I}\lbrak{\alpha,|F_3|,\cE\evd,-\dele,-\delz}\uncR\rk],
\nonumber\\
\eea
For $F_3=0$ which occurs only for $I=1/2$ we label the
states as defined in (\ref{eqA.9}) with $+(-)$ if $F+l$ is
even (odd).
In (\ref{eq2.29}) R denotes the reflection (\ref{eqA.1}) on the 1--3-plane. For $\dele=\delz=0$ the states (\ref{eq2.29}) are simultaneous eigenstates of $\unM$ (\ref{eq1.5.30}) and $\uncR$.

As one can easily check, the definition of the
relative phases between the two linearly combined
$F_3$-eigenstates as well as the normalizations
are chosen in such a way that the orthonormality relations
\be
   \levlRvar{\alp,|F_3|}{r}{\dele,\delz}\,
   \bet,|F'_3|,r',{\cal E}\evd,\dele,\delz)
   = \del_{\alp\bet}\del_{|F_3||F'_3|}\del_{rr'}, 
\label{eqA.13}
\ee
with $r,r'\in\{+,-\}$, are fulfilled. The
transformation under T, R, R$\after$T of the states
for the case $I=0$ is given by
\bea
   \mbox{T}:\qquad
   \Big(\unT^\dagger
               \revrRvar{\alp,|F_3|}{\pm}{ \dele}\Big)^T
   &=& \mp i\, \levlRvar{\alp,|F_3|}{\mp}{ \dele},\nonumber\\
          \Big(\levlRvar{\alp,|F_3|}{\pm}{ \dele}\unT\Big)^T
   &=& \pm i\, \revrRvar{\alp,|F_3|}{\mp}{ \dele},
\label{eqA.14a}\\
   \mbox{R}:\qquad\quad\
   \unR^\dagger\revrRvar{\alp,|F_3|}{\pm}{ \dele}
   &=& \pm i\, \revrRvar{\alp,|F_3|}{\pm}{-\dele},\nonumber\\
               \levlRvar{\alp,|F_3|}{\pm}{ \dele}\unR
   &=& \mp i\, \levlRvar{\alp,|F_3|}{\pm}{-\dele},
\label{eqA.14}\\
   \mbox{R$\after$T}:\qquad\quad
   \Big(       \revrRvar{\alp,|F_3|}{\pm}{ \dele}\Big)^T
   &=&         \levlRvar{\alp,|F_3|}{\mp}{-\dele},\nonumber\\
   \Big(       \levlRvar{\alp,|F_3|}{\pm}{ \dele}\Big)^T
   &=&         \revrRvar{\alp,|F_3|}{\mp}{-\dele},
\label{eqA.15}
\eea
and for the case  $I=1/2$  by
\bea
   \mbox{T}:\qquad\Big(
   \unT^\dagger\revrRvar{\alp,|F_3|}{\pm}{ \dele, \delz}\Big)^T
   &=& \pm\,   \levlRvar{\alp,|F_3|}{\pm}{ \dele, \delz},\nonumber\\
          \Big(\levlRvar{\alp,|F_3|}{\pm}{ \dele, \delz}\unT\Big)^T
   &=& \pm\,   \revrRvar{\alp,|F_3|}{\pm}{ \dele, \delz},
\label{eqA.16a}\\
   \mbox{R}:\qquad\quad\
   \unR^\dagger\revrRvar{\alp,|F_3|}{\pm}{ \dele, \delz}
   &=& \pm\,   \revrRvar{\alp,|F_3|}{\pm}{-\dele,-\delz},\nonumber\\
               \levlRvar{\alp,|F_3|}{\pm}{ \dele, \delz}\unR
   &=& \pm\,   \levlRvar{\alp,|F_3|}{\pm}{-\dele,-\delz},
\label{eqA.16}\\
   \mbox{R$\after$T}:\qquad\quad
   \Big(       \revrRvar{\alp,|F_3|}{\pm}{ \dele, \delz}\Big)^T
   &=&         \levlRvar{\alp,|F_3|}{\pm}{-\dele,-\delz},\nonumber\\
   \Big(       \levlRvar{\alp,|F_3|}{\pm}{ \dele, \delz}\Big)^T
   &=&         \revrRvar{\alp,|F_3|}{\pm}{-\dele,-\delz}.
\label{eqA.17}
\eea

We note that for given $\alpha=(2\lhat,j,F)$
and $|F_3|\not=0$ the corresponding $+$ and $-$ states
are linearly independent and eigenstates of $\unM$ to the same
complex energy eigenvalue. For $\alpha=(2\lhat,j,F)$ and
$F_3=0$ we have only the $+(-)$ state for $l+F$ even (odd) as
explained above, and there is no energy degeneracy.

%
%
\section{The continuity matrix $M_c$}
\label{appB}
\renewcommand{\theequation}{B.\arabic{equation}}
\setcounter{equation}{0}
In this appendix we give details of the calculation which leads to the results of Section \ref{sec2.1}. We consider first, as in Sect.\ \ref{sec2.1.1}, a neutral atom with an $I=0$ nucleus.

Let us have a closer look at the eigenstates $\rket{\ikvar{\sig},\beta}$ of the internal motion (cf.\ (\ref{eq2.11})).

In Sect.\ \ref{sec1.2} we have recalled the properties of the right and left eigenvectors of the mass matrix (\ref{eq1.5.30}) for an external electric field in 3-direction. 
Let us now consider such electric field vectors of length
$\cE^\iKsig=|\cEv^\iKsig|\ (\sig=1,2)$, rotated into the 3-direction. As constructed in (\ref{eq1.10}), we denote the right eigenvectors of
  $\unMn-\unDd\cE^\iKsig$
 by
\be
\label{eq2.24}
  \rket{\ikvar{\sig}',\beta,\dele,\delz}
  :=\rket{\alp,F_3,\cE^\iKsig\evd,\dele,\delz},
\ee
where $\beta$ is a shorthand notation for $(\alpha,F_3)$, and  we indicate
explicitly the dependence on $\delez$. The corresponding
left eigenvectors are
\[
  \lbrak{\ikvar{\sig}',\beta,\dele,\delz}.
\]
By a proper rotation $\mbox{Ro}$ we can bring the pair of vectors
$\cEv^\iKe$ and $\cEv^\iKz$ into the 1-3-plane. By
further suitable rotations around the 2-axis by angles
$\tet^\iKe$ or $\tet^\iKz$ we can make either
$\cEv^\iKe$ or $\cEv^\iKz$ to point along the positive 3-axis.
With $F_i\ (i=1,2,3)$ the generators of rotations around
the coordinate axes this is formally expressed as follows:
\bea
\label{eq2.25}
   \exp\lk[i\tet^\iKe F_2\rk]\cdot \mbox{Ro}:&&\,\cEv^\iKe\to|\cEv^\iKe|\evd,
\nonumber\\
   \exp\lk[i\tet^\iKz F_2\rk]\cdot \mbox{Ro}:&&\,\cEv^\iKz\to|\cEv^\iKz|\evd.
\eea
For the eigenvectors defined by (\ref{eq2.11}) we can then set
\be
\label{eq2.26}
  \rket{\ikvar{\sig},\beta,\dele,\delz}=
  \uncD(\mbox{Ro}^{-1})\cdot\exp\lk[-i\tet^\iKsig\unFz\rk]
  \rket{\ikvar{\sig}',\beta,\dele,\delz},\quad
  (\sig=1,2),
\ee
and correspondingly
\be
\label{eq2.27}
  \lbrak{\ikvar{\sig},\beta,\dele,\delz}=
  \lbrak{\ikvar{\sig}',\beta,\dele,\delz}
  \cdot\exp\lk[i\tet^\iKsig\unFz\rk]\cdot\uncD
  (\mbox{Ro}),\quad
  (\sig=1,2).
\ee
Here $\uncD$ are the representation matrices of
the rotation group in the space of the $n=2$ atomic states.
From (\ref{eq2.26}), (\ref{eq2.27}) we can write the matrices (\ref{eq2.13}) as:
\bea
\label{eq2.28}
   U^\iKvar{\sig,\tau}_{\gam,\beta}(\dele,\delz)
   &=& \lbrak{\ikvar{\sig}',\gam,\dele,\delz}
   \exp\lk[i\lk(\tet^\iKsig-\tet^\iKtau\rk)\unFz\rk]
   \rket{\ikvar{\tau}',\beta,\dele,\delz},\nonumber\\
   &&(1\leq\sig,\tau\leq2).
\eea
Thus we have expressed $\unU^\iKvar{\sig,\tau}$
by the eigenstates (\ref{eq2.24}) for electric fields pointing
in 3-direction and the relative angle $\tet^\iKsig-\tet^\iKtau$
between the fields $\cEv^\iKsig$ and $\cEv^\iKvar{\tau}$.\\

We will now show that (\ref{eq2.31}) is equivalent to an equation having no linear terms in $\dele$ (\ref{eq2.42}). In order to achieve this we choose the states defined in (\ref{eq2.29}) of Appendix \ref{appA} as a basis. In the notation $|n\hat l,F,|F_3|, r, \cE\evd,\dele) \equiv
|n\hat l_F,|F_3|,r,\cE\evd,\dele),\quad (r=\pm)$, for $I=0$, $j=F$, as explained in (\ref{eqA.9}) ff. of Appendix \ref{appA}, we
write the basis of right eigenvectors as:
\bea
\label{eq2.32}
   &&|2\hat S_{1/2},\eh,r,|\cE^\iKsig|\evd,\dele),\nonumber\\
   &&|2\hat P_{1/2},\eh,r,|\cE^\iKsig|\evd,\dele),\nonumber\\
   &&|2\hat P_{3/2},\eh,r,|\cE^\iKsig|\evd,\dele),\nonumber\\
   &&|2\hat P_{3/2},\dh,r,|\cE^\iKsig|\evd,\dele),\nonumber\\
   &&\quad (r=\pm,\sig=1,2)
\eea
and the analogous set for the left eigenvectors.
We now write $|\ikvar{\sig}',\alp,r,\dele)$, $\alp=1,2,3,4$, as
shorthand for the states (\ref{eq2.32}) and define the following 4$\times$4 matrices (cf.\ (\ref{eq2.28})):
\bea
\label{eq2.33}
   \lk(\Ustpmpm(\dele)\rk)_{\alp\beta} &:=& \lbrak{\ikvar{\sig}',\alp,\pm,\dele}
   e^{i(\sig-\tau)\theta\unFz}\rket{\ikvar{\tau}',\beta,\pm,\dele},\\
\label{eq2.34}
   \lk(\Ustpmmp(\dele)\rk)_{\alp\beta} &:=& \lbrak{\ikvar{\sig}',\alp,\pm,\dele}
   e^{i(\sig-\tau)\theta\unFz}\rket{\ikvar{\tau}',\beta,\mp,\dele},
\eea
where $\sig, \tau\in\{1,2\}$ and
\be
\label{eq2.35}
   \theta=\theta^\iKz-\theta^\iKe
\ee
is the relative angle between the fields $\cEv^\iKz$
and $\cEv^\iKe$.
From the transformation properties of $\unFz$ and of the states
(\ref{eq2.32}) under R (cf.\ (\ref{eqA.15})) we find
\be
\label{eq2.35a}
   \Ustrs(\dele)=r\cdot s\cdot\Ustrs(-\dele)
\ee
and from the R$\after$T transformation (\ref{eqA.15}) we
get
\be
\label{eq2.35b}
   \lk[\Ustrs(\dele)\rk]^T=\Utsmsmr(-\dele).
\ee
Therefore, as a consequence of the T-transformation, the $\unU$-matrices obey
\be
\label{eq2.35c}
   \lk[\Ustrs(\dele)\rk]^T=r\cdot s\cdot\Utsmsmr(\dele).
\ee
From (\ref{eq2.35a}) we see that $\Ustpmpm$ in (\ref{eq2.33})
is even, $\Ustpmmp$ in (\ref{eq2.34})
is odd under  $\dele\to-\dele$.

Let us furthermore define 4$\times$4 matrices $\unS^\iKsig$,
$\unC^\iKsig$ as in (\ref{eq2.21}). Note that the values
$Q^\iKsig_\beta$ defined in (\ref{eq2.15}) are the same for
the $+$ and $-$ states. Putting everything together we find for
$M_c(E,\dele)$, in the basis of the $\pm$ states
and with a suitable arrangement of rows and columns:
\be
\label{eq2.36}
   M_c(E,\dele)=M_c'(E,\dele)\cdot M_c''(E,\dele),
\ee
where
\be
\label{eq2.37}
   M_c'=\lk(\ba{cccc}
        \Uzepp  & \unTg^\iKz & \Uzepm & 0  \\
        \unCt^\iKe & \Uezpp & 0  & \Uezpm \\
        \Uzemp  & 0  & \Uzemm & \unTg^\iKz \\
        0  & \Uezmp & \unCt^\iKe & \Uezmm
        \ea\rk),
\ee
\bea
\label{eq2.38}
   \unTg^\iKz&=&\unS^\iKz\cdot\unC^{\iKz -1},\nonumber\\
   \unCt^\iKe&=&\unC^\iKe\cdot\unS^{\iKe -1},
\eea
\be
\label{eq2.39}
   M_c''(E,\dele)
   =\lk(\ba{cccc}
        \unS^\iKe & 0  & 0  & 0  \\
   0  & -\unC^\iKz & 0  & 0  \\
    0  & 0  & \unS^\iKe & 0  \\
    0  & 0  & 0  & -\unC^\iKz
    \ea\rk).
\ee
From (\ref{eq2.36}) we get
\be
\label{eq2.40}
   \det M_c(E,\dele)=\det M_c'(E,\dele)\cdot \det M_c''(E,\dele).
\ee
For general fields $\cEv^\iKe,\cEv^\iKz$ we
have $\det M_c''\not=0$ and therefore $\det M_c(E,\dele)=0$ implies
$\det M_c'(E,\dele)=0$ and vice versa. Now we know that in
$M_c'$ (\ref{eq2.37}) the submatrices
$\Ustpmpm$ are independent of, and the $\Ustpmmp$ are linear
in $\dele$, neglecting terms of order $\dele^2$. Thus
we have
\be
\label{eq2.41}
   \det M_c'(E,\dele)
   =\det\lk(\ba{cc}       A & \dele D \\
                    \dele C & B  \ea\rk)
   =\det(CAC^{-1}B-\dele^2 CD),
\ee
where $A,...,D$ are square matrices depending on $E$ but not on
$\dele$. Therefore we can write the equation determining $E$ as in (\ref{eq2.42}).\\

%
%
Now we will discuss the properties of the matrix
$M_c'$ (\ref{eq2.37}) and give the proofs of (\ref{eq2.43})--(\ref{eq2.45}).

After rearranging rows and columns of $M_c'$ and multiplying some
columns with $(-1)$, we find
\be
\label{eqB.1}
   \det M'_c(E,\delta_1) = \det M_c^{asym}(E,\delta_1),
\ee
\be 
   M_c^{asym}:=\lk(\ba{cccc}
                0               & -\Uzepm &
                \Uzepp          & \unTg^\iKz   \\
                -\Uezpm         & 0        &
                \unCt^\iKe      &  \Uezpp \\
                -\Uezmm         & -\unCt^\iKe  &
                0               &  \Uezmp \\
                -\unTg^\iKz     & -\Uzemm &
                \Uzemp          & 0
                \ea\rk).
\label{eqB.2}
\ee
Using now the T relations (\ref{eq2.35b}) and the definitions (\ref{eq2.21}), (\ref{eq2.38}) of the matrices $\unTg^\iKsig$, $\unCt^\iKsig$,
it follows immediately that the above matrix is
antisymmetric. Note that $M_c^{asym}$ is a 16$\times$16-matrix.\\

A basic theorem of linear algebra --- that the
determinant of an antisymmetric matrix $M$ with
an even number of rows is a square --- may
be applied now. In order to prove this
theorem, one writes the matrix $M=-M^T$,
assuming $M\not\equiv0$ (after possibly
necessary rearrangements of rows and columns) as
\be
\label{eqB.3}
   M  = \lk(\ba{cc} A & B \\ -B^T & C \ea\rk),\quad\mbox{with}\quad
   A := \lk(\ba{cc} 0 & a \\ -a   & 0 \ea\rk),\ a\not=0,
\ee
being a 2$\times$2 matrix and $C=-C^T$ antisymmetric. Then, with
\be
\label{eqB.4}
   \det M = \det\lk\{
              \lk(\ba{cc} 1 & 0        \\ B^T A^{-1} & 1 \ea\rk)\,M\,
              \lk(\ba{cc} 1 & -A^{-1}B \\ 0          & 1 \ea\rk)\rk\}
          = \det\lk(\ba{cc} A & 0 \\ 0 & C+B^T A^{-1} B \ea\rk),
\ee
it follows that $\det M = a^2\det M'$, with $M'=C+B^T A^{-1} B$ being  
antisymmetric. The theorem now follows by mathematical
induction over the number of rows of M.\\

Applying this theorem to $M_c^{asym}$ (\ref{eqB.2}) and using (\ref{eq2.42}) we find:
\be
\label{eqB.5}
   \det M_c(E,\delta_1)=\det M_c^{asym}(E,\delta_1)
   =[g(E,\delta_1)]^2=f_0(E)+\delta_1^2f_2(E)+0(\delta_1^4),
\ee
\be
\label{eqB.6}
   g(E,\delta_1)=g_0(E)+\delta_1^2g_2(E)+0(\delta^4_1).
\ee
From (\ref{eqB.5}) we see that the energy eigenvalues $E$ are at
least twofold degenerate for all fields $\cEv^\iKvar{1},\cEv^\iKvar{2}$
up to order $\delta_1$ (Terms of higher order in $\dele$ have been neglected in previous stages of the calculations). We know that for $\cEv^\iKvar{1}=\cEv^\iKvar{2}$
there is at most a twofold degeneracy, thus, apart from possible
special values of $\cEv^\iKvar{1},\cEv^\iKvar{2}$ the degeneracy will
not be bigger than twofold. It follows that for generic values
of $\cEv^\iKvar{1},\cEv^\iKvar{2}$ $g_0(E)$ will have only
non-degenerate zeroes. The energy eigenvalues are now obtained
from
\be
\label{eqB.7}
   g(E,\delta_1)=0
\ee
and using the implicit function theorem we find from
(\ref{eqB.6}) that the solutions $E=E(\delta_1)$ of (\ref{eqB.7})
have no terms linear in $\delta_1$:
\be
\label{eqB.8}
   \lk.\frac{\partial E(\delta_1)}{\partial \delta_1}\rk|_{\dele=0}=0.
\ee
This completes the proof of the assertions made in Sect.\ \ref{sec2.1.1}
for the case of nuclear spin $I=0$.\\

%
%
Finally we have to deal with the case $I=1/2$ and to prove that (\ref{eq2.46}) is equivalent to (\ref{eq2.56}) (cf.\ Sect.\ \ref{sec2.1.2}).

Using the $\pm$ eigenstates as defined in (\ref{eq2.29}) and the subsequent text of Appendix \ref{appA} we introduce matrices
$\Ustrs(\dele,\delz)\ (r,s\in\{+,-\})$, as in
(\ref{eq2.33}), (\ref{eq2.34}), for which the
transformation properties under R, R$\after$T and T analogous
to (\ref{eq2.35a})--(\ref{eq2.35c}) now read:
\bea
\label{eq2.47}
   \mbox{R}:&&
   \Ustrs (\dele,\delz)           = r\cdot s\cdot\Ustrs (-\dele,-\delz),\\
\label{eq2.48}
   \mbox{R$\after$T}:&&
   \lk[\Ustrs (\dele,\delz)\rk]^T = \Utssr (-\dele,-\delz),\\
\label{eq2.49}
   \mbox{T}:&&
   \lk[\Ustrs (\dele,\delz)\rk]^T = r\cdot s\cdot\Utssr (\dele,\delz).
\eea
From (\ref{eq2.47}) we see again that $\Ustrs$ is symmetric
(antisymmetric) under $(\dele,\delz)\to(-\dele,
-\delz)$ for $r=s$ $(r\not=s)$.
The matrix $M_c$ factorizes as follows:
\be
\label{eq2.50}
   M_c(E,\dele,\delz)=M_c'(E,\dele,\delz)
   \cdot M_c''(E,\dele,\delz),
\ee
with
\be
\label{eq2.51}
   M_c'=\lk(\ba{cccc}
    \Uzepp  & \unTg^{\iKz+} & \Uzepm & 0  \\
    \unCt^{\iKe+} & \Uezpp & 0  & \Uezpm \\
    \Uzemp  & 0  & \Uzemm & \unTg^{\iKz-} \\
    0  & \Uezmp & \unCt^{\iKe-} & \Uezmm \ea
   \rk),
\ee
\be
\label{eq2.52}
   M_c''
   =\lk(\ba{cccc}
        \unS^\ivar{(1)+}& 0  & 0  & 0  \\
   0  & -\unC^\ivar{(2)+}& 0  & 0  \\
    0  & 0  & \unS^\ivar{(1)-}& 0  \\
    0  & 0  & 0  & -\unC^\ivar{(2)-}
    \ea\rk),
\ee
\bea
\label{eq2.53}
   \unTg^{\iKz r} &=& \unS^{\iKz r}(\unC^{\iKz r})^{-1},\nonumber\\
   \unCt^{\iKe r} &=& \unC^{\iKe r}(\unS^{\iKe r})^{-1},\nonumber\\
   r   &\in&  \{+,-\}.
\eea
Here $\unS^{\iKz r}$, $\unC^{\iKe r}$ are the diagonal matrices
as defined in (\ref{eq2.21}) for the $r=+,-$ states.
Now the important difference to the $I=0$ case is that for
$I=1/2$ the $r=+$ and $r=-$ sectors are not simply related
since the $F_3=0$ states
with $r=\pm$ are {\it not} energy degenerate (cf.\ Appendix A).
Thus the corresponding values $Q_\beta$ in (\ref{eq2.21}) are
different and also $\unS^\ivar{(2)+}\not=\unS^\ivar{(2)-},
\unC^\ivar{(1)+}\not=\unC^\ivar{(1)-}$.

We can draw the following conclusions for the eigenenergies
$E$ obtained from
\be
\label{eq2.54}
   \det\ M_c'(E,\dele,\delz)=0.
\ee
For $\dele=\delz=0$ the matrix $M_c'$ is block-diagonal
since
\be
\label{eq2.55}
   \unU^\ivar{(2,1)\pm\mp}(0,0)=\unU^\ivar{(1,2)\pm\mp}(0,0)=0.
\ee
But for general fields $\cEv^\iKe,\cEv^\iKz$ this does not
lead to an energy degeneracy, since the $++$ and $-$ blocks of $M_c'$
have different determinants due to the non-degeneracy of the $F_3=0$ plus and minus states. For $\dele,\delz$ unequal zero, we see from
(\ref{eq2.51}) that we can apply the same arguments as in
(\ref{eq2.41}) and (\ref{eq2.42}) in order to obtain (\ref{eq2.56}).

%
%
\section{A modified version of Bloch's method\\ for degenerate perturbation theory}
\label{appC}
\renewcommand{\theequation}{C.\arabic{equation}}
\setcounter{equation}{0}
In order to calculate eigenvalues and eigenvectors of a time independent operator $\Oh$ in a perturbative approach, where $\Oh$ is a sum
\be
\label{eqC.1}
   \Oh = \Oh_0 + \lam\Oh_1
\ee
of an operator $\Oh_0$, whose eigenvalues and (orthonormalized, complete set of) eigenvectors are known, and a perturbation $\lam\Oh_1$ with a small parameter $\lam$, one uses time independent perturbation theory to calculate perturbative corrections to these eigenvalues and eigenvectors. If (some of) the unperturbed eigenvectors of $\Oh_0$ have the same eigenvalues, which e.g.\ is the case in most of our problems, one has to use degenerate perturbation theory (cf.\ e.g.\  \cite{Messiah}). For this case of degeneracies, several methods have been given to write down the explicit perturbative expansion of the eigenvalues of $\Oh$. In Chap.\ \ref{sec3} we have used both Kato's (cf.\ (\ref{eq2.92})--(\ref{eq2.96})) and Bloch's (cf.\ (\ref{eq2.102})--(\ref{eq2.106})) method. The advantage of Bloch's method is that by modifying $\Hh$ (cf.\ (\ref{eq2.102})) one obtains an eigenvalue problem (\ref{eq2.106}), where the sought eigenvalue $E$ is multiplied only by a unit operator.\footnote{
Since the discussion here is general we drop the indeces $\balp$ or $(\balp\bbet)$ on the operators $\Hh$, $\Kh$, as well as on the eigenvalues $E$, which in the main text are used to indicate the subspace of unperturbed states in which the diagonalization takes place.}
This eventually yields simpler expressions (compare (\ref{eq2.103})\,ff. with (\ref{eq2.92})\,ff.). However, if $\Hh$ and $\Kh$ are hermitian $\Hhh$ in general is not because of the asymmetric multiplication with $\Kh$ in (\ref{eq2.102}):
\be
\label{eqC.1.1}
   \Hhh\Kh = \Hh.
\ee
In order to obtain both, an eigenvalue problem of the form (\ref{eq2.106}) and hermiticity, we introduce a modified version of Bloch's method (cf.\ (\ref{eq2.147a})\,f.), which we describe in more detail in this appendix. 

The method is similar to the well-known treatment of small oscillations in classical mechanics, where the equations of motion, after a Fourier transformation, lead to the eigenvalue problem
\be
\label{eqC.2}
   (\unV-\omega^2_k\unrT)\cdot\av_k = 0.
\ee
Here $\unV=(V_{ij})$ and $\unrT=(T_{ij})$ are the second order coefficients of the expansion of the potential and the kinetic energy about the minimum of the potential w.r.t.\ small displacements of the generalized coordinates. The elements of the eigenvectors $\av_k$ give the Fourier coefficients of the tranformed solutions of the equations of motion with eigenfrequencies $\omega_k$ (cf.\ e.g.\ \cite{Goldstein}).\\
If $\unrT$ is hermitian and positive, it possesses a hermitian and positive inverse $\unrT^{-1}$ and square root $\unrT^{1/2}$ with $(\unrT^{1/2})^2=\unrT$.
Then (\ref{eqC.2}) is equivalent to
\be
\label{eqC.3}
   (\unrT^{-1/2}\unV\unrT^{-1/2}-\omega^2_k)\cdot\av'_k = 0,
\ee
with $\av'_k=\unrT^{1/2}\av_k$. Therefore it suffices to find a set of eigenvectors $\av'_k$ of $\unU:=\unrT^{-1/2}\unV\unrT^{-1/2}$.

We extend this method to our case of the (in general) non-hermitian matrices $\Hh$ and $\Kh$ (the analogues of $\unV$ and $\unrT$ respectively) and define a matrix $\Htil$ by
\bea
\label{eqC.4}
   \Kh^{1/2}\Htil\Kh^{1/2} 
   &=& \Hh,\nonumber\\
   \Htil\,\Pro^\iKn 
   &=& \Htil,\nonumber\\
   \Pro^\iKn \Htil
   &=& \Htil.
\eea
Our eigenvalue problem (\ref{eq2.96}) which is similar to (\ref{eqC.2}) is then equivalent to (\ref{eq2.147b}):
\be
\label{eqC.5}
   (\,\Htil - E\,)\,|\,\,\ra = 0,
\ee
where $|\,\,\ra \in \cR^\iKn$.

From the expansions of $\Kh$ and $\Hh$, (\ref{eq2.92})\,ff., we find the expansion of $\Kh^{-1/2}$, the inverse of $\Kh^{1/2}$ in the space $\cR^\iKn$. This inverse exists for sufficiently small $\cE'$, since $\Kh$ is then arbitrarily close to $\Pro^\iKn$ which acts as the unit matrix over $\cR^\iKn$ (cf.\ (\ref{eq2.92})). Up to fourth order in $\cE'$ we get:
\bea
\label{eqC.8}
   \Kh^{-1/2}
      &=& \Pro^\iKn + \eh (0,2,0)\nonumber\\
      & & +\ \eh\lk[(0,1,2,0) + (0,2,1,0) + (0,3,0,0) + (0,0,3,0)\rk]\nonumber\\
      & & +\ \eh{\sum}_{(4)}(0,k_1,k_2,k_3,0)-\da(0,2,0,2,0)\nonumber\\
      & & +\ \cO(\cE'^5).
\eea
From (\ref{eq2.92})\,f. and (\ref{eqC.8}) we calculate the expansion of $\Htil$ as
\bea
\label{eqC.9}
   \Htil &=& \Kh^{-1/2}\Hh\Kh^{-1/2}\nonumber\\
         &=& E^\iKn\Pro^\iKn 
           + \Kh^{-1/2}\lk(\sum_{n=1}^\infty {B'}^\iKen\rk)\Kh^{-1/2}\nonumber\\
         &=:& E^\iKn\Pro^\iKn + \sum_{n=1}^\infty \Ctil^\iKen,
\eea
with the terms of order $n=1,2,3,4$:
\bea
\label{eqC.9.1}
   \Ctil^\iKe &=& (0,0),\nonumber\\
   \Ctil^\iKz &=& (0,1,0),\nonumber\\
   \Ctil^\iKd &=& (0,1,1,0) + \eh\lk[(0,2,0,0) + (0,0,2,0)\rk],\nonumber\\
   \Ctil^\iKv &=& (0,1,1,1,0)\nonumber\\
              & &\ +\ \eh\Big[(0,0,1,2,0) + (0,1,2,0,0) + (0,2,0,1,0)\nonumber\\
              & &\ +\ \ \     (0,0,2,1,0) + (0,2,1,0,0) + (0,1,0,2,0)\nonumber\\
              & &\ +\ \ \     (0,0,0,3,0) + (0,3,0,0,0)\Big].
\eea

%
%
Finally we will discuss the T-invariance properties of Kato's eigenvalue problem and show that the matrices $\unHtil_{\balp\balp}$, $\unHtil_{\bbet\bbet}$ and $\unHtil_{\balp\bbet}\unHtil_{\bbet\balp}$ obtained by our modified version of Bloch's method (cf.\ (\ref{eq2.148}), (\ref{eq2.151}) and Section \ref{sec3.1}) are all multiples of the 2$\times$2 unit matrix.

Let us consider first Kato's operator $\Kh$. The matrix
elements of $\Kh_{(\balp\bbet)}$ read:
\be
\label{eq2.144}
   (\Kh_{(\balp\bbet)})_{(\bgam',m'),(\bgam,m)}
   = \lbrak{\bgam',m'}\Kh_{(\balp\bbet)}\rket{\bgam,m},
\ee
where $\bgam',\bgam\in\{\balp,\bbet\};m',m\in\{+1,-1\}$.
For the diagonal block matrices with $\bgam'=\bgam$, the
same T-invariance arguments as in (\ref{eq2.98})--(\ref{eq2.100})
apply, which yield that they are multiples of the unit matrices in
$\cR^\iKn_{\balp},\cR^\iKn_{\bbet}$; cf.\ (\ref{eq2.101}). As far as the off-diagonal matrices $\bgam'\not=\bgam$
are concerned, it follows from (\ref{eqA.11}) that
\be
\label{eq2.145}
   \lbrak{\bgam',m'}\Kh_{(\balp\bbet)}|\bar\gam,m)
   = (-1)^{F'+F+l'+l+(m'+m)|F_3|}
     \lbrak{\bgam,-m}\Kh_{(\balp\bbet)}|\bgam',-m').
\ee
Analogous relations apply for the matrix elements of $\Hh_{(\balp
\bbet)}$. 

We can now explicitly check that e.g.\ $\unHtil_{\balp\bbet}\unHtil_{\bbet\balp}$ is a multiple of the 2$\times$2 unit matrix. From (\ref{eq2.145}) and the analogous relation for $\Hh_{(\balp\bbet)}$, together with (\ref{eq2.147a}), we get:
\bea
\label{eq2.151a}
   \lbrak{\balp,m}\unHtil_{\balp\bbet}\unHtil_{\bbet\balp}\rket{\balp,m''}
   &=& \sum_{m'=\pm1}\lbrak{\balp,m} \Htil_{(\balp\bbet)}|\bbet,m')
                     \lbrak{\bbet,m'}\Htil_{(\balp\bbet)}|\balp,m'')
     \\
   &=& \sum_{m'=\pm1}(-1)^{2(F+F'+m'|F_3|)+(m+m'')|F_3|}
     \nonumber\\
   &&\qquad\times\ 
     \lbrak{\balp,-m''}\Htil_{(\balp\bbet)}|\bbet,-m')
     \lbrak{\bbet,-m'} \Htil_{(\balp\bbet)}|\balp,-m )
     \nonumber
\eea
Since $2(F'+m'|F_3|)$ is even, it follows that $2F+(m+m'')|F_3|$
is even/odd for $m''=m\,/\,m''=-m$, respectively. Because we sum over
$m'$ in (\ref{eq2.151a}) this completes the proof. To show that $\unHtil_{\balp\bbet}(\unHtil_{\bbet\bbet}-E)\unHtil_{\bbet\balp}$ is also a multiple of the unit matrix, we still have to use the analogue of (\ref{eq2.100}) for $\unHtil_{\bbet\bbet}$.

%
%
\section{General proof of the existence of $\wdele$-resonances}
\label{appD}
\renewcommand{\theequation}{D.\arabic{equation}}
\setcounter{equation}{0}
In this appendix we give a proof that for sufficiently small $\cE'$ there
exist ``resonance'' values for the vectors $\etav^\iKsig$ where energy shifts proportional to $\wdele$ occur (cf.\ (\ref{eq2.161a})--(\ref{eq2.162})).

Let us again first consider the case of a ``hydrogen'' atom with nuclear spin $I=0$.
We expand $a$ and $b$ (\ref{eq2.156}) in powers of $\cE'$ up to second order and $c$, which has the dimension of $a^2, b^2$, up to third order, where $\dele$-linear contributions of lowest order in $\cE'$ show up. This gives
\bea
\label{eq2.163}
   a &=& a_0^\iKn+a_0^\iKe\cE' + a_0^\iKz\cE'^2 + \cO(\cE'^3),\nonumber\\
   b &=& b_0^\iKn+b_0^\iKe\cE' + b_0^\iKz\cE'^2 + \cO(\cE'^3),\nonumber\\
   c &=& c_0^\iKn+c_0^\iKe\cE' + c_0^\iKz\cE'^2 + 
        (c_0^\iKd+\dele c_1^\iKd)\cE'^3 + \cO(\cE'^4).
\eea
All coefficients $a_i^\iKen,b_i^\iKen,c_i^\iKen$, $i=0,1;\,n=0,1,2,...$, depend still on the $\etav^\iKsig$ and of course on the unperturbed configuration parameters (\ref{eq2.131}), (\ref{eq2.136c}), (\ref{eq2.136d}), but not on $\cE'$.
From the definition (\ref{eq2.147a}) of $\Htil_{(\balp\bbet)}$ together with (\ref{eq2.142}) and (\ref{eq2.143}) it follows, that (cf.\ (\ref{eq2.152}))
\bea
\label{eq2.164}
   a_0^\iKn &=& E_{\balp}^\iKn,\nonumber\\
   b_0^\iKn &=& E_{\bbet}^\iKn,\nonumber\\
   c_0^\iKn &=& c_0^\iKe = 0.
\eea
The series for $c$ in (\ref{eq2.163}) implies
\be
\label{eq2.163a}
   \sqrt{c} = \sqrt{c_0^\iKz}\cE' 
            + \frac{c_0^\iKd+\dele c_1^\iKd}{2\sqrt{c_0^\iKz}}\cE'^2 
            + \cO(\cE'^3).
\ee
As one can see, the third order term of $c$ is needed in order to obtain a consistent expansion of $\pm i(a-b)+2\sqrt{c}$ up to second order in $\cE'$.
Since we $E_{\balp}^\iKn=E_{\bbet}^\iKn$ ((\ref{eq2.136a}), (\ref{eq2.136b})), the lowest order contribution to the radicand in (\ref{eq2.155}) will be $\prop\cE'^2$. Therefore we can write the l.h.s.\ of (\ref{eq2.159}), (\ref{eq2.159.1}) as follows:
\bea
\label{eq2.165}
   \pm i(a_0-b_0)+2\sqrt{c_0} 
   &=:& \cE'\cdot F_\pm(\{\etav^\iKsig\},\cE')
       \nonumber\\
   &=& \lk[\pm i(a_0^\iKe-b_0^\iKe)+2\sqrt{c_0^\iKz} +\cO(\cE')\rk]\cE'.
\eea
Suppose now a set of perturbation fields (\ref{eq2.162}) can be found satisfying (\ref{eq2.159}) or (\ref{eq2.159.1}) at order $\cE'$, i.e.
\bea
\label{eq2.166}
   F_+(\{\etav^\iKsig\},\cE'=0) &=& 0\quad\mbox{for}\quad
   \etav^\iKsig=\etavb^\iKsig_{res,+},\\
\label{eq2.166.1}
   F_-(\{\etav^\iKsig\},\cE'=0) &=& 0\quad\mbox{for}\quad
   \etav^\iKsig=\etavb^\iKsig_{res,-},
\eea
where also (\ref{eq2.160}) holds. Considering e.g.\ (\ref{eq2.166}) we can then apply the implicit function theorem in order to show that a solution to (\ref{eq2.159}), i.e.
\be
\label{eq2.166.2}
   F_+(\{\etav^\iKsig\},\cE') = 0
\ee
for some values $\etav^\iKsig=\etav^\iKsig_{res,+}$ exists to all orders in $\cE'$ for sufficiently small $\cE'$.

Now the condition that the implicit function theorem may be used is that (a) $F_+$ is a continous function and that (b) the functional determinant
\be
\label{eq2.168}
   \frac{\cD F_+}{\cD(\eta_i^\iKsig,\eta_j^\iKtau)} :=
   \lk|\ba{cc}
   \ptdxndy{\Re F_+}{\eta_i^\iKsig} & \ptdxndy{\Im F_+}{\eta_i^\iKsig} 
   \vspace{1ex}\\
   \ptdxndy{\Re F_+}{\eta_j^\iKtau} & \ptdxndy{\Im F_+}{\eta_j^\iKtau} \ea
   \rk|_{\cE'=0,
         \eta_i^\iKsig=\etab_{i,res,+}^\iKsig,
         \eta_j^\iKtau=\etab_{j,res,+}^\iKtau}
   \not= 0
\ee
with respect to some pair of coefficients $\eta_i^\iKsig\not=\eta_j^\iKtau$ does not vanish at the zero of $F_+(\cE'=0)$ which is determined by (\ref{eq2.166}). 
Similar statements hold for $F_-$. We emphasize that we can in this way give a general {\it existence proof} of P-violating energy shifts proportional to $\wdele$ using only the lowest relevant orders of perturbation theory.

For the case of our ``hydrogen'' atom in an unperturbed configuration described by (\ref{eq2.131}), (\ref{eq2.136c}), (\ref{eq2.136d}), the following perturbation field configurations satisfy (\ref{eq2.166}), (\ref{eq2.166.1}):
\be
\label{eq2.168.1}
  \etavb^\iKe_{res,\pm}=\lk( \ba{c} \eta_1^\iKe   \\ 
                                    \eta_2^\iKe   \\
                                    \pm  77.08\,\etatil \ea\rk),\quad
  \etavb^\iKz_{res,\pm}=\lk( \ba{c} 0   \\ 
                                    0   \\
                                    \mp 480.58\,\etatil \ea\rk),\quad
  \etavb^\iKd_{res,\pm}=\lk( \ba{c} \eta_1^\iKd   \\ 
                                    \eta_2^\iKd   \\
                                    \pm  77.30\,\etatil \ea\rk),
\ee
where
\be
\label{eq2.168.2}
   \etatil := \sqrt{(\eta_1^\iKe-\eta_1^\iKd)^2+(\eta_2^\iKe-\eta_2^\iKd)^2}.
\ee
In general, there will be a continuous set of solutions of (\ref{eq2.166}), (\ref{eq2.166.1}). In order to simplify the calculation, we have chosen our solutions such, that they also satisfy
\be
\label{eq2.168.2.1}
   \Re(a_0^\iKe-b_0^\iKe)=0.
\ee
The functional determinant (\ref{eq2.168}) with respect to $\eta_3^\iKe-\eta_3^\iKd$ and $\eta_3^\iKe+\eta_3^\iKd$ evaluates to 
\be
\label{eq2.168.3}
   \frac{\cD F_\pm}{\cD(\eta_3^\iKe-\eta_3^\iKd,\eta_3^\iKe+\eta_3^\iKd)}
   = \mp\,2.63\cdot10^{-5} ({\mathrm MHz\,}h{\mathrm \,V^{-1} cm})^2.
\ee
This completes the proof, that to any order in $\cE'$, there exist non-trivial solutions of (\ref{eq2.159}) and (\ref{eq2.159.1}). In general one expects, that for such a solution also (\ref{eq2.160}) holds. We will now give explicit solutions of (\ref{eq2.159}), (\ref{eq2.159.1}) to second order in $\cE'$, which yield a P-violating energy shift proportional to $\sqrt{\dele}$ (\ref{eq2.161}).

We expand the two P-even factors of the radicand in (\ref{eq2.155}), i.e.\ $\pm i(a_0-b_0)+2\sqrt{c_0}$ up to second order in $\cE'$ (cf.\ (\ref{eq2.163}), (\ref{eq2.163a})):
\bea
\label{eq2.168.4}
   \pm i(a_0-b_0)+2\sqrt{c_0} 
   &=:& \cE'\cdot\lk[F_\pm^\iKn(\{\etav^\iKsig\})
                    +F_\pm^\iKe(\{\etav^\iKsig\})\cE'
                    +\cO({\cE'}^2)\rk]
       \nonumber\\
   &=& \lk[\pm i(a_0^\iKe-b_0^\iKe)+2\sqrt{c_0^\iKz}\,\rk]\cE'
       \nonumber\\
   & &\ +\lk[\pm i(a_0^\iKz-b_0^\iKz)
             +c_0^\iKd/\sqrt{c_0^\iKz}\,\rk]{\cE'}^2
       \nonumber\\
   & &\ +\ \cO({\cE'}^3).
\eea
Here $F_\pm^\iKn(\{\etav^\iKsig\})\equiv F_\pm(\{\etav^\iKsig\},\cE'=0)$ and $F_\pm^\iKe(\{\etav^\iKsig\})$ are the zero and first order coefficients of the expansion of $F_\pm$ (\ref{eq2.165}) in terms of $\cE'$.
Now, for $\cE'\not=0$ and to second order in $\cE'$, (\ref{eq2.159}) and (\ref{eq2.159.1}) imply
\be
\label{eq2.168.5}
   F_\pm^\iKn(\{\etav^\iKsig\})+F_\pm^\iKe(\{\etav^\iKsig\})\cdot\cE'=0.
\ee
Again, for our ``hydrogen'' atom in an unperturbed configuration described by (\ref{eq2.131}), (\ref{eq2.136c}), (\ref{eq2.136d}), the perturbation field configuration (\ref{eq2.132}) satisfies (\ref{eq2.168.5}) with either:
\be
\label{eq2.168.7}
   \etav^\iKe_{res,+} = \lk( \ba{c} 1   \\ 
                                    0   \\
                                    117.00 \ea\rk),\quad
   \etav^\iKz_{res,+} = \lk( \ba{c} 0   \\ 
                                    0   \\
                                    -702.30 \ea\rk),\quad
   \etav^\iKd_{res,+} = \lk( \ba{c} 0   \\ 
                                    1   \\
                                    117.21 \ea\rk)
\ee
for
\be
\label{eq2.168.8}
  \cE' = 0.3\,\mathrm{V/cm}
\ee
or
\be
\label{eqD.168.9}
   \etav^\iKe_{res,-} = \lk( \ba{c} 1   \\ 
                                    0   \\
                                    -61.31 \ea\rk),\quad
   \etav^\iKz_{res,-} = \lk( \ba{c} 0   \\ 
                                    0   \\
                                    574.93 \ea\rk),\quad
   \etav^\iKd_{res,-} = \lk( \ba{c} 0   \\ 
                                    1   \\
                                    -63.31 \ea\rk),
\ee
for
\be
\label{eq2.168.10}
  \cE' = 3\,\mathrm{V/cm}.
\ee
Note that the perturbation fields (\ref{eq2.168.8}), (\ref{eq2.168.10}) have been chosen in a way that they fulfill the condition $|\cE'|<\cE'_{max}$, where $\cE'_{max}$ is obtained by an argument as in (\ref{eq2.134}) adapted for our case here. This gives an estimate of the maximum for which the perturbation series is reliable. $\cE'_{max}$ is found to be
\be
\label{eq2.168.11}
  \cE'_{max,+} =  9\,{\mathrm V/cm},\qquad\qquad
  \cE'_{max,-} = 12\,{\mathrm V/cm}
\ee
for the ``$+$''- and ``$-$''-solutions (\ref{eq2.168.7}) and (\ref{eqD.168.9})
respectively.
Within these limits, we looked for solutions which for simplicity we required to satisfy:
\be
\label{eq2.168.12}
   \Re\lk(a_0^\iKe-b_0^\iKe+\cE'[a_0^\iKz-b_0^\iKz]\rk)=0.
\ee
This led to further restrictions on $\cE'$, and the values given in (\ref{eq2.168.8}), (\ref{eq2.168.10}) represent approximately the maximal ones compatible with (\ref{eq2.168.12}).
Then for the ``$+$''-solution (\ref{eq2.168.7}) $\cE'$ turned out to be an order of magnitude smaller than for the ``$-$''-solution (\ref{eqD.168.9}).\\

To finish we turn to the case of an ion with $I=0$. For the case of a $\vzHep$ ion in a box described by (\ref{eq2.131}), (\ref{eq2.176}), (\ref{eq2.177}),
the following perturbation field configurations satisfy (\ref{eq2.166}), (\ref{eq2.166.1}):
\bea
\label{eq2.178}
   \etavb^\iKe_{res,\pm}
   &=& \lk( \ba{c} 
       \eta_1^\iKe   \\ 
       \eta_2^\iKe   \\
       0.02\,\etatil_1
       \mp (62.02 \mp 0.11)\,\sqrt{\etatil_1^2+\etatil_2^2} \ea\rk),\nonumber\\
   \etavb^\iKz_{res,\pm}
   &=& \lk( \ba{c} 
       0   \\ 
       0   \\
       -0.06\,\etatil_1
       \pm 445.77\sqrt{\etatil_1^2+\etatil_2^2} \ea\rk),\nonumber\\
   \etavb^\iKd_{res,\pm}
   &=& \lk( \ba{c} 
       \eta_1^\iKd   \\ 
       \eta_2^\iKd   \\
       0.02\etatil_1
       \mp (62.02 \pm 0.11)\,\sqrt{\etatil_1^2+\etatil_2^2} \ea\rk),
\eea
where
\bea
\label{eq2.178.1}
   \etatil_1 &:=& \eta_1^\iKe-\eta_1^\iKd,\nonumber\\
   \etatil_2 &:=& \eta_2^\iKe-\eta_2^\iKd.
\eea
In general, as for $q=0$, there will be continuous sets of solutions of (\ref{eq2.166}), (\ref{eq2.166.1}). In order to simplify the calculation, we have again chosen our solutions such, that they also satisfy (\ref{eq2.168.2.1}).\\
The functional determinants (\ref{eq2.168}) with respect to $\eta_3^\iKe-\eta_3^\iKd$ and $\eta_3^\iKe+\eta_3^\iKd$ evaluate to 
\be
\label{eq2.178.2}
   \frac{\cD F_\pm}{\cD(\eta_3^\iKe-\eta_3^\iKd,\eta_3^\iKe+\eta_3^\iKd)}
   = \mp 4.70\cdot10^{-8} 
   \lk({\mathrm MHz\,}h{\mathrm\,V^{-1} cm\,}\rk)^2.
\ee
This completes the proof for $\vzHep$, that to any order in $\cE'$, there exist non-trivial solutions of (\ref{eq2.159}), (\ref{eq2.159.1}) (cf.\ the discussion concerning the implicit function theorem in \ref{sec3.1}). For such a solution one in general expects, that (\ref{eq2.160}) holds as well. As in the previous section, we will now give explicit solutions of (\ref{eq2.159}), (\ref{eq2.159.1}) to second order in $\cE'$, which yield P-violating energy shifts proportional to $\sqrt{\dele}$ (\ref{eq2.161}). We proceed as in (\ref{eq2.168.4})\,f. The following perturbation field configurations satisfy (\ref{eq2.168.5}):
\be
\label{eq2.178.3}
   \etav^\iKe_{res,+} = \lk( \ba{c} 1   \\ 
                                    0   \\
                                    94.45 \ea\rk),\quad
   \etav^\iKz_{res,+} = \lk( \ba{c} 0   \\ 
                                    0   \\
                                    -653.30 \ea\rk),\quad
   \etav^\iKd_{res,+} = \lk( \ba{c} 0   \\ 
                                    1   \\
                                    94.62 \ea\rk)
\ee
for
\be
\label{eq2.178.4}
  \cE' = 9.6\,\mathrm{V/cm} 
\ee
and
\be
\label{eqD.178.5}
   \etav^\iKe_{res,-} = \lk( \ba{c} 1   \\ 
                                    0   \\
                                    -47.37 \ea\rk),\quad
   \etav^\iKz_{res,-} = \lk( \ba{c} 0   \\ 
                                    0   \\
                                    356.97 \ea\rk),\quad
   \etav^\iKd_{res,-} = \lk( \ba{c} 0   \\ 
                                    1   \\
                                    -49.73 \ea\rk),
\ee
for
\be
\label{eqD.178.6}
  \cE' = 96\,\mathrm{V/cm}.
\ee
As in the $q=0$ case the perturbation fields (\ref{eq2.178.4}), (\ref{eqD.178.6}) satisfy the condition $|\cE'|<\cE'_{max}$, where $\cE'_{max}$ is given by (\ref{eq2.134}) and is an estimate of the maximum for which the perturbation series can be used. $\cE'_{max}$ is found to be
\be
\label{eq2.178.6a}
  \cE'_{max,+} = 192\,{\mathrm V/cm},\qquad\qquad
  \cE'_{max,-} = 320\,{\mathrm V/cm}
\ee
for the ``$+$''- and ``$-$''-solutions (\ref{eq2.178.3}) and (\ref{eqD.178.5})
respectively.
Within these limits, (\ref{eq2.178.4}) and (\ref{eqD.178.6}) are chosen such that solutions exist, which for simplicity we again  chose to be restricted to those which satisfy (\ref{eq2.168.12}).

\end{appendix}

\newpage
%
\section*{Table Captions}
\small
\begin{description}
\item[{\sc Table} {\rm I.}]
The P-conserving as well as the P-violating perturbative contributions to the eigenenergies $E_{\balp,\pm}$ (\ref{eq2.201})\,ff.\ up to 2nd order perturbation theory for different levels $\balp=(\alp,|F_3|;\nv=(1,1,1))$ of $\eeH$ at different electric fields $\cE$ for the corresponding maximally allowed perturbation field strengths $\cE'_{max}$.

\end{description}
\normalsize

\section*{Figure Captions}
\small
\begin{description}
\item[{\sc Fig.} {\rm 1.}]
Sketch of the electric field configuration in a box divided into three segments (1)--(3). The electric field is homogeneous within every segment and varies suddenly along the 1-direction.

\item[{\sc Fig.} {\rm 2 a,b.}]
The energy levels $E_{\balp}^\iKn=E^\iKn(\alp,|F_3|,\cE\evd;\nv)$ (\ref{eq2.85}) for a ``hydrogen'' atom with $I=0$ in an external electric field $\cE\evd$ and $\nv=(n_1,n_2,n_3)$. The real parts for $\alp=2S_{1/2}$ and $2P_{1/2}$ and $n_2=n_3=1$ are shown in {\bf a}, the decay widths (twice the negative of the imaginary parts in units of s$^{-1}$) for $\alp=2S_{1/2}$, $2P_{1/2}$ and $2P_{3/2}$ in {\bf b.} In our calculation the imaginary parts are independent of $\nv$.

\item[{\sc Fig.} {\rm 3.}]
The real and imaginary parts of the P-violating energy-difference $\Delta E_{\balp}^\iKd\cdot\cE^3/(\eta\cE^{'3}\dele h)$ (cf.\ (\ref{eq2.133})) as a function of $\cE$ for the state $\balp=(\alp,|F_3|;\nv)=(2S_{1/2},1/2;(1,1,1))$ within a range of $\cE$, where extrema due to a level crossing occur (cf.\ Fig.\ 2a).  

\item[{\sc Fig.} {\rm 4.}]
The field strength value $\cE'_{max}$ characterizing the limit of applicability of our perturbative treatment (cf.\ (\ref{eq2.134})) as a function of the field strength $\cE$ for the same level $\balp$ as in Fig.\ 3.

\item[{\sc Fig.} {\rm 5 a,b.}]
The P-violating energy shift $\Delta E_{\balp}^\iKd$ (\ref{eq2.133}) for $\cE'=\cE'_{max}$. The real part is plotted in {\bf a}, twice the imaginary part in {\bf b.}

\item[{\sc Fig.} {\rm 6 a,b.}]
The decadic logarithms of the moduli of ({\bf a}) the real and ({\bf b}) twice the imaginary parts of the P-violating energy-difference $E_{\balp\bbet,+}(\{\cEv^{'\iKsig}\})
-E_{\balp\bbet,+}(\{\cEvR^{'\iKsig}\})$ (\ref{eq2.168.13})
for our ``hydrogen'' vs.\ the deviation $\eta_3^\iKz-\eta_{3,res,-}^\iKz$ of $\eta_3^\iKz$ from the resonance value $\eta_{3,res,-}^\iKz=574.9$ for which the P-even splitting of levels is removed, revealing a P-violating splitting of the order of $\sqrt{\dele}$. 

\item[{\sc Fig.} {\rm 7.}]
The lowest energy levels of the c.m.\ motion of a $\vzHep$ ion moving in one dimension in a linearily rising electric field potential of a field strength $\cE$ within an interval of length $A=1\,$nm. These are the eigenenergies of the Airy states (cf.\ (\ref{eq2.63})). For $\cE=0$ they show the quadratic behaviour of the eigenenergies of sinus states (cf.\ (\ref{eq2.9})).

\item[{\sc Fig.} {\rm 8.}]
The field strength value $\cE'_{max}$ characterizing the limit of applicability of our perturbative treatment of $\vzHep$ as a function of the field strength $\cE$ for the level $\balp=(2S_{1/2},1/2;(1,1,1))$.

\item[{\sc Fig.} {\rm 9 a,b.}]
The P-violating energy shift $\Delta E_{\balp}^\iKd$ in $\vzHep$ for $\cE'=\cE'_{max}$ for the same level $\balp$ as in Fig.\ 8. The real part is plotted in {\bf a.}, the imaginary part in {\bf b.}

\item[{\sc Fig.} {\rm 10 a,b.}]
The decadic logarithms of the moduli of ({\bf a}) the real and ({\bf b}) twice the imaginary parts of the P-violating energy-difference  $E_{\balp\bbet,+}(\{\cEv^{'\iKsig}\})
-E_{\balp\bbet,+}(\{\cEvR^{'\iKsig}\})$ (\ref{eq2.178.7})
for $\vzHep$ vs.\ the deviation $\eta_3^\iKd-\eta_{3,res,-}$ of $\eta_3^\iKd$ from the resonance value $\eta_{3,res,-}^\iKz=525.4$ for which the P-even splitting of levels is removed in 2nd order perturbation theory, revealing a P-violating splitting of the order of $\sqrt{\dele}$. 

\item[{\sc Fig.} {\rm 11 a,b.}]
The decadic logarithm of the moduli squared of ({\bf a}) the real and ({\bf b}) twice the imaginary parts of the P-violating energy-difference $E_{\balp,\pm}(\{\cEv^{'\iKsig}\})-E_{\balp,\pm}(\{\cEvR^{'\iKsig}\})$ (\ref{eq2.213}) for $\eeH$ vs.\ the deviation $\cE_2^{'\iKd}-\cE_{2,res}^{'\iKd}$ of $\cE_2^{'\iKd}$ from the resonance value $\cE_{2,res}^{'\iKd}=-1.64\,$V/cm for which the P-even splitting of levels is removed in 2nd order perturbation theory, revealing a P-violating splitting of the order of $\sqrt{\delz}$.

\end{description}

%
\pagestyle{empty}
\newpage
%
%
%
%

\renewcommand{\arraystretch}{1.5}
\renewcommand{\tabcolsep}{0.1cm}
\begin{center}
TABLE I
\vspace*{.5cm}
  
\begin{tabular}{ccc|cccc|cccc}
\hline 
\hline 
 $\alp,|F_3|$ & 
 $\cE$ & 
 $\cE'_{max}$ &
 \multicolumn{2}{c}{$E_{\balp}^\iKvar{2,0}/h$}   	& 
 \multicolumn{2}{c|}{$\Del E_{\balp}^\iKvar{2,0}/h$}   	& 
 \multicolumn{2}{c}{$\Del E_{\balp}^\iKvar{2,1}/h$}	&
 \multicolumn{2}{c}{$\Del E_{\balp}^\iKvar{2,2}/h$}      
\\
 & 
 [V/cm] &
 [V/cm] &
 \multicolumn{2}{c}{[MHz]} &
 \multicolumn{2}{c|}{[MHz]} &
 \multicolumn{2}{c}{[$10^{4}\,$Hz]} &
 \multicolumn{2}{c}{[$10^{4}\,$Hz]} 
\\
 & & &
 Re & Im &
 Re & Im &
 Re & Im &
 Re & Im 
\\
\hline
$2S\eh 1,1$ & 
$340	 	$ & $ 84	$ &
$-13 		$ & $- 0,\!89	$ &
$  0,\!091 	$ & $  0,\!082	$ &
$  0,\!12 	$ & $- 0,\!085	$ &
$- 2,\!0 	$ & $  1,\!3	$ 
\\
%
$2P\eh1,1	$ & 
$380	 	$ & $ 228	$ &
$-44	 	$ & $   0,\!57	$ &
$  0,\!34	$ & $   0,\!0026$ &
$- 0,\!027 	$ & $   0,\!0045$ &
$- 0,\!70	$ & $-  0,\!041	$ 
\\
%
$2P\dh1,1	$ & 
$ 370	 	$ & $ 284	$ &
$- 15	 	$ & $   3,\!5	$ &
$   6,\!5	$ & $-  1.5	$ &
$  25 	 	$ & $   1,\!0	$ &
$-450	 	$ & $- 14	 $ 
\\
%
$2P\dh2,1 	$ & 
$ 350	 	$ & $ 206	$ &
$   3,\!9	$ & $   1,\!9 	$ &
$   0,\!20 	$ & $-  0,\!064	$ &
$  80	 	$ & $- 12	$ &
$-1250	 	$ & $ 240	$ 
\\
\hline
\hline 
\end{tabular}
\vspace{1cm}\ \\
\end{center}
\renewcommand{\arraystretch}{1.0}


%
\setlength{\unitlength}{1.4mm}
\begin{picture}(100,80)
\put(35,78){\bf Figure 1}
\put( 5,15){\line(3,1){60}}
\put(35, 5){\line(3,1){60}}
\put( 5,45){\line(3,1){60}}
\put(35,35){\line(3,1){60}}
\put( 5,15){\line(3,-1){30}}
\put(65,35){\line(3,-1){30}}
\put( 5,45){\line(3,-1){30}}
\put(65,65){\line(3,-1){30}}
\put( 5,15){\line(0, 1){30}}
\put(35, 5){\line(0, 1){30}}
\put(65,35){\line(0, 1){30}}
\put(95,25){\line(0, 1){30}}
\multiput(55,11.66)(0,3){10}{\circle*{0.2}}
\multiput(75,18.33)(0,3){10}{\circle*{0.2}}
\multiput(25,21.66)(0,3){10}{\circle*{0.2}}
\multiput(45,28.33)(0,3){10}{\circle*{0.2}}
\multiput(25,21.66)(3,-1){10}{\circle*{0.2}}
\multiput(45,28.33)(3,-1){10}{\circle*{0.2}}
\multiput(25,51.66)(3,-1){10}{\circle*{0.2}}
\multiput(45,58.33)(3,-1){10}{\circle*{0.2}}
\put(12,10.66){\vector(-3, 1){10}}
\put(22, 7.33){\vector( 3,-1){10}}
\put(48, 7.33){\vector(-3,-1){10}}
\put(88,20.66){\vector( 3, 1){10}}
\put( 2,24   ){\vector( 0,-1){10}}
\put( 2,34   ){\vector( 0, 1){10}}
\put(15,45){(1)}
\put(35,51.66){(2)}
\put(55,58.33){(3)}
\put(68,13){$A_1$}
\put(14, 7){$A_2$}
\put( 0,29){$A_3$}
\put(39, 8){$\eve$}
\put(29, 8){$\evz$}
\put(36,11){$\evd$}

\thicklines
\put(24,15.33){\vector(1,2){12}}
\put(50,20   ){\vector(0,1){30}}
\put(64,24.66){\vector(1,4){8}}
\put(35, 5){\vector( 3, 1){6}}
\put(35, 5){\vector(-3, 1){6}}
\put(35, 5){\vector( 0, 1){8}}
\thinlines
\put(53,49){$\vec{\cal E}$}
\end{picture}
\newpage
\input{Fig2a-b}
\newpage
\input{Fig3-4}
\newpage
\input{Fig5a-b}
\newpage
\input{Fig6a-b}
\newpage
\input{Fig7-8} 
\newpage
\input{Fig9a-b}
\newpage
\input{Fig10a-b}
\newpage
\input{Fig11a-b}

}
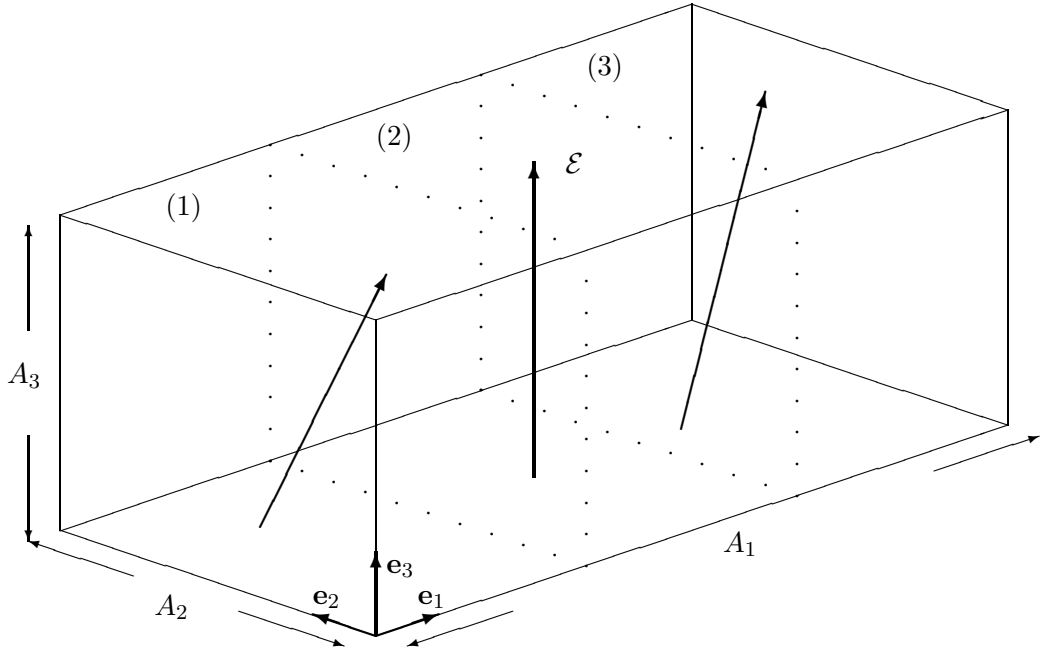
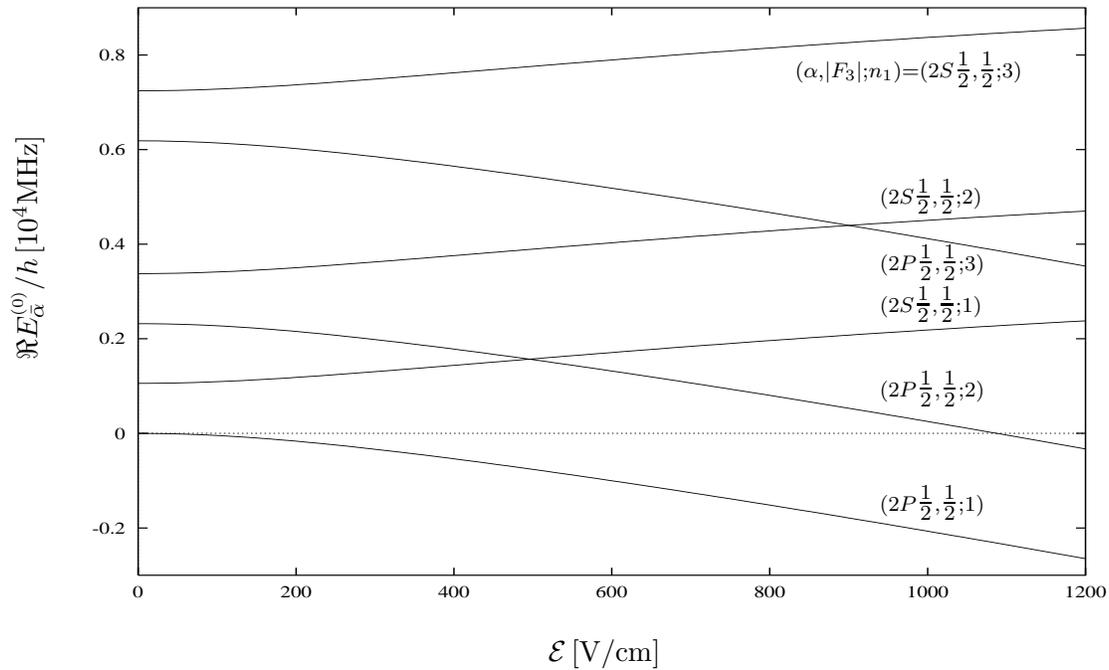
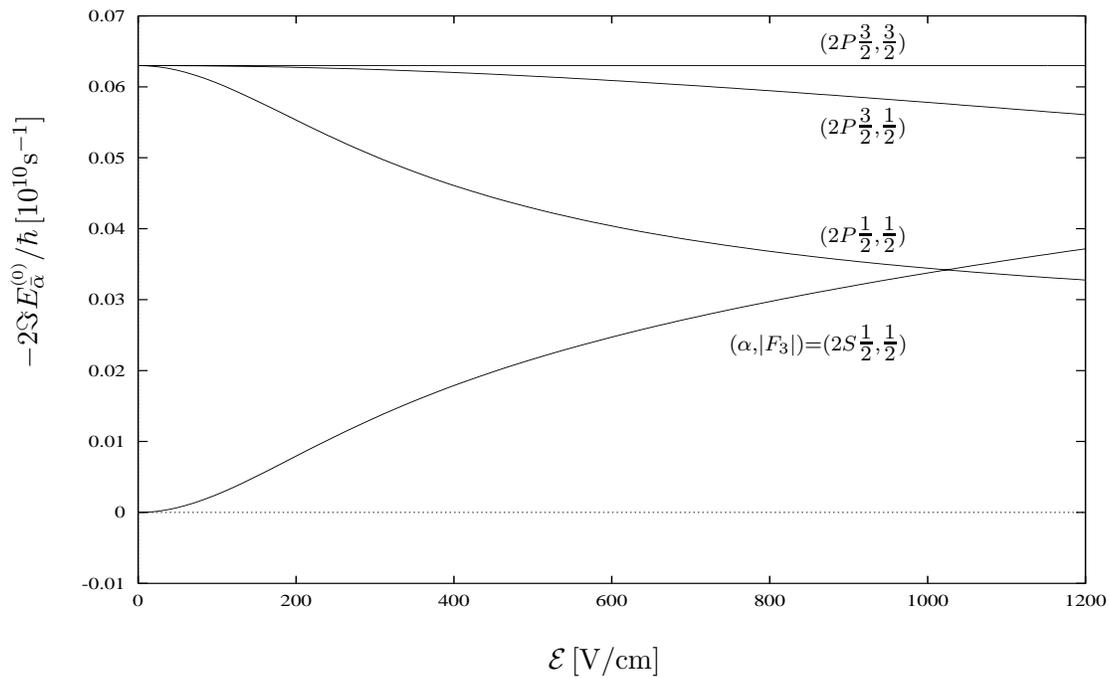
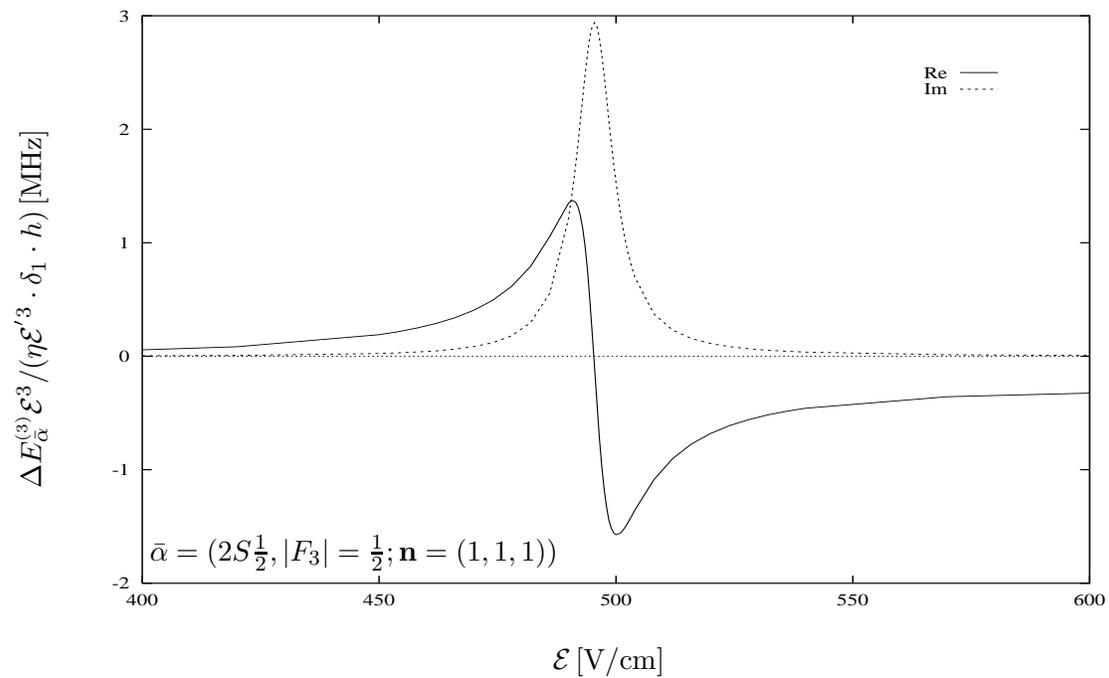
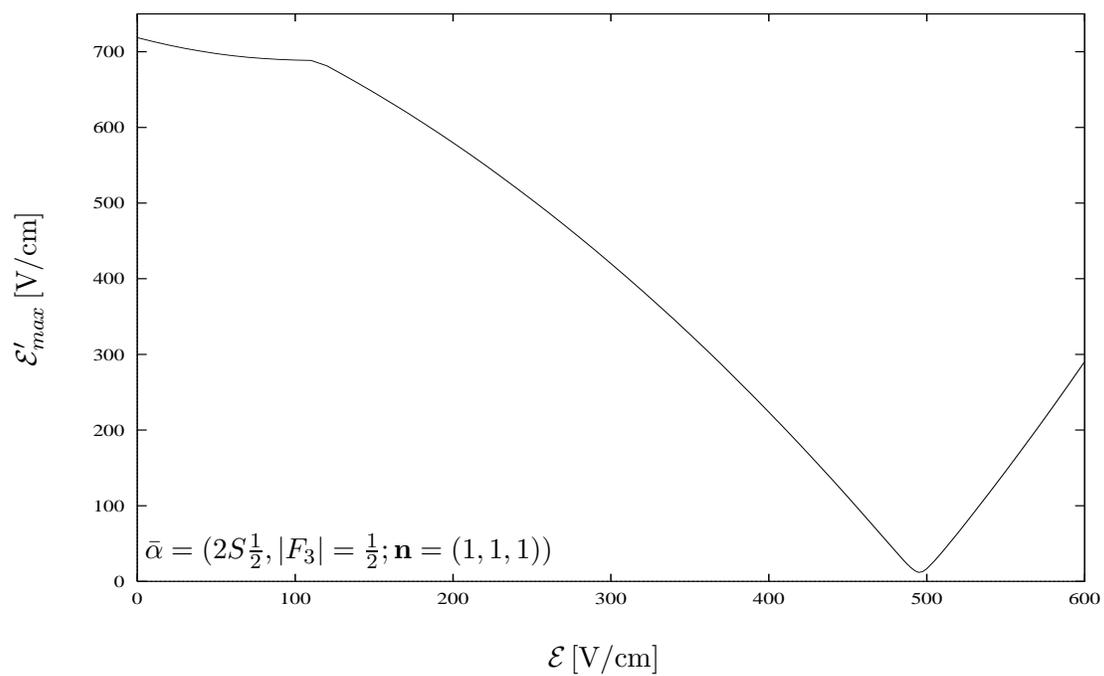
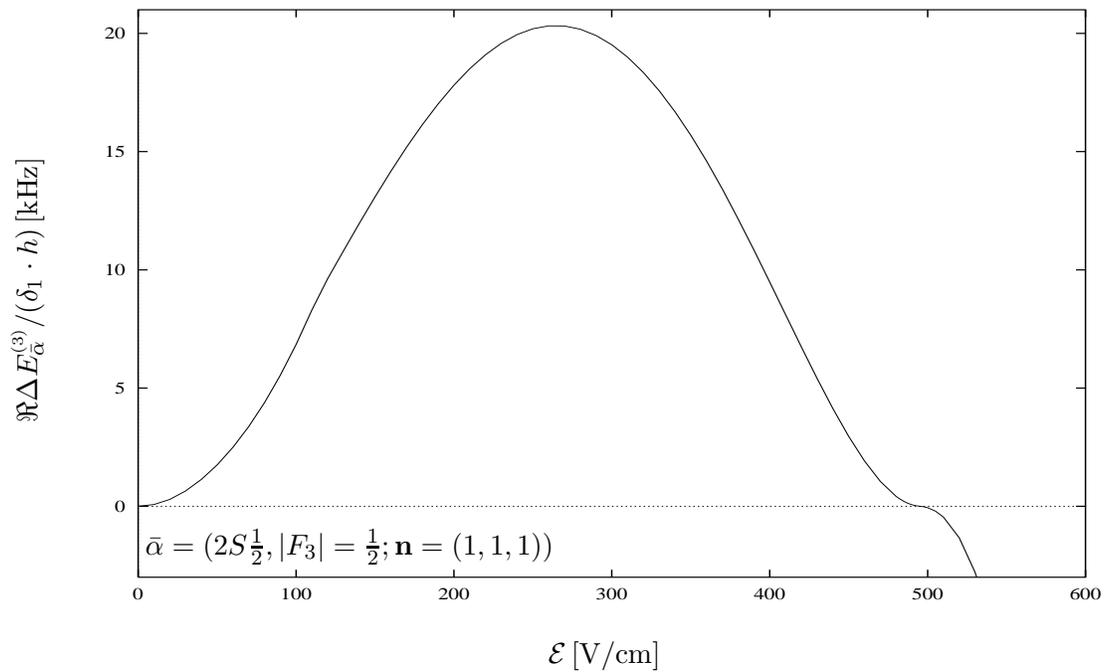
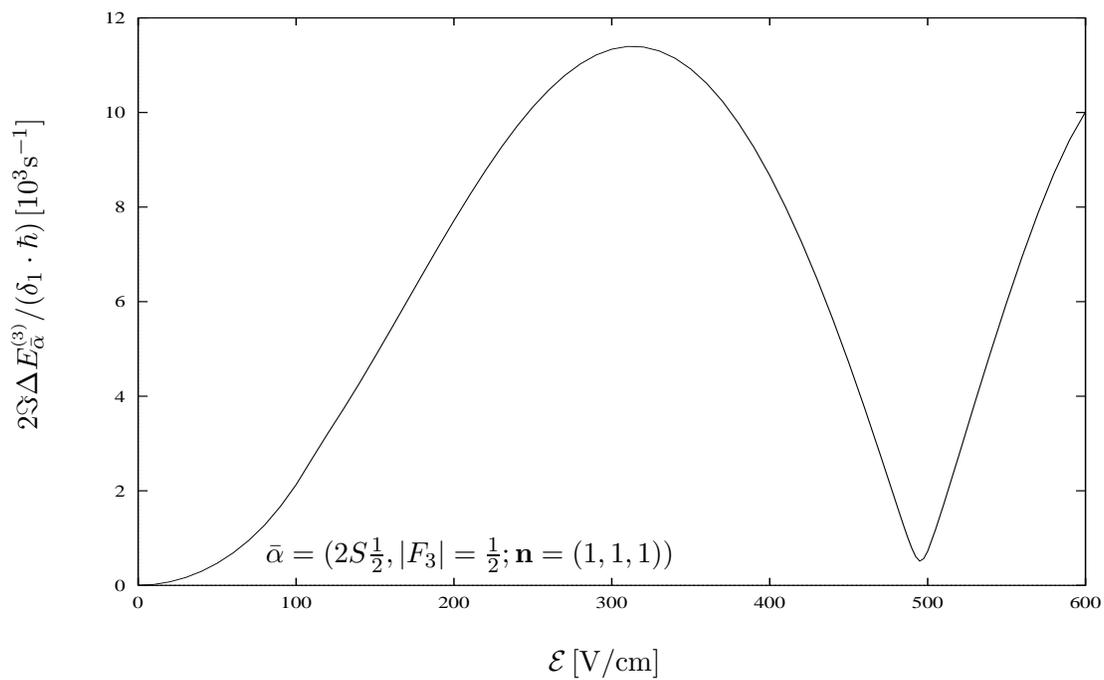
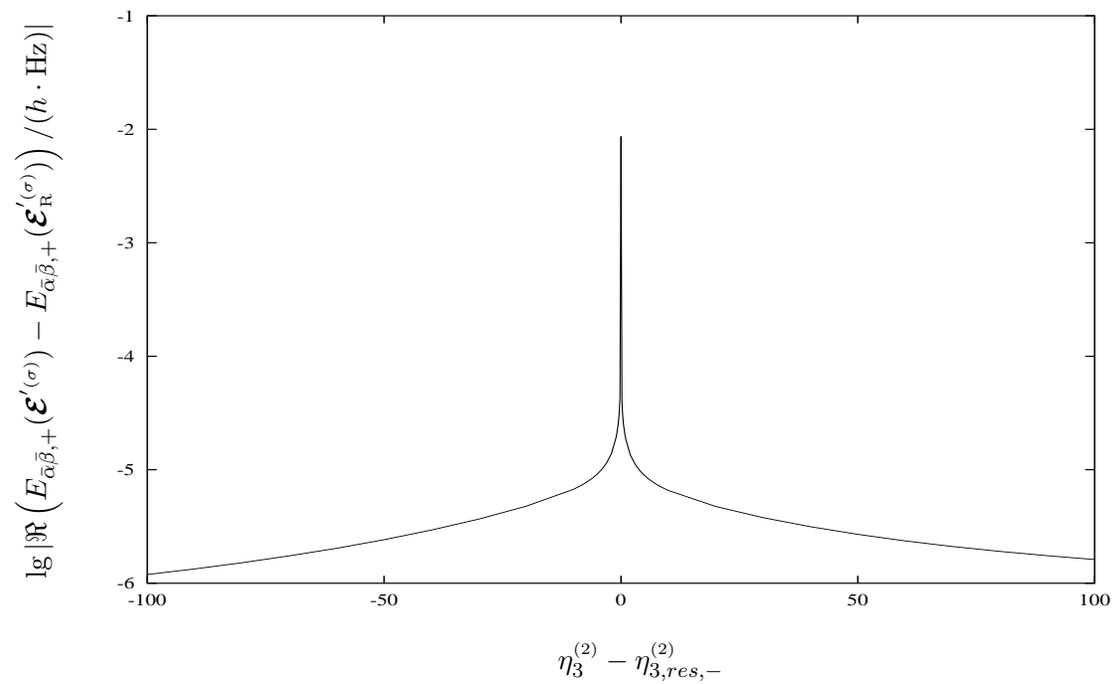
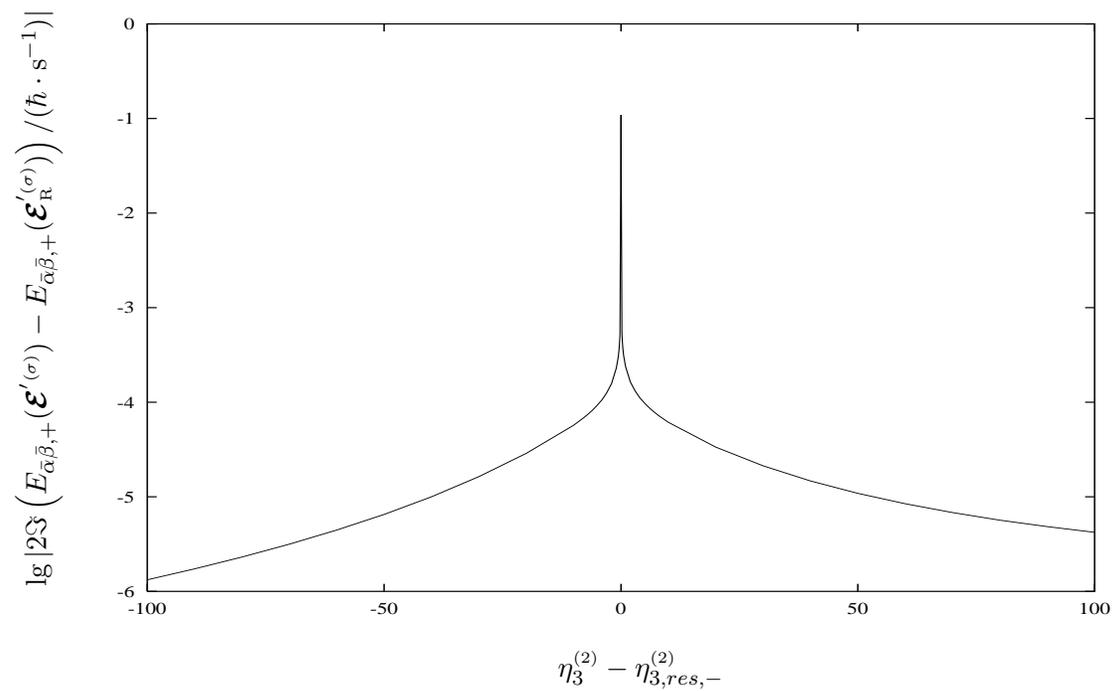
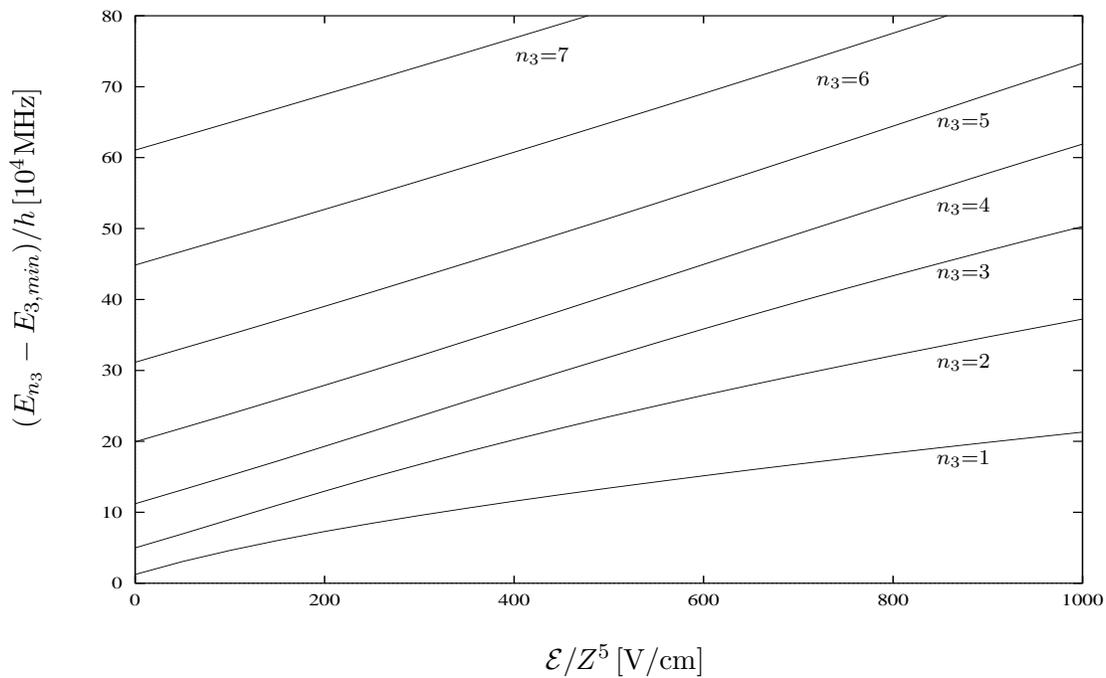
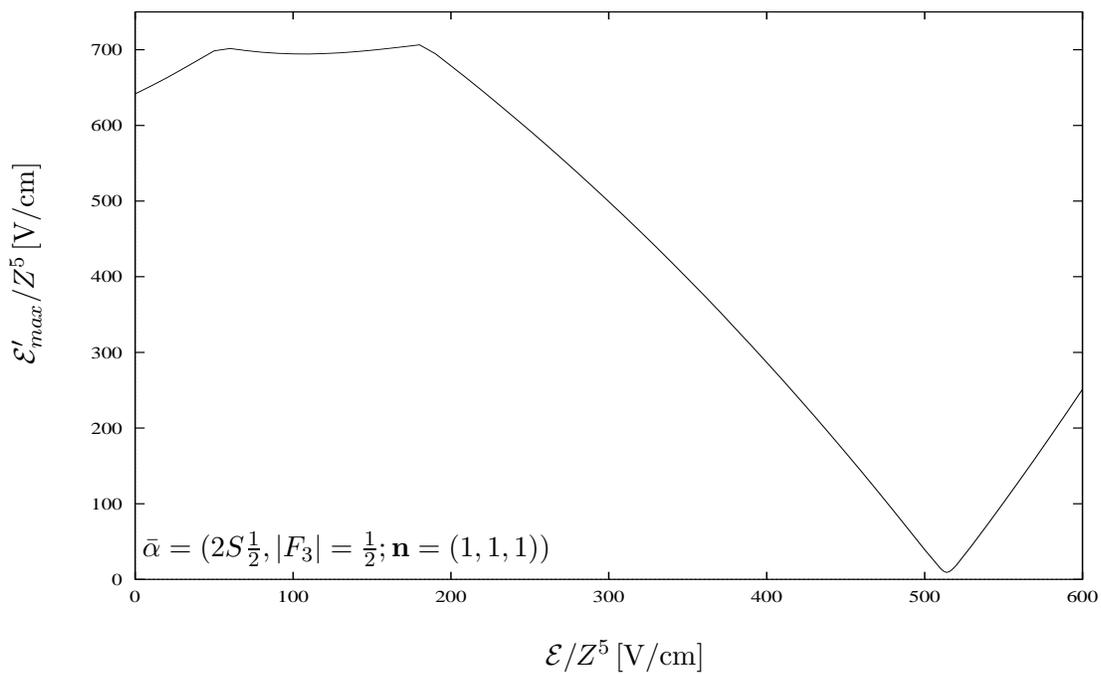
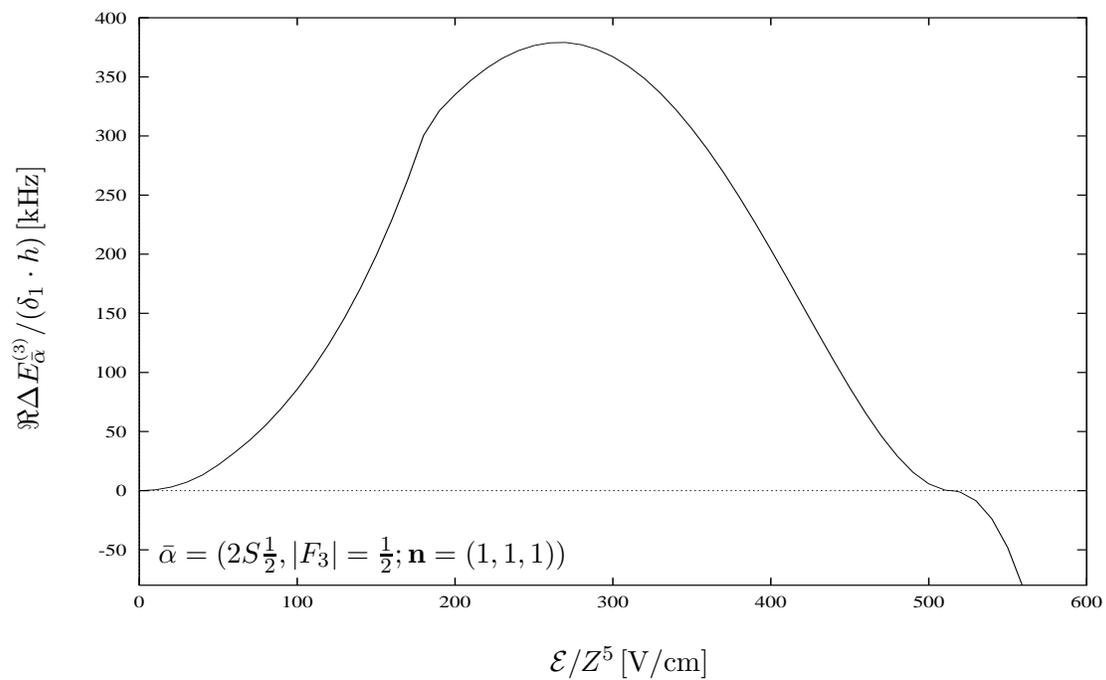
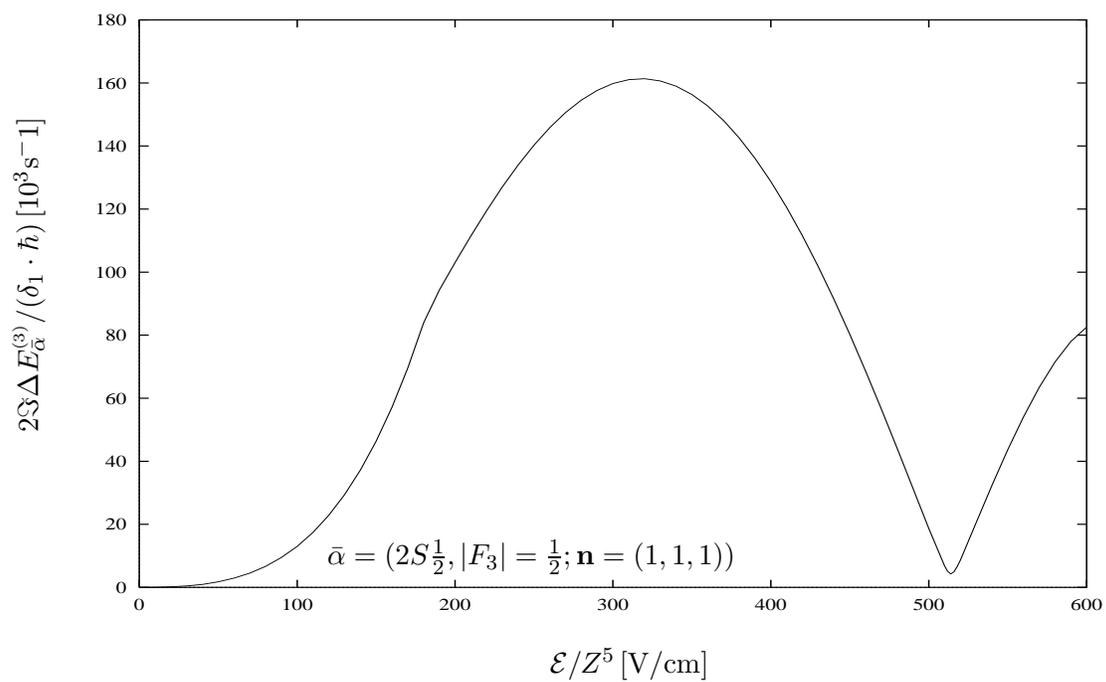
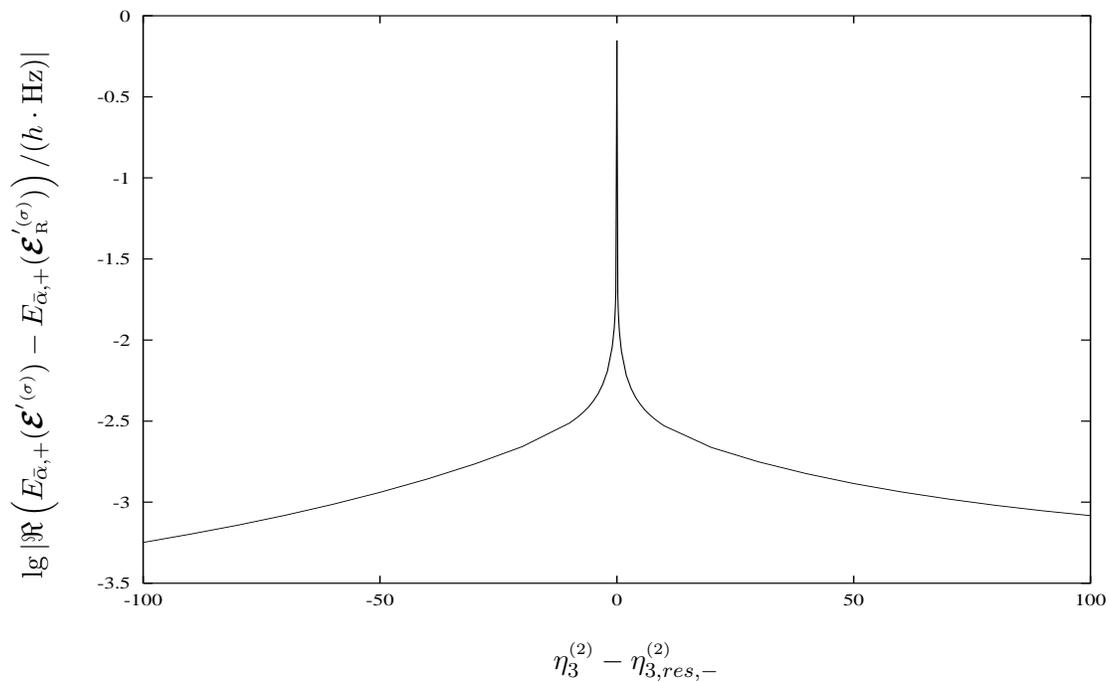
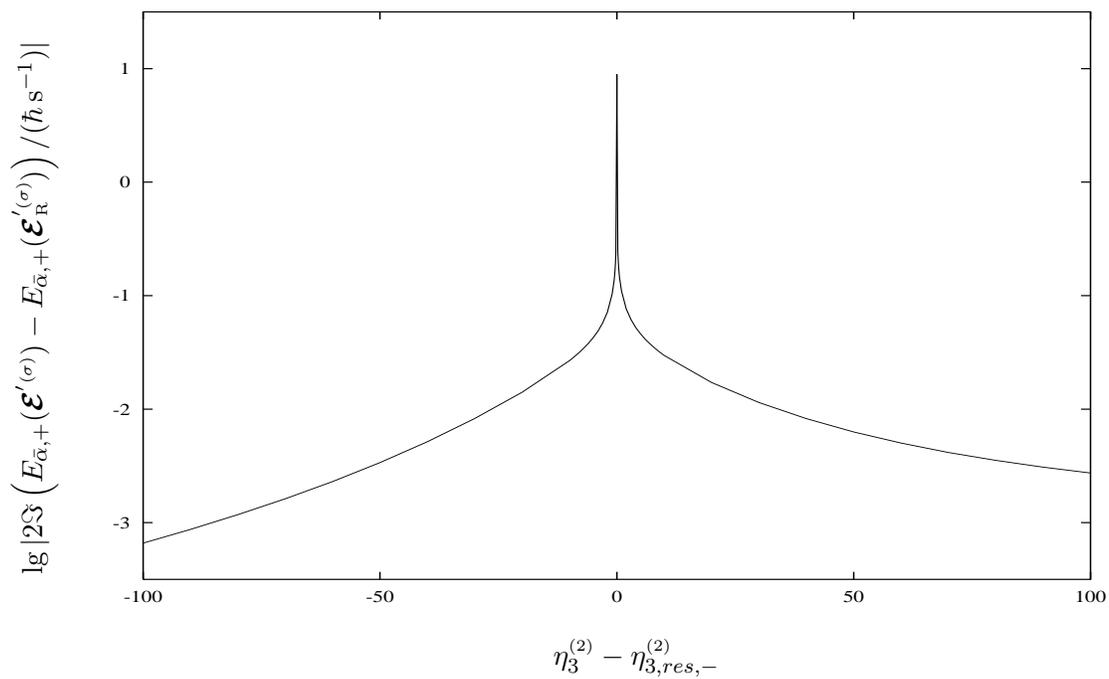
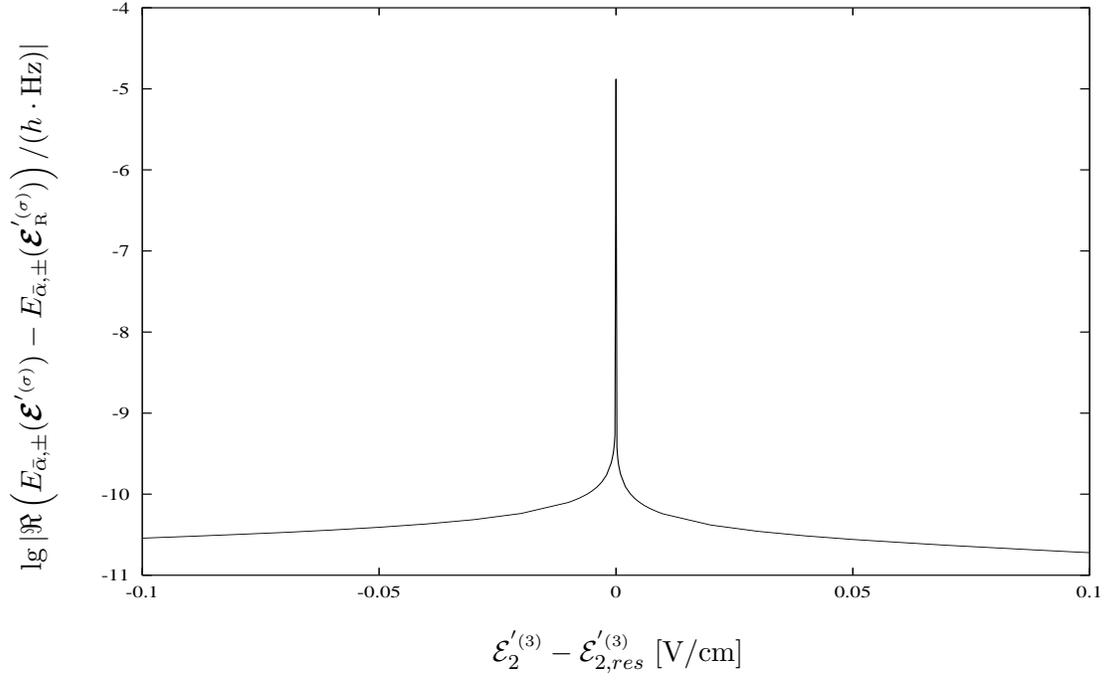
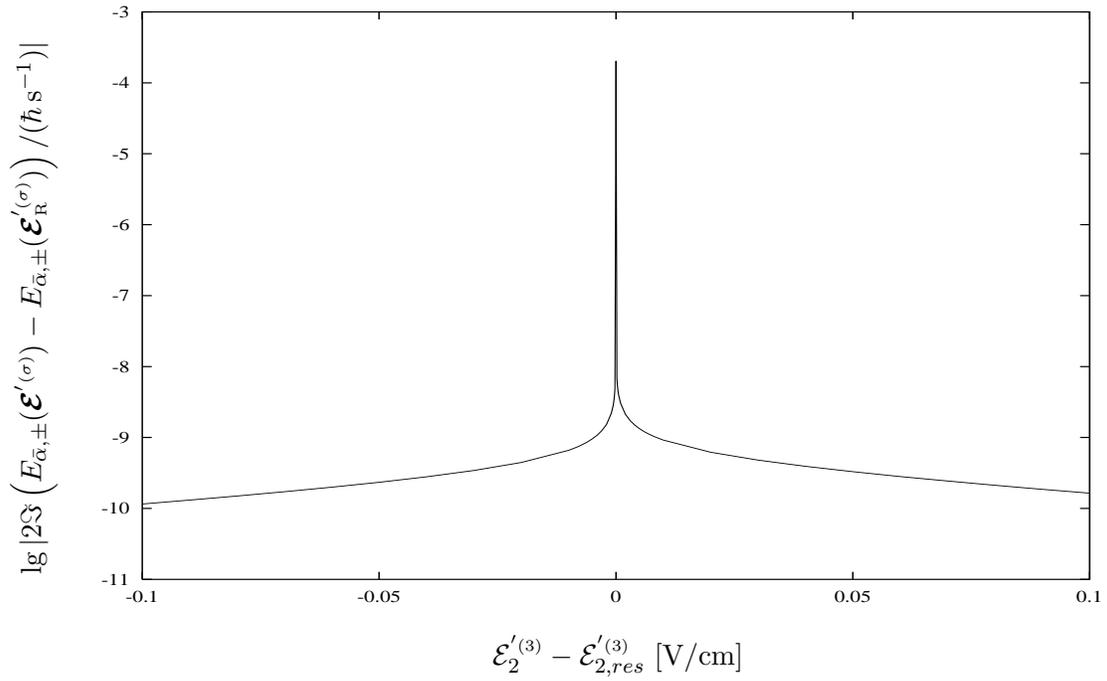
\end{document}